\documentclass[twocolumn]{aastex62}

\usepackage{url}
\usepackage{hyperref}
\usepackage{amsmath, amssymb}
\usepackage{natbib}
\usepackage{ulem}
\usepackage{color}

\submitjournal{\apj}

\begin{document}

\newcommand{\tjmax}{\ensuremath{t_{J {\rm max}}}}
\newcommand{\thmax}{\ensuremath{t_{H {\rm max}}}}
\newcommand{\tkmax}{\ensuremath{t_{K {\rm max}}}}

\newcommand{\trest}{\ensuremath{t_{\rm rest}}}
\newcommand{\tobs}{\ensuremath{t_{\rm obs}}}

\newcommand{\dmlcs}{\ensuremath{\Delta}}

\newcommand{\threesig}{$3$-$\sigma$}
\newcommand{\vdag}{(v)^\dagger}

\newcommand{\OMG}   {\ensuremath{\vec{\Omega}}}
\newcommand{\OM}   {\ensuremath{{\Omega}_{\rm M}}}
\newcommand{\OL}   {\ensuremath{{\Omega}_{\Lambda}}}
\newcommand{\ODM}   {\ensuremath{{\Omega}_{DM}}}
\newcommand{\OB}   {\ensuremath{{\Omega}_{b}}}
\newcommand{\OMT}   {\ensuremath{\Omega}}
\newcommand{\dHo}   {\ensuremath{ \sigma_{\Ho} }}
\newcommand{\jbmax}{\ensuremath{J_{\rm B max}}}
\newcommand{\hbmax}{\ensuremath{H_{\rm B max}}}
\newcommand{\kbmax}{\ensuremath{K_{\rm B max}}}
\newcommand{\mjbmax}{\ensuremath{M_{J_{\rm B max}}}}
\newcommand{\mhbmax}{\ensuremath{M_{H_{\rm B max}}}}
\newcommand{\mkbmax}{\ensuremath{M_{K_{\rm B max}}}}
\newcommand{\bvri}{\ensuremath{BVRI}}
\newcommand{\jhkbmax}{\ensuremath{\jhk_{\rm B max}}}

\newcommand{\snIax}{\mbox{SN~Iax}}
\newcommand{\snIbc}{\mbox{SN~Ib/c}}
\newcommand{\snII}{\mbox{SN~II}}
\newcommand{\chisqnu} {\ensuremath{\chi^2/{\rm DoF}}}
\newcommand{\chisq}   {\ensuremath{\chi^2}}
\newcommand{\kms}     {{}km~s$^{-1}$}
\newcommand{\kmsmpc}  {{}km~s$^{-1}$~Mpc$^{-1}$}
\newcommand{\lcdm}    {\ensuremath{\Lambda{\rm CDM}}}

\newcommand{\ndeg}{\ensuremath{N_{\rm degree}}}

\newcommand{\scale}{1}

\newcommand{\inprep}{$in\ prep.$}
\newcommand{\subm}{$submitted$}
\newcommand{\etal}{et al.}

\newcommand{\rv}{r_{V}}
\newcommand{\Rv}{R_{V}}
\newcommand{\Av}{A_{V}}
\newcommand{\sRv}{\sigma_{R_{V}}}
\newcommand{\sAv}{\sigma_{A_{V}}}

\newcommand{\eps}{ \epsilon_{o} }
\newcommand{\datao}{ {\bf{O}} }

\newcommand{\sn}{SN~}

\newcommand{\WV}{WV08}
\newcommand{\F}{CfAIR2}
\newcommand{\AF}{F14}
\newcommand{\PTL}{PAIRITEL}
\newcommand{\swarp}{SWarp}
\newcommand{\dophot}{DoPHOT}
\newcommand{\mlcs}{MLCS2k2}

\newcommand{\jhcsp}{ (J-H)_{CSP} }
\newcommand{\jhptel}{ (J-H)_{PTEL} }

\newcommand{\fcsp}{\F{}/CSP}
\newcommand{\wcsp}{WV08/CSP}
\newcommand{\fw}{\F{}/WV08}
\newcommand{\fdofdi}{\F{}fdo/fdi}

\newcommand{\nt}{N_{\rm T}}
\newcommand{\nsn}{N_{\rm SN}}
\newcommand{\nnt}{\rm NNT}
\newcommand{\nntwo}{\rm NN2}
\newcommand{\snt}{\rm SNTEMP}

\newcommand{\ptel}{PTL}

\newcommand{\ptl}{{\rm P}}
\newcommand{\twomass}{{\rm 2M}}

\newcommand{\bmath}{\bm}

\newcommand{\colordir}{./}


\newcommand{\myemail}{aavelino@cfa.harvard.edu}

\newcommand{\yjhk}{\ensuremath{YJHK_{s}}}
\newcommand{\yjh}{\ensuremath{YJH}}
\newcommand{\jhk}{\ensuremath{JHK_{s}}}

\newcommand{\newIa}{{\bf YY}}
\newcommand{\Ho}   {\ensuremath{ H_{0} }}

\newcommand{\La}{\Lambda}

\newcommand{\vhel}{\ensuremath{v_{\rm helio}}}
\newcommand{\vcmb}{\ensuremath{v_{\rm CMB}}}
\newcommand{\vcmbf}{\ensuremath{v_{\rm CMB,flow}}}

\newcommand{\zhel}{\ensuremath{z_{\rm helio}}}
\newcommand{\zcmb}{\ensuremath{z_{\rm CMB}}}
\newcommand{\snIa}{\mbox{SN~Ia}}

\newcommand{\TBmaxx}{t_{B {\rm max}}}
\newcommand{\TBmax}{t_{B {\rm max}, s}}

\newcommand{\tbmaxx}{t_{B {\rm max}}}
\newcommand{\tbmax}{t_{B {\rm max}, s}}
\newcommand{\tnirmaxx}{t_{\rm NIR max}}
\newcommand{\tnirmax}{t_{{\rm NIR max}, s}}

\newcommand{\tmaxx}{t_{\rm max}}
\newcommand{\tmax}{t_{{\rm max}, s}}

\newcommand{\appmagmaxHat}{\hat{m}_{{\rm max},s}}

\newcommand{\appmagBmaxHat}{\hat{m}_{B{\rm max},s}}
\newcommand{\sappmagmaxHat}{\hat{\sigma}_{{\rm max},s}}

\newcommand{\appmagNIRmax}{m_{{\rm NIR max},s}}
\newcommand{\appmagNIRmaxHat}{\hat{m}_{{\rm NIR max},s}}
\newcommand{\sappmNIRmax}{\sigma^2_{{\rm NIR max}, s}}
\newcommand{\sappmNIRmaxHat}{\hat{\sigma}_{{\rm NIR max}, s}}

\newcommand{\snoopy}{SNooPy}
\newcommand{\bayesn}{\textsc{BayeSN}}
\newcommand{\dm}{\ensuremath{\Delta m_{15}(B)}}
\newcommand{\nvpecfid}{150}

\newcommand{\ebvmw}{E(B-V)_{\rm MW}}

\newcommand{\ebvhost}{E(B-V)_{\rm host}}

\newcommand{\figs}{Figs.}
\newcommand{\eq}{Eq.}
\newcommand{\eqs}{Eqs.}

\newcommand{\sigmaint}{\sigma_{\rm int}}
\newcommand{\sigmainthat}{\hat{\sigma}_{\rm int}}

\newcommand{\sigmafit}{\sigma_{{\rm fit},s}}
\newcommand{\sigmafithat}{\hat{\sigma}_{{\rm fit},s}}

\newcommand{\sigmamuphoto}{\sigma_{\mu, s}}
\newcommand{\sigmamuphotohat}{\hat{\sigma}_{\mu, s}}

\newcommand{\sigmaResidual}{\sigma_{\Delta,s}}
\newcommand{\sigmaResidualHar}{\hat{\sigma}_{\Delta,s}}

\newcommand{\mupec}{\mu_{\rm pec}}
\newcommand{\smupec}{\sigma_{\mupec,s}}
\newcommand{\smupecNoS}{\sigma_{\mupec}}

\newcommand{\sigmaVpec}{\sigma_{\rm pec}}

\newcommand{\muind}{\mu_{\rm eff}}
\newcommand{\dmuind}{\sigma_{\muind}}

\newcommand{\muLCDM}{\mu_{\Lambda{\rm CDM}}}

\newcommand{\likelihood}{\mathcal{L}}
\newcommand{\GaussDistr}{\mathcal{N}}

\newcommand{\Ivector}{\mathbf{1}}

\newcommand{\mean}{\mathbb{E}}

\newcommand{\cov}{\mathrm{Cov}}

\newcommand{\var}{\mathrm{Var}}

\newcommand{\data}{\mathcal{D}}

\newcommand{\nLCs}{n}

\newcommand{\nGPgrid}{n^*}

\newcommand{\nSNHD}{N_{\rm SN}}

\newcommand{\nSNT}{N_{\rm T}}

\newcommand{\nSNTast}{N^*_{\rm T}}

\newcommand{\latMVec}{\boldsymbol{\mathcal{M}}}
\newcommand{\latM}{\mathcal{M}}

\newcommand{\normaLCVec}{\mathbf{L}}
\newcommand{\normaLC}{L}

\newcommand{\postGPmeanVec}{\boldsymbol{\mu}^{\rm post}}
\newcommand{\postGPmean}{\mu_i^{\rm post}}

\newcommand{\meanNormaLCVec}{\boldsymbol{\mu}^{\rm L}}
\newcommand{\meanNormaLC}{\mu^{\rm L}}

\newcommand{\postGPcovMatrix}{\mathbf{\Sigma}^{\rm post}}
\newcommand{\postGPcov}{\Sigma_{ij}^{\rm post}}

\newcommand{\covNormaLCMatrix}{\mathbf{\Sigma}^{\rm L}}
\newcommand{\covNormaLC}{\Sigma_{ij}^{\rm L}}

\newcommand{\sigmakGPkernel}{\sigma_K}

\newcommand{\MeanAbsMag}{\eta_s}
\newcommand{\AbsMagTilde}{\MeanAbsMag}

\newcommand{\AbsMagTildeSigmaSq}{\sigma^2_{\eta,s}}
\newcommand{\AbsMagTildeSigma}{\sigma_{\eta,s}}

\newcommand{\hypermeanHBM}{\theta}

\newcommand{\hyperStdDevHBM}{\sigma_{\theta}}

\newcommand{\residualAbsMags}{M_{{\rm r}, s}}
\newcommand{\residualAbsMagsVec}{\mathbf{M}_{{\rm r}, s}}
\newcommand{\residualAbsMagsMean}{\bar{M}_{{\rm r}, s}}
\newcommand{\residualAbsMagsMeanVec}{\bar{\mathbf{M}}_{{\rm r}, s}}

\newcommand{\nCepheid}{7}

\newcommand{\nSBFTF}{2}

\newcommand{\nTotalInitialSample}{177}

\newcommand{\nsnIa}{89}
\newcommand{\nY}{$44$}
\newcommand{\nJ}{$87$}
\newcommand{\nH}{$81$}
\newcommand{\nK}{$32$}
\newcommand{\nKgp}{14}

\newcommand{\Y}{$Y$}
\newcommand{\J}{$J$}
\newcommand{\HH}{$H$}
\newcommand{\K}{$K_s$}
\newcommand{\AnyNIR}{any $YJHK_s$}
\newcommand{\JH}{$JH$}
\newcommand{\YJH}{$YJH$}
\newcommand{\JHK}{$JHK_s$}
\newcommand{\SALTminusY}{SALT2 $-Y$}
\newcommand{\SALTminusJ}{SALT2 $-J$}
\newcommand{\SALTminusH}{SALT2 $-H$}
\newcommand{\SALTminusK}{SALT2 $-K_s$}
\newcommand{\SALTminusAnyNIR}{SALT2 $-$ any $YJHK_s$}
\newcommand{\SALTminusJH}{SALT2 $-JH$}
\newcommand{\SALTminusYJH}{SALT2 $-YJH$}
\newcommand{\SALTminusJHK}{SALT2 $-YJHK_s$}
\newcommand{\SNOOPYminusY}{\snoopy$-Y$}
\newcommand{\SNOOPYminusJ}{\snoopy$-J$}
\newcommand{\SNOOPYminusH}{\snoopy$-H$}
\newcommand{\SNOOPYminusK}{\snoopy$-K_s$}
\newcommand{\SNOOPYminusAnyNIR}{\snoopy$-$any $YJHK_s$}
\newcommand{\SNOOPYminusJH}{\snoopy$-JH$}
\newcommand{\SNOOPYminusYJH}{\snoopy$-YJH$}
\newcommand{\SNOOPYminusJHK}{\snoopy$-YJHK_s$}

\newcommand{\nAnyYJHKgp}{56}
\newcommand{\sigmaIntAnyYJHKgp}{0.047 \pm 0.018}
\newcommand{\sigmaIntAnyYJHKgpAtTBmaxGPsubsample}{0.066 \pm 0.016}
\newcommand{\nAnyYJHKtemp}{89}
\newcommand{\sigmaIntAnyYJHKTemp}{0.123 \pm 0.014}
\newcommand{\sigmaIntAnyYJHKTempGPsubsample}{0.112 \pm 0.016}
\newcommand{\sigmaIntSALT}{0.133 \pm 0.022}
\newcommand{\sigmaIntSnoopy}{0.128 \pm 0.018}
\newcommand{\sigmaIntHgp}{0.032 \pm 0.027}
\newcommand{\sigmaIntYtemp}{0.105 \pm 0.018}

\newcommand{\wrmsAnyYJHKgp}{0.100 \pm 0.013}
\newcommand{\wrmsAnyYJHKgpAtTBmaxGPsubsample}{0.106 \pm 0.010}
\newcommand{\wrmsAnyYJHKTempGPsubsample}{0.140 \pm 0.016}
\newcommand{\wrmsSALT}{0.174 \pm 0.020}
\newcommand{\wrmsSnoopy}{0.159 \pm 0.019}
\newcommand{\wrmsHgp}{0.095 \pm 0.010}
\newcommand{\wrmsYtemp}{0.139 \pm 0.013}

\newcommand{\rmsSALT}{0.179 \pm 0.018}
\newcommand{\rmsSnoopy}{0.174 \pm 0.021}
\newcommand{\rmsAnyYJHKgp}{0.117 \pm 0.014}
\newcommand{\rmsAnyYJHKgpBmax}{0.115 \pm 0.011}
\newcommand{\rmsAnyYJHKTempGPsample}{0.138 \pm 0.014}

\newcommand{\rmsGPNIRmaxSmallest}{0.087 \pm 0.013}
\newcommand{\rmsGPNIRmaxSmallestBand}{\YJH}
\newcommand{\rmsGPNIRmaxLargest}{0.179 \pm 0.029}
\newcommand{\rmsGPNIRmaxLargestBand}{\K}

\newcommand{\rmsGPNIRmaxOneBandSmallest}{0.111 \pm 0.018}
\newcommand{\rmsGPNIRmaxOneBandSmallestBand}{\Y}
\newcommand{\rmsGPNIRmaxOneBandLargest}{0.179 \pm 0.029}
\newcommand{\rmsGPNIRmaxOneBandLargestBand}{\K}

\newcommand{\rmsTempSmallest}{0.137 \pm 0.018}
\newcommand{\rmsTempSmallestBand}{\YJH}
\newcommand{\rmsTempLargest}{0.207 \pm 0.020}
\newcommand{\rmsTempLargestBand}{\K}

\newcommand{\rmsTempOneBandSmallest}{0.152 \pm 0.016}
\newcommand{\rmsTempOneBandSmallestBand}{\Y}
\newcommand{\rmsTempOneBandLargest}{0.207 \pm 0.020}
\newcommand{\rmsTempOneBandLargestBand}{\K}

\newcommand{\rmsGPBmaxSmallest}{0.098 \pm 0.014}
\newcommand{\rmsGPBmaxSmallestBand}{\YJH}
\newcommand{\rmsGPBmaxLargest}{0.170 \pm 0.027}
\newcommand{\rmsGPBmaxLargestBand}{\K}

\newcommand{\rmsGPBmaxOneBandSmallest}{0.118 \pm 0.017}
\newcommand{\rmsGPBmaxOneBandSmallestBand}{\Y}
\newcommand{\rmsGPBmaxOneBandLargest}{0.170 \pm 0.027}
\newcommand{\rmsGPBmaxOneBandLargestBand}{\K}

\newcommand{\optminusnirsigmabest}{3.1}
\newcommand{\optminusnirsigmabestBand}{\SNOOPYminusAnyNIR}

\newcommand{\saltminusnirgp}{0.086 \pm 0.028}
\newcommand{\wrmsdiffSALTAnyYJHK}{0.074 \pm 0.024}
\newcommand{\saltminusnirgpAnyYJHKrms}{0.062 \pm 0.023}

\newcommand{\saltminusnirgpAnyYJHKrmsNS}{2.7}

\newcommand{\snoopyminusnirgp}{0.080 \pm 0.026}
\newcommand{\wrmsdiffSnoopyAnyYJHK}{0.059 \pm 0.023}
\newcommand{\snoopyminusnirgpAnyYJHKrms}{0.057 \pm 0.025}

\newcommand{\snoopyminusnirgpAnyYJHKrmsNS}{2.3}

\newcommand{\optminusnirsigma}{$\sim 2.5$-$3.1\sigma$}

\newcommand{\optminusnirsigmaGeneral}{$\sim 0.8$-$3.1\sigma$}

\newcommand{\optminusnirRMSnsSmallest}{1.3}
\newcommand{\optminusnirRMSnsSmallestBand}{\SNOOPYminusJ}

\newcommand{\optminusnirRMSnsLargest}{4.1}
\newcommand{\optminusnirRMSnsLargestBand}{\SALTminusYJH}
\newcommand{\optminusnirRMSdeltaLargest}{0.09}

\newcommand{\optminusnirRMSnsSmallestCombined}{2.3}
\newcommand{\optminusnirRMSnsSmallestCombinedBands}{\SNOOPYminusAnyNIR}

\newcommand{\optminusnirRMSnsLargestCombined}{4.1}
\newcommand{\optminusnirRMSnsLargestCombinedBands}{\SALTminusYJH}

\newcommand{\wrmsNIRmaxK}{0.163}

\newcommand{\optminusnirsigmaJWorstwRMS}{0.9}

\newcommand{\optminusnirsigmaJWorstRMS}{1.3}

\newcommand{\nirgpbestscatter}{$\sim 0.03$-$0.11$}

\newcommand{\BmaxminusNIRmaxLargestNS}{0.99}
\newcommand{\BmaxminusNIRmaxLargestNSBand}{\JH}

\newcommand{\BmaxminusNIRmaxLargestNSrms}{0.58}
\newcommand{\BmaxminusNIRmaxLargestNSrmsBand}{\YJH}
\newcommand{\BmaxminusNIRmaxLargestDeltarms}{0.01}

\newcommand{\MBfidSALT}{-19.44}

\newcommand{\Hounits}{km\,s$^{-1}$\,Mpc$^{-1}$}

\newcommand{\vskipps}{-0.8cm}

\newcommand{\lc}{LC}
\newcommand{\lcs}{LCs}

\newcommand{\andy}[1]{\textcolor{blue}{\bf [ASF: #1]}}
\newcommand{\andyin}[1]{\textcolor{blue}{#1}}

\newcommand{\artin}[1]{\textcolor{black}{#1}}
\newcommand{\artinb}[1]{\textcolor{black}{#1}}

\newcommand{\orcidwidth}{0.1in}

\title{\textsc{Type Ia Supernovae are Excellent Standard Candles in the Near-Infrared}}
\shorttitle{Type Ia Supernovae are Excellent Standard Candles in the Near-Infrared}
\shortauthors{Avelino et al.}

\author{Arturo~Avelino
\href{https://orcid.org/0000-0002-2938-7822}{\includegraphics[width=\orcidwidth]{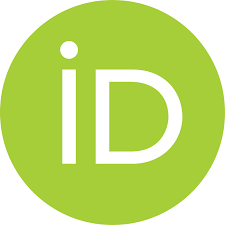}}}
\affiliation{Harvard-Smithsonian Center for Astrophysics, 60 Garden Street, Cambridge, MA 02138}
\email{aavelino@cfa.harvard.edu}

\author{Andrew~S.~Friedman
\href{https://orcid.org/0000-0003-1334-039X}{\includegraphics[width=\orcidwidth]{./figures/orcid.png}}}
\affiliation{University of California, San Diego, La Jolla, California 92093, USA}
\email{asf@ucsd.edu}

\author{Kaisey~S.~Mandel
\href{https://orcid.org/0000-0001-9846-4417}{\includegraphics[width=\orcidwidth]{./figures/orcid.png}}}
\affiliation{Institute of Astronomy and Kavli Institute for Cosmology, Madingley Road, Cambridge, CB3 0HA, UK}
\affiliation{Statistical Laboratory, DPMMS, University of Cambridge, Wilberforce Road, Cambridge, CB3 0WB, UK}

\author{David~O.~Jones
\href{https://orcid.org/0000-0002-6230-0151}{\includegraphics[width=\orcidwidth]{./figures/orcid.png}}}
\affiliation{University of California, Santa Cruz, Santa Cruz, California 95064, USA}

\author{Peter~J.~Challis}
\affiliation{Harvard-Smithsonian Center for Astrophysics, 60 Garden Street, Cambridge, MA 02138}

\author{Robert~P.~Kirshner
\href{https://orcid.org/0000-0002-1966-3942}{\includegraphics[width=\orcidwidth]{./figures/orcid.png}}}
\affiliation{Harvard-Smithsonian Center for Astrophysics, 60 Garden Street, Cambridge, MA 02138}
\affiliation{Gordon and Betty Moore Foundation, 1661 Page Mill Road, Palo Alto, CA 94304}

\date{\today}

\begin{abstract}
We analyze a set of \nsnIa{} Type Ia supernovae (\snIa) that have both optical and near-infrared (NIR) photometry to derive distances and construct low redshift ($z \le 0.04$) Hubble diagrams. We construct mean light curve (LC) templates using a hierarchical Bayesian model. We explore both Gaussian process (GP) and template methods for fitting the \lcs{} and estimating distances, while including peculiar velocity and photometric uncertainties. For the \nAnyYJHKgp{} \snIa{} with both optical and NIR observations near maximum light, the GP method yields a NIR-only Hubble-diagram with a RMS of $\rmsAnyYJHKgp $ mag when referenced to the NIR maxima.
For each NIR band, a comparable GP method RMS is obtained when referencing to NIR-max or $B$-max. Using NIR \lc{} templates referenced to $B$-max yields a larger RMS value of $\rmsAnyYJHKTempGPsample$ mag. Fitting the corresponding optical data using standard \lc{} fitters that use \lc{} shape and color corrections yields larger RMS values of $\rmsSALT $ mag with SALT2 and $\rmsSnoopy $ mag with \snoopy{}.
Applying our GP method to subsets of \snIa{} NIR \lcs{} at NIR maximum light, even without corrections for \lc{} shape, color, or host-galaxy dust reddening, provides smaller RMS in the inferred distances, at the $\sim \optminusnirRMSnsSmallestCombined$-$\optminusnirRMSnsLargestCombined \sigma$ level, than standard optical methods that do correct for those effects. Our ongoing RAISIN program on the Hubble Space Telescope will exploit this promising infrared approach to limit systematic errors when measuring the expansion history of the universe to constrain dark energy.
\end{abstract}


\section{Introduction}
\label{sec:intro}
The increasing sample of high quality, low-redshift (low-$z$), near-infrared (NIR) light curves (\lcs) of Type Ia supernovae (\snIa) provides an opportunity to further investigate their utility as cosmological standard candles.
Optical samples of \snIa{} are large enough now that systematic uncertainties are major limitation to accurate cosmological constraints. Infrared observations of \snIa{} can help in that essential way because supernovae are more nearly standard candles in the NIR and the effects of dust are diminished. This paper explores ways to use NIR observations of \snIa{} to measure distances.  This investigation is for a low-$z$ sample, but we are working to extend this technique to cosmologically-interesting distances with the Hubble Space Telescope (HST).

Before NIR photometry became practical for large samples of \snIa, photometry and spectroscopy of \snIa{} at optical wavelengths enabled the unexpected 1998 discovery of cosmic acceleration (\citealt{riess98,schmidt98,perlmutter99}). Since then, a suite of independent cosmological methods has confirmed the \snIa{} results (see \citealt{frieman08,weinberg13} for reviews). The prevailing view is that the mechanism behind cosmic acceleration is some form of dark energy. The constraints on cosmological parameters from the \snIa{} Pantheon sample (\citealt{scolnic18}) combined with the Planck 2015/2018 Cosmic Microwave Background data (\citealt{ade16,planck18}), as well as Baryon Acoustic Oscillations (\citealt{alam17}) and local Hubble constant measurements (\citealt{riess16,riess18a,riess18b,riess18c}) are consistent with this view. Among the major cosmological techniques, \snIa{} provide precise measurements of extragalactic distances and the most direct evidence for cosmic acceleration (see \citealt{goobar11,kirshner13,goobar15,davis16,riess18a} for reviews).

Optical \snIa{} \lcs{} are known to be excellent {\it standardizable} candles that exploit correlations between intrinsic luminosity and \lc{} shape and color (\citealt{phillips93,phillips99,hamuy96,riess96,riess98,perlmutter97,goldhaber01,tonry03,wang03,prieto06,jha06,jha07,astier06,takanashi08,conley08,mandel09,guy05,guy07,guy10,mandel11,mandel17}). Recent work has demonstrated that \snIa{} in the NIR
are more nearly {\it standard} candles, even before correction for \lc{} shape or host galaxy dust reddening (e.g.~\citealt{krisciunas04a,woodvasey08,mandel09,krisciunas09,friedman12,kattner12}). NIR \lcs{} are $\sim$5--11 times less sensitive to dust extinction than optical $B$-band data (\citealt{cardelli89}). When constructing \snIa{} Hubble diagrams using NIR data, the distance errors produced by extinction are small: ignoring dust would be fatal for optical studies, but nearly not as serious for NIR studies like \citealt{woodvasey08} or the present work.  An improved approach would use optical and infrared data simultaneously to determine the extinction (\citealt{mandel11}).

Optical-only samples yield typical Hubble diagram intrinsic scatter of $\sigmaint \sim 0.12$ mag and a RMS of 0.141 mag after applying light-curve shape, host-galaxy dust, and host-galaxy mass corrections, assuming a peculiar-velocity uncertainty of 250 km s$^{-1}$ (e.g. \citealt{scolnic18}). For simplicity, we adopt a conservative peculiar-velocity uncertainty for the host galaxies in our sample of 150 km s$^{-1}$.  If the typical redshifts in the sample were large enough, this would be of no consequence, but for our nearby sample, the inferred intrinsic scatter of the supernova luminosities depends on the value we choose. As a result, though we have confidence when comparing the RMS and intrinsic scatter for various subsamples containing the same SN with both optical and infrared data, the real value of the scatter should be determined from observations that are securely in the Hubble flow beyond 10,000 km s$^{-1}$.

When including a peculiar-velocity uncertainty of 150 km s$^{-1}$, our best method yields intrinsic scatters as small as \nirgpbestscatter{} mag, depending on the NIR filter subset, and a RMS of $\sim \rmsGPNIRmaxSmallest$ mag for the best NIR $YJH$-band subset, confirming and strengthening previous results for NIR methods (\citealt{meikle00,krisciunas04a,krisciunas05a,krisciunas07,folatelli10,BurnsEtal2011Snoopy,woodvasey08,mandel09,mandel11,kattner12,dhawan15}). Assuming a larger peculiar-velocity uncertainty, such as 250 km s$^{-1}$, makes our estimated intrinsic scatter even {\it smaller}. In addition, our best NIR method using any of the $\yjhk$ bands yields an RMS of only $\rmsAnyYJHKgp$ mag, compared to $\rmsSALT $ mag and $\rmsSnoopy$ mag for SALT2 and \snoopy{} fits to optical $BVR$ data for the same \nAnyYJHKgp{} \snIa, respectively. While using \lc{} shape, color, and host galaxy dust corrections would likely lead to improvements, the simpler approaches in this paper are still remarkably effective.

Overall, a substantial body of evidence indicates that rest-frame {\lcs{} of} \snIa{} in NIR are both better standard candles than at optical wavelengths and less sensitive to the confounding effects of dust. When NIR data are combined with $UBVRI$ photometry, this yields accurate and precise distance estimates (\citealt{krisciunas04b,krisciunas07,woodvasey08,folatelli10,BurnsEtal2011Snoopy,friedman12,phillips12,kattner12,burns14,mandel09,mandel11,mandel14a,mandel17}).

This is significant for supernova cosmology because, along with photometric-calibration uncertainties (\citealt{scolnic15,foley18}), uncertain dust extinction estimates and the intrinsic variability of \snIa{} colors present challenging and important systematic problems for dark energy measurements (\citealt{wang06,jha07,woodvasey07,hicken09a,kessler09,guy07,guy10,conley07,conley11,komatsu11,campbell13,rest13,scolnic13,narayan13,BetouleEtal2014_JLA,rest14,mosher14,scolnic14,scolnic14b,scolnic15,narayan16a,scolnic17,mandel17,foley18,scolnic18,DES_Brout_2019,DES_Kessler_2019}).
Combining optical and NIR \lcs{} promises to reduce these systematic distance uncertainties
(\citealt{folatelli10,BurnsEtal2011Snoopy,kattner12,mandel11,mandel14a}).

This work is organized as follows. In \S\ref{sec:prev}, we review previous results with \snIa{} in NIR, detail our analysis selection criteria, and discuss host galaxy redshifts. In \S\ref{SecTemplate}, we outline our Gaussian process (GP) procedure to fit \lcs{} and our hierarchical Bayesian model to construct mean $\yjhk$ \lc{} templates. In \S\ref{SecHubbleDiagram}, we use these templates and GP fits to individual \lcs{} to construct Hubble diagrams in each NIR band, as well as a combined \yjhk{} NIR Hubble diagram. We compare this to optical $BVR$ Hubble diagrams for the very same set of \nAnyYJHKgp{} supernovae that use the SALT2 and \snoopy{} LC fitters.  We end with \S\ref{sec_disc} by documenting how, even without correcting for LC shape or dust, \snIa{} in the NIR using our GP fits at NIR maximum are better standard candles than optical \snIa{} observations corrected for these effects.  Mathematical details of the Gaussian process, the hierarchical Bayesian model, and the method for determining the intrinsic scatter are presented in the Appendices.

\section{\snIa{} in NIR as Standard Candles}
\label{sec:prev}

Pioneering studies by \citet{meikle00} and \citet{krisciunas04a} demonstrated that \snIa{} have smaller luminosity variation in the NIR \jhk{} bands than in the optical $BV$ bands at the time of $B$-band maximum light ({$t_{\rm Bmax}$}). \citet{krisciunas04a} found that optical \lc{} shape and intrinsic NIR luminosity were uncorrelated in a sample of 16 \snIa{}, while measuring a NIR absolute magnitude scatter of $\sigma_J=0.14$, $\sigma_H=0.18$, and $\sigma_{K_s}=0.12$~mag. Following this, \citet{woodvasey08} used a homogeneously-observed sample of 18 spectroscopically-normal \snIa{} in the \jhk{} bands, with intrinsic root-mean-square (RMS) absolute magnitudes of $0.15$ mag in the $H$-band, {\it without applying any reddening or \lc{} shape corrections}. By combining these 18 objects with 23 \snIa{} from the literature, the sample in \citet{woodvasey08} yielded an $H$-band RMS of $0.16$ mag, strengthening the evidence that normal \snIa{} are excellent NIR standard candles. In the present work, we show that \snIa{} in NIR yield a narrow distribution of $\yjhk$ peak magnitudes with RMS Hubble Diagram scatter as small as $\rmsGPNIRmaxSmallest $ mag for the combined $\yjh$ bands and as large as $\rmsGPNIRmaxLargest$ mag for the $K_s$ band, consistent with previous results.

Following \citealt{woodvasey08}, \citealt{mandel09} developed a new hierarchical Bayesian model (\bayesn{}) and a template model to account for $J$-band \lc{} shape variation to the existing \snIa{} in NIR sample, finding a marginal scatter in the peak absolute magnitudes of $0.17$, $0.11$, and $0.19$ mag, in \jhk{}, respectively, while finding that $J$-band \lc{} shape does correlate with NIR intrinsic luminosity. Subsequent work by \citealt{folatelli10} applied a different \lc{} shape correction method, but found scatters of $0.12$--$0.16$ mag in $\yjhk$, consistent with the results of \cite{mandel09}.

Additional work by \cite{kattner12} found an absolute magnitude scatter of $0.12$, $0.12$, and $0.09$ mag in \yjh, respectively, by analyzing a subset of 13 well-sampled normal NIR \snIa{} \lcs{} with relatively little host galaxy dust extinction. \citealt{kattner12} also showed evidence for a correlation between the $JH$-band absolute magnitudes at $t_{\rm Bmax}$ and, {$\dm$}, the light-curve decline rate parameter in $B$-band after 15 days of $t_{\rm Bmax}$  (\citealt{phillips93}), with no evidence for strong correlation in the $Y$-band. This is also consistent with the results of \citealt{mandel09}, who found that $J$-band \lc{} shape and luminosity are correlated.

Using a small data set of 12 \snIa{} $JH$-band \lcs, each with only $3$-$5$ data points, \citealt{barone12,barone13} find a scatter of $0.116$ mag and $0.085$ mag in the $J$ and $H$-bands, respectively. In the first data release of the SweetSpot survey, \citealt{weyant14} present a similarly small sample of 13 low-$z$ \snIa{}, each with $1$-$3$ \lc{} points, finding an $H$-band scatter of $0.164$ mag. This was followed by a second SweetSpot data release, which included a total of 33 \snIa{} with 168 \jhk{} observations in the redshift range $0.02 < z < 0.09$, well into the smooth Hubble flow, but which did not yet include NIR Hubble diagrams (\citealt{weyant18}).

By analyzing 45 NIR LCs with data near NIR-maximum, \citealt{stanishev18} find an intrinsic Hubble diagram scatter of $\sim 0.10$ mag, after accounting for potential new correlations between light curve shape, color excess, and $J-H$ color at NIR-max. \citealt{stanishev18} also present single-epoch $JH$ photometry for 16 new \snIa{} with $z > 0.037$. The Carnegie Supernova Project (CSP) final data release (CSP-I; \citealt{krisciunas17}), was recently analyzed in \cite{Burns2018_Ho}, which found peculiar velocity corrected Hubble diagram dispersions of $\sim 0.08-0.15$ mag, depending on the subset of the 120 \snIa{} they considered. Additional CSP-II photometric data, to be published in 2019, was recently described in \citealt{phillips19}. \citet{hsiao19} present an overview of the NIR \snIa{} spectroscopy obtained by the CSP and the Center for Astrophysics (CfA) Supernova Group.

While the current sample of optical \snIa{} \lcs{} exceeds 1000 (\citealt{scolnic18}), and will be increased by orders of magnitude by ongoing and future surveys including the Dark Energy Survey (DES; \citealt{DES_Abbott_2019b,DES_Abbott_2019a,DES_Brout2018b,DES_DAndrea_2018}), the Zwicky Transient Facility (ZTF; \citealt{smithrm14}), and the Large Synoptic Survey Telescope (LSST; \citealt{ivezic08,zhan17}), the number of normal \snIa{} with published NIR \lcs{} is still less than $250$. Nevertheless, the NIR sample has the potential to improve systematics compared to optical-only \snIa{} cosmology samples, which are already systematics limited
(\citealt{scolnic18}).

Overall, the growing sample of photometric data suggests that NIR observations of \snIa{} present a promising path to standardize \snIa{} for distance estimates (\citealt{dhawan15,shariff16,Burns2018_Ho,stanishev18}), Hubble constant estimates (\citealt{cartier14,efstathiou14,riess16,cardona17,dhawan18,Burns2018_Ho}), and eventually, cosmological parameter estimates, when the nearby and
high-$z$ samples are combined as in the HST RAISIN program
({\bf RAISIN}: {\it Tracers of cosmic expansion with {\bf SN IA} in the {\bf IR}}, PI. R. Kirshner, HST GO-13046, GO-14216).

\subsection{Nearby \snIa{} in NIR Sample and Data Cuts}
\label{sec:cuts}

This work analyzes a suitable~subset including \nsnIa{} objects from the current sample of low-redshift photometric data for \snIa{} NIR $\yjhk$-band \lcs{} including data releases 1 and 2 from the Carnegie Supernova Project (\citealt{schweizer08,contreras10,stritzinger10,stritzinger11,taddia12}), now superseded by CSP data release 3 (\citealt{krisciunas17}), the CfA (\citealt{woodvasey08,friedman12,friedman15}), and other groups (e.g. \citealt{krisciunas00,krisciunas04b,krisciunas04c,krisciunas05a,krisciunas07}). We limit our analysis to spectroscopically normal \snIa{} from Table~3 of \citealt{friedman15}, plus the definitive version of the CSP-I DR3 sample of low-$z$ \snIa{} (\citealt{krisciunas17}), and other groups. Additional CSP-II photometric data, to be published in 2019, was recently described in \citealt{phillips19} and will be analyzed in future work.
We apply the following data cuts to analyze a subset of \nsnIa{} \snIa{} with NIR data. Table \ref{Table_DataCuts} shows how the initial sample of \nTotalInitialSample{} \snIa{} decreases after applying the different cuts, and Table~\ref{Table_LC_params} lists the general properties of the remaining \nsnIa{} \snIa{}. \artin{We determine $\dm$ and $\ebvhost$ with \snoopy{}}.

\begin{itemize}
\item Optical light curve shape parameter $0.8 < \dm < 1.6$, to consider {\it normal} \snIa{} only (\citealt{hicken09b}). Objects must have accompanying $B$-band optical data to measure $\dm$.
\item Host galaxy reddening: $-0.15<\ebvhost < 0.4$. \artin{This cut is inspired by} the standard SALT2 cut in color, $-0.3< c <0.3$, in optical-only analysis (\citealt{BetouleEtal2014_JLA,scolnic18}) but with a less stringent cut considering that \snIa{} in the NIR are less sensitive to dust.
\item One advantage of the relative NIR insensitivity to dust reddening is that it also allows us to set a large threshold for Milky Way color excess: $\ebvmw < 1$ mag, to exclude highly reddened \snIa{}. All \nTotalInitialSample{} \snIa{} in the sample passed this cut. SN2006lf with $\ebvmw= 0.8135$ mag has the largest color excess in the initial sample.
\item Redshift range: $z < 0.04$. The maximum redshift cut limits the effects of Malmquist bias. Section~\ref{sec:z} describes corrections to deal with \snIa{} at $z<0.01$, that suffer from peculiar velocity bias.
\item Duplicates: For a given supernova observed by multiple surveys, we use the CSP data (\citealt{krisciunas17}), which typically has smaller photometric uncertainties than the CfA PAIRITEL data (\citealt{friedman15}).
\item We include only spectroscopically normal \snIa{} as identified by the Supernova Identification Code (SNID) \cite{blondin07}.
\item At least 3 photometric points in a given band for each \snIa{} LC. A large fraction of the NIR data from \citet{barone12}, \citet{stanishev18}, and the SweetSPOT survey with WIYN (\citealt{weyant14,weyant18}) did not meet this criterion, so we chose not to analyze these data in this work.
\end{itemize}

\begin{table}
\begin{center}
\caption{Data cuts}
\begin{tabular}{l c}
\hline \hline
Cuts & \# SN Ia after cuts  \\
\hline
Initial sample & 177 \\
$0.8<\Delta m_{15}<1.6$ & 138  \\
$-0.15<\ebvhost < 0.4$ & 122  \\
$\ebvmw < 1$ & 122  \\
$z_{\rm CMB} < 0.04$  & 111    \\
Remove duplicates & 100   \\
Normal spectrum  & 95   \\
$\ge$ 3 LC points  & 89   \\
\hline
\multicolumn{2}{l}{Reduction of the initial sample based on data cuts}\\
\end{tabular}
\label{Table_DataCuts}
\end{center}
\end{table}

\renewcommand{\arraystretch}{0.001}
\renewcommand{\baselinestretch}{0.8}
\begin{table*}[h!]
\begin{center}
\caption{\snIa{} Light Curve Parameters}
\label{Table_LC_params}
\tiny
\begin{tabular}{l cccc rrrr}
\hline \hline
\multicolumn{1}{c}{SN name} & \multicolumn{1}{c}{$z_{\rm (helio)}^a$}  & \multicolumn{1}{c}{$\zcmb^b$} & \multicolumn{1}{c}{$\sigma_{\rm pec}^c$} & \multicolumn{1}{c}{LC Data} & \multicolumn{1}{c}{$\TBmaxx^e$} & \multicolumn{1}{c}{$\dm^f$} & \multicolumn{1}{c}{$E(B-V)_{\rm host}^g$}  & \multicolumn{1}{c}{$E(B-V)_{\rm MW}^h$} \\
\multicolumn{1}{c}{} & \multicolumn{1}{c}{} & \multicolumn{1}{c}{} & \multicolumn{1}{c}{(mag)}  & \multicolumn{1}{c}{Source$^d$} & \multicolumn{1}{c}{(MJD days)} & \multicolumn{1}{c}{(mag)} & \multicolumn{1}{c}{(mag)}  & \multicolumn{1}{c}{(mag)} \\
\hline
SN1998bu & 0.0030 $\pm$ 0.000003 & 0.0025 $\pm$ 0.00023 & 0.475 & CfA & 50953.11 $\pm$ 0.08 & 1.076 $\pm$ 0.012 &  0.351 $\pm$ 0.006 &  0.022 $\pm$ 0.0002 \\ 
SN1999ee & 0.0114 $\pm$ 0.000010 & 0.0112 $\pm$ 0.00050 & 0.137 & CSP & 51469.61 $\pm$ 0.04 & 0.802 $\pm$ 0.007 &  0.384 $\pm$ 0.004 &  0.017 $\pm$ 0.0001 \\ 
SN1999ek & 0.0176 $\pm$ 0.000007 & 0.0178 $\pm$ 0.00050 & 0.086 & K04c & 51482.60 $\pm$ 0.19 & 1.113 $\pm$ 0.031 &  0.277 $\pm$ 0.014 &  0.479 $\pm$ 0.0187 \\ 
SN2000bh & 0.0229 $\pm$ 0.000027 & 0.0240 $\pm$ 0.00050 & 0.064 & CSP & 51636.16 $\pm$ 0.25 & 1.055 $\pm$ 0.019 &  0.065 $\pm$ 0.012 &  0.047 $\pm$ 0.0064 \\ 
SN2000ca & 0.0236 $\pm$ 0.000200 & 0.0239 $\pm$ 0.00050 & 0.064 & CSP & 51666.25 $\pm$ 0.18 & 0.917 $\pm$ 0.019 & -0.033 $\pm$ 0.010 &  0.057 $\pm$ 0.0025 \\ 
SN2000E & 0.0047 $\pm$ 0.000003 & 0.0056 $\pm$ 0.00050 & 0.273 & V03 & 51577.20 $\pm$ 0.13 & 1.041 $\pm$ 0.027 &  0.217 $\pm$ 0.011 &  0.319 $\pm$ 0.0086 \\ 
SN2001ba & 0.0296 $\pm$ 0.000033 & 0.0302 $\pm$ 0.00050 & 0.051 & CSP & 52034.47 $\pm$ 0.17 & 0.997 $\pm$ 0.020 & -0.072 $\pm$ 0.009 &  0.054 $\pm$ 0.0017 \\ 
SN2001bt & 0.0146 $\pm$ 0.000033 & 0.0142 $\pm$ 0.00050 & 0.108 & K04c & 52064.69 $\pm$ 0.07 & 1.199 $\pm$ 0.009 &  0.216 $\pm$ 0.008 &  0.056 $\pm$ 0.0007 \\ 
SN2001cn & 0.0152 $\pm$ 0.000127 & 0.0154 $\pm$ 0.00050 & 0.100 & K04c & 52071.93 $\pm$ 0.19 & 1.044 $\pm$ 0.012 &  0.176 $\pm$ 0.008 &  0.051 $\pm$ 0.0008 \\ 
SN2001cz & 0.0155 $\pm$ 0.000027 & 0.0171 $\pm$ 0.00050 & 0.090 & K04c & 52104.10 $\pm$ 0.10 & 0.956 $\pm$ 0.014 &  0.136 $\pm$ 0.008 &  0.079 $\pm$ 0.0005 \\ 
SN2001el & 0.0039 $\pm$ 0.000007 & 0.0045 $\pm$ 0.00014 & 0.000 & K03 & 52182.38 $\pm$ 0.10 & 1.080 $\pm$ 0.019 &  0.277 $\pm$ 0.010 &  0.012 $\pm$ 0.0003 \\ 
SN2002dj & 0.0094 $\pm$ 0.000003 & 0.0083 $\pm$ 0.00152 & 0.421 & P08 & 52451.04 $\pm$ 0.14 & 1.111 $\pm$ 0.019 &  0.093 $\pm$ 0.013 &  0.082 $\pm$ 0.0009 \\ 
SN2003du & 0.0064 $\pm$ 0.000013 & 0.0094 $\pm$ 0.00035 & 0.000 & St07 & 52766.01 $\pm$ 0.09 & 1.010 $\pm$ 0.015 & -0.033 $\pm$ 0.010 &  0.008 $\pm$ 0.0008 \\ 
SN2003hv & 0.0056 $\pm$ 0.000037 & 0.0049 $\pm$ 0.00034 & 0.267 & L09 & 52891.49 $\pm$ 0.11 & 1.501 $\pm$ 0.006 & -0.092 $\pm$ 0.007 &  0.013 $\pm$ 0.0008 \\ 
SN2004ef & 0.0310 $\pm$ 0.000017 & 0.0301 $\pm$ 0.00050 & 0.051 & CSP & 53264.90 $\pm$ 0.05 & 1.422 $\pm$ 0.011 &  0.116 $\pm$ 0.006 &  0.046 $\pm$ 0.0013 \\ 
SN2004eo & 0.0156 $\pm$ 0.000003 & 0.0152 $\pm$ 0.00050 & 0.101 & CSP & 53278.90 $\pm$ 0.04 & 1.318 $\pm$ 0.006 &  0.077 $\pm$ 0.005 &  0.093 $\pm$ 0.0010 \\ 
SN2004ey & 0.0158 $\pm$ 0.000003 & 0.0154 $\pm$ 0.00050 & 0.100 & CSP & 53304.81 $\pm$ 0.04 & 1.025 $\pm$ 0.011 &  0.006 $\pm$ 0.004 &  0.120 $\pm$ 0.0139 \\ 
SN2004gs & 0.0274 $\pm$ 0.000007 & 0.0287 $\pm$ 0.00050 & 0.054 & CSP & 53356.75 $\pm$ 0.05 & 1.546 $\pm$ 0.006 &  0.189 $\pm$ 0.006 &  0.026 $\pm$ 0.0006 \\ 
SN2004S & 0.0093 $\pm$ 0.000003 & 0.0107 $\pm$ 0.00050 & 0.143 & K07 & 53040.00 $\pm$ 0.29 & 1.052 $\pm$ 0.021 &  0.112 $\pm$ 0.014 &  0.086 $\pm$ 0.0014 \\ 
SN2005bo & 0.0139 $\pm$ 0.000027 & 0.0144 $\pm$ 0.00050 & 0.107 & CfA & 53479.63 $\pm$ 0.15 & 1.310 $\pm$ 0.020 &  0.272 $\pm$ 0.007 &  0.044 $\pm$ 0.0006 \\ 
SN2005cf & 0.0064 $\pm$ 0.000017 & 0.0069 $\pm$ 0.00036 & 0.000 & CfA & 53534.31 $\pm$ 0.06 & 1.072 $\pm$ 0.023 &  0.088 $\pm$ 0.010 &  0.084 $\pm$ 0.0013 \\ 
SN2005el & 0.0149 $\pm$ 0.000017 & 0.0153 $\pm$ 0.00050 & 0.101 & CSP & 53647.42 $\pm$ 0.04 & 1.370 $\pm$ 0.006 & -0.102 $\pm$ 0.005 &  0.098 $\pm$ 0.0004 \\ 
SN2005iq & 0.0340 $\pm$ 0.000123 & 0.0336 $\pm$ 0.00050 & 0.046 & CSP & 53688.14 $\pm$ 0.06 & 1.280 $\pm$ 0.012 & -0.049 $\pm$ 0.006 &  0.018 $\pm$ 0.0007 \\ 
SN2005kc & 0.0151 $\pm$ 0.000003 & 0.0145 $\pm$ 0.00050 & 0.106 & CSP & 53698.31 $\pm$ 0.08 & 1.112 $\pm$ 0.023 &  0.350 $\pm$ 0.012 &  0.114 $\pm$ 0.0023 \\ 
SN2005ki & 0.0195 $\pm$ 0.000010 & 0.0203 $\pm$ 0.00050 & 0.076 & CSP & 53706.01 $\pm$ 0.04 & 1.365 $\pm$ 0.004 & -0.065 $\pm$ 0.004 &  0.027 $\pm$ 0.0009 \\ 
SN2005lu & 0.0320 $\pm$ 0.000037 & 0.0317 $\pm$ 0.00050 & 0.048 & CSP & 53712.08 $\pm$ 0.23 & 0.834 $\pm$ 0.008 &  0.324 $\pm$ 0.011 &  0.022 $\pm$ 0.0009 \\ 
SN2005na & 0.0263 $\pm$ 0.000083 & 0.0272 $\pm$ 0.00050 & 0.056 & CfA & 53739.37 $\pm$ 0.30 & 1.027 $\pm$ 0.014 & -0.050 $\pm$ 0.012 &  0.068 $\pm$ 0.0025 \\ 
SN2006ac & 0.0231 $\pm$ 0.000010 & 0.0237 $\pm$ 0.00050 & 0.065 & CfA & 53781.55 $\pm$ 0.10 & 1.189 $\pm$ 0.008 &  0.066 $\pm$ 0.010 &  0.014 $\pm$ 0.0006 \\ 
SN2006ax & 0.0167 $\pm$ 0.000020 & 0.0180 $\pm$ 0.00050 & 0.085 & CSP & 53827.78 $\pm$ 0.04 & 1.058 $\pm$ 0.012 & -0.009 $\pm$ 0.005 &  0.041 $\pm$ 0.0019 \\ 
SN2006bh & 0.0108 $\pm$ 0.000013 & 0.0107 $\pm$ 0.00050 & 0.143 & CSP & 53834.14 $\pm$ 0.06 & 1.408 $\pm$ 0.007 & -0.043 $\pm$ 0.004 &  0.023 $\pm$ 0.0004 \\ 
SN2006bt & 0.0321 $\pm$ 0.000007 & 0.0307 $\pm$ 0.00050 & 0.050 & CSP & 53859.29 $\pm$ 0.26 & 1.093 $\pm$ 0.042 &  0.313 $\pm$ 0.023 &  0.042 $\pm$ 0.0013 \\ 
SN2006cp & 0.0223 $\pm$ 0.000003 & 0.0223 $\pm$ 0.00050 & 0.069 & CfA & 53897.45 $\pm$ 0.15 & 1.023 $\pm$ 0.046 &  0.134 $\pm$ 0.022 &  0.022 $\pm$ 0.0011 \\ 
SN2006D & 0.0085 $\pm$ 0.000017 & 0.0090 $\pm$ 0.00050 & 0.171 & CfA & 53757.84 $\pm$ 0.08 & 1.460 $\pm$ 0.013 &  0.062 $\pm$ 0.009 &  0.039 $\pm$ 0.0004 \\ 
SN2006ej & 0.0204 $\pm$ 0.000007 & 0.0205 $\pm$ 0.00050 & 0.075 & CSP & 53977.24 $\pm$ 0.25 & 1.394 $\pm$ 0.013 &  0.016 $\pm$ 0.011 &  0.030 $\pm$ 0.0008 \\ 
SN2006kf & 0.0200 $\pm$ 0.000010 & 0.0194 $\pm$ 0.00050 & 0.079 & CSP & 54041.86 $\pm$ 0.05 & 1.517 $\pm$ 0.008 &  0.007 $\pm$ 0.006 &  0.210 $\pm$ 0.0020 \\ 
SN2006lf & 0.0132 $\pm$ 0.000017 & 0.0121 $\pm$ 0.00050 & 0.127 & CfA & 54045.56 $\pm$ 0.06 & 1.406 $\pm$ 0.010 & -0.054 $\pm$ 0.010 &  0.814 $\pm$ 0.0503 \\ 
SN2006N & 0.0143 $\pm$ 0.000083 & 0.0145 $\pm$ 0.00050 & 0.106 & CfA & 53761.48 $\pm$ 0.15 & 1.457 $\pm$ 0.013 & -0.030 $\pm$ 0.007 &  0.083 $\pm$ 0.0010 \\ 
SN2007A & 0.0176 $\pm$ 0.000087 & 0.0172 $\pm$ 0.00050 & 0.089 & CSP & 54113.67 $\pm$ 0.13 & 1.037 $\pm$ 0.034 &  0.225 $\pm$ 0.014 &  0.063 $\pm$ 0.0016 \\ 
SN2007af & 0.0055 $\pm$ 0.000013 & 0.0056 $\pm$ 0.00018 & 0.000 & CSP & 54174.97 $\pm$ 0.04 & 1.116 $\pm$ 0.010 &  0.183 $\pm$ 0.005 &  0.034 $\pm$ 0.0008 \\ 
SN2007ai & 0.0317 $\pm$ 0.000137 & 0.0327 $\pm$ 0.00050 & 0.047 & CSP & 54174.03 $\pm$ 0.26 & 0.844 $\pm$ 0.021 &  0.339 $\pm$ 0.013 &  0.286 $\pm$ 0.0035 \\ 
SN2007as & 0.0176 $\pm$ 0.000460 & 0.0184 $\pm$ 0.00050 & 0.084 & CSP & 54181.15 $\pm$ 0.23 & 1.120 $\pm$ 0.023 &  0.138 $\pm$ 0.010 &  0.123 $\pm$ 0.0007 \\ 
SN2007bc & 0.0208 $\pm$ 0.000007 & 0.0211 $\pm$ 0.00050 & 0.073 & CSP & 54200.82 $\pm$ 0.09 & 1.282 $\pm$ 0.012 &  0.039 $\pm$ 0.006 &  0.019 $\pm$ 0.0006 \\ 
SN2007bd & 0.0304 $\pm$ 0.000100 & 0.0311 $\pm$ 0.00050 & 0.049 & CSP & 54207.43 $\pm$ 0.06 & 1.270 $\pm$ 0.012 & -0.018 $\pm$ 0.010 &  0.029 $\pm$ 0.0009 \\ 
SN2007ca & 0.0141 $\pm$ 0.000010 & 0.0145 $\pm$ 0.00050 & 0.106 & CSP & 54228.20 $\pm$ 0.14 & 1.037 $\pm$ 0.024 &  0.376 $\pm$ 0.012 &  0.057 $\pm$ 0.0016 \\ 
SN2007co & 0.0270 $\pm$ 0.000110 & 0.0274 $\pm$ 0.00050 & 0.056 & CfA & 54264.91 $\pm$ 0.23 & 1.040 $\pm$ 0.040 &  0.208 $\pm$ 0.017 &  0.096 $\pm$ 0.0037 \\ 
SN2007cq & 0.0260 $\pm$ 0.000080 & 0.0252 $\pm$ 0.00050 & 0.061 & CfA & 54280.90 $\pm$ 0.10 & 1.062 $\pm$ 0.021 &  0.051 $\pm$ 0.011 &  0.092 $\pm$ 0.0020 \\ 
SN2007jg & 0.0371 $\pm$ 0.000013 & 0.0380 $\pm$ 0.00050 & 0.040 & CSP & 54366.64 $\pm$ 0.25 & 1.088 $\pm$ 0.034 &  0.150 $\pm$ 0.017 &  0.090 $\pm$ 0.0020 \\ 
SN2007le & 0.0067 $\pm$ 0.000003 & 0.0065 $\pm$ 0.00050 & 0.237 & CSP & 54399.85 $\pm$ 0.07 & 1.027 $\pm$ 0.016 &  0.379 $\pm$ 0.008 &  0.029 $\pm$ 0.0003 \\ 
SN2007qe & 0.0240 $\pm$ 0.000050 & 0.0236 $\pm$ 0.00050 & 0.065 & CfA & 54429.59 $\pm$ 0.10 & 0.988 $\pm$ 0.023 &  0.069 $\pm$ 0.014 &  0.033 $\pm$ 0.0008 \\ 
SN2007sr & 0.0055 $\pm$ 0.000030 & 0.0044 $\pm$ 0.00025 & 0.000 & CSP & 54449.73 $\pm$ 0.19 & 1.084 $\pm$ 0.015 &  0.173 $\pm$ 0.009 &  0.040 $\pm$ 0.0010 \\ 
SN2007st & 0.0212 $\pm$ 0.000030 & 0.0211 $\pm$ 0.00050 & 0.073 & CSP & 54455.09 $\pm$ 0.32 & 1.486 $\pm$ 0.019 &  0.101 $\pm$ 0.018 &  0.014 $\pm$ 0.0004 \\ 
SN2008af & 0.0334 $\pm$ 0.000007 & 0.0340 $\pm$ 0.00050 & 0.045 & CfA & 54499.69 $\pm$ 0.43 & 1.178 $\pm$ 0.010 & -0.028 $\pm$ 0.023 &  0.029 $\pm$ 0.0012 \\ 
SN2008ar & 0.0262 $\pm$ 0.000010 & 0.0290 $\pm$ 0.00050 & 0.053 & CSP & 54535.22 $\pm$ 0.07 & 1.032 $\pm$ 0.014 &  0.081 $\pm$ 0.008 &  0.031 $\pm$ 0.0011 \\ 
SN2008bc & 0.0151 $\pm$ 0.000120 & 0.0156 $\pm$ 0.00050 & 0.098 & CSP & 54550.41 $\pm$ 0.08 & 1.015 $\pm$ 0.019 &  0.003 $\pm$ 0.008 &  0.225 $\pm$ 0.0042 \\ 
SN2008bf & 0.0235 $\pm$ 0.000167 & 0.0254 $\pm$ 0.00050 & 0.061 & CSP & 54555.31 $\pm$ 0.06 & 0.967 $\pm$ 0.012 & -0.013 $\pm$ 0.006 &  0.030 $\pm$ 0.0027 \\ 
SN2008C & 0.0166 $\pm$ 0.000013 & 0.0175 $\pm$ 0.00050 & 0.088 & CSP & 54466.60 $\pm$ 0.23 & 1.075 $\pm$ 0.019 &  0.239 $\pm$ 0.010 &  0.072 $\pm$ 0.0023 \\ 
SN2008fl & 0.0199 $\pm$ 0.000103 & 0.0199 $\pm$ 0.00050 & 0.077 & CSP & 54721.85 $\pm$ 0.13 & 1.328 $\pm$ 0.006 &  0.080 $\pm$ 0.005 &  0.157 $\pm$ 0.0058 \\ 
SN2008fr & 0.0390 $\pm$ 0.002001 & 0.0384 $\pm$ 0.00050 & 0.040 & CSP & 54733.93 $\pm$ 0.26 & 0.920 $\pm$ 0.014 & -0.002 $\pm$ 0.011 &  0.040 $\pm$ 0.0012 \\ 
SN2008fw & 0.0085 $\pm$ 0.000017 & 0.0086 $\pm$ 0.00050 & 0.178 & CSP & 54732.29 $\pm$ 0.15 & 0.844 $\pm$ 0.009 &  0.112 $\pm$ 0.008 &  0.112 $\pm$ 0.0030 \\ 
SN2008gb & 0.0370 $\pm$ 0.000167 & 0.0381 $\pm$ 0.00050 & 0.040 & CfA & 54748.22 $\pm$ 0.34 & 1.183 $\pm$ 0.014 &  0.080 $\pm$ 0.018 &  0.171 $\pm$ 0.0035 \\ 
SN2008gg & 0.0320 $\pm$ 0.000023 & 0.0311 $\pm$ 0.00050 & 0.049 & CSP & 54750.93 $\pm$ 0.34 & 1.036 $\pm$ 0.028 &  0.155 $\pm$ 0.013 &  0.019 $\pm$ 0.0010 \\ 
SN2008gl & 0.0340 $\pm$ 0.000117 & 0.0332 $\pm$ 0.00050 & 0.046 & CSP & 54768.70 $\pm$ 0.09 & 1.319 $\pm$ 0.010 &  0.030 $\pm$ 0.006 &  0.024 $\pm$ 0.0008 \\ 
SN2008gp & 0.0330 $\pm$ 0.000070 & 0.0335 $\pm$ 0.00050 & 0.046 & CSP & 54779.62 $\pm$ 0.04 & 1.017 $\pm$ 0.008 & -0.018 $\pm$ 0.004 &  0.104 $\pm$ 0.0051 \\ 
SN2008hj & 0.0379 $\pm$ 0.000130 & 0.0372 $\pm$ 0.00050 & 0.041 & CSP & 54802.26 $\pm$ 0.12 & 1.055 $\pm$ 0.027 &  0.038 $\pm$ 0.012 &  0.030 $\pm$ 0.0008 \\ 
SN2008hm & 0.0197 $\pm$ 0.000077 & 0.0210 $\pm$ 0.00050 & 0.073 & CfA & 54804.74 $\pm$ 0.21 & 0.993 $\pm$ 0.025 &  0.182 $\pm$ 0.014 &  0.380 $\pm$ 0.0085 \\ 
SN2008hs & 0.0174 $\pm$ 0.000070 & 0.0189 $\pm$ 0.00004 & 0.058 & CfA & 54812.94 $\pm$ 0.14 & 1.531 $\pm$ 0.015 &  0.122 $\pm$ 0.024 &  0.050 $\pm$ 0.0003 \\ 
SN2008hv & 0.0126 $\pm$ 0.000007 & 0.0140 $\pm$ 0.00050 & 0.110 & CSP & 54817.65 $\pm$ 0.04 & 1.328 $\pm$ 0.006 & -0.065 $\pm$ 0.006 &  0.028 $\pm$ 0.0008 \\ 
SN2008ia & 0.0219 $\pm$ 0.000097 & 0.0225 $\pm$ 0.00050 & 0.068 & CSP & 54813.67 $\pm$ 0.09 & 1.340 $\pm$ 0.009 &  0.003 $\pm$ 0.007 &  0.195 $\pm$ 0.0050 \\ 
SN2009aa & 0.0273 $\pm$ 0.000047 & 0.0287 $\pm$ 0.00050 & 0.054 & CSP & 54878.81 $\pm$ 0.04 & 1.172 $\pm$ 0.008 &  0.020 $\pm$ 0.005 &  0.029 $\pm$ 0.0009 \\ 
SN2009ab & 0.0112 $\pm$ 0.000020 & 0.0103 $\pm$ 0.00050 & 0.149 & CSP & 54883.89 $\pm$ 0.08 & 1.288 $\pm$ 0.016 &  0.050 $\pm$ 0.010 &  0.184 $\pm$ 0.0028 \\ 
SN2009ad & 0.0284 $\pm$ 0.000003 & 0.0287 $\pm$ 0.00050 & 0.054 & CSP & 54886.91 $\pm$ 0.07 & 0.949 $\pm$ 0.013 &  0.020 $\pm$ 0.007 &  0.095 $\pm$ 0.0011 \\ 
SN2009ag & 0.0086 $\pm$ 0.000007 & 0.0102 $\pm$ 0.00050 & 0.151 & CSP & 54890.23 $\pm$ 0.16 & 1.088 $\pm$ 0.019 &  0.343 $\pm$ 0.009 &  0.218 $\pm$ 0.0012 \\ 
SN2009al & 0.0221 $\pm$ 0.000080 & 0.0234 $\pm$ 0.00050 & 0.066 & CfA & 54897.20 $\pm$ 0.18 & 1.079 $\pm$ 0.033 &  0.236 $\pm$ 0.020 &  0.021 $\pm$ 0.0004 \\ 
SN2009an & 0.0092 $\pm$ 0.000007 & 0.0107 $\pm$ 0.00050 & 0.144 & CfA & 54898.56 $\pm$ 0.09 & 1.327 $\pm$ 0.010 &  0.063 $\pm$ 0.010 &  0.016 $\pm$ 0.0003 \\ 
SN2009bv & 0.0366 $\pm$ 0.000017 & 0.0385 $\pm$ 0.00050 & 0.040 & CfA & 54927.07 $\pm$ 0.20 & 0.948 $\pm$ 0.033 & -0.026 $\pm$ 0.019 &  0.008 $\pm$ 0.0008 \\ 
SN2009cz & 0.0212 $\pm$ 0.000010 & 0.0218 $\pm$ 0.00050 & 0.070 & CSP & 54943.50 $\pm$ 0.09 & 0.899 $\pm$ 0.014 &  0.102 $\pm$ 0.007 &  0.022 $\pm$ 0.0003 \\ 
SN2009D & 0.0250 $\pm$ 0.000033 & 0.0243 $\pm$ 0.00050 & 0.063 & CSP & 54841.65 $\pm$ 0.11 & 1.025 $\pm$ 0.024 &  0.054 $\pm$ 0.009 &  0.044 $\pm$ 0.0012 \\ 
SN2009kk & 0.0129 $\pm$ 0.000150 & 0.0122 $\pm$ 0.00050 & 0.126 & CfA & 55126.37 $\pm$ 0.20 & 1.189 $\pm$ 0.006 & -0.055 $\pm$ 0.011 &  0.116 $\pm$ 0.0025 \\ 
SN2009kq & 0.0117 $\pm$ 0.000020 & 0.0126 $\pm$ 0.00050 & 0.122 & CfA & 55154.81 $\pm$ 0.17 & 0.992 $\pm$ 0.025 &  0.089 $\pm$ 0.010 &  0.035 $\pm$ 0.0005 \\ 
SN2009Y & 0.0093 $\pm$ 0.000027 & 0.0094 $\pm$ 0.00050 & 0.163 & CSP & 54877.10 $\pm$ 0.10 & 1.063 $\pm$ 0.023 &  0.169 $\pm$ 0.010 &  0.087 $\pm$ 0.0005 \\ 
SN2010ai & 0.0184 $\pm$ 0.000123 & 0.0239 $\pm$ 0.00018 & 0.048 & CfA & 55277.50 $\pm$ 0.08 & 1.421 $\pm$ 0.016 & -0.075 $\pm$ 0.016 &  0.008 $\pm$ 0.0010 \\ 
SN2010dw & 0.0381 $\pm$ 0.000150 & 0.0392 $\pm$ 0.00050 & 0.039 & CfA & 55358.25 $\pm$ 0.35 & 0.844 $\pm$ 0.058 &  0.177 $\pm$ 0.028 &  0.080 $\pm$ 0.0009 \\ 
SN2010iw & 0.0215 $\pm$ 0.000007 & 0.0228 $\pm$ 0.00050 & 0.067 & CfA & 55497.14 $\pm$ 0.26 & 0.876 $\pm$ 0.019 &  0.084 $\pm$ 0.012 &  0.047 $\pm$ 0.0006 \\ 
SN2010kg & 0.0166 $\pm$ 0.000007 & 0.0171 $\pm$ 0.00050 & 0.090 & CfA & 55543.96 $\pm$ 0.10 & 1.194 $\pm$ 0.011 &  0.183 $\pm$ 0.014 &  0.131 $\pm$ 0.0022 \\ 
SN2011ao & 0.0107 $\pm$ 0.000003 & 0.0120 $\pm$ 0.00050 & 0.128 & CfA & 55639.61 $\pm$ 0.11 & 1.012 $\pm$ 0.018 &  0.035 $\pm$ 0.019 &  0.017 $\pm$ 0.0001 \\ 
SN2011B & 0.0047 $\pm$ 0.000003 & 0.0056 $\pm$ 0.00050 & 0.276 & CfA & 55583.38 $\pm$ 0.06 & 1.174 $\pm$ 0.005 &  0.112 $\pm$ 0.008 &  0.026 $\pm$ 0.0011 \\ 
SN2011by & 0.0028 $\pm$ 0.000003 & 0.0051 $\pm$ 0.00020 & 0.000 & CfA & 55690.95 $\pm$ 0.05 & 1.053 $\pm$ 0.008 &  0.067 $\pm$ 0.005 &  0.012 $\pm$ 0.0002 \\ 
SN2011df & 0.0145 $\pm$ 0.000020 & 0.0150 $\pm$ 0.00050 & 0.102 & CfA & 55716.40 $\pm$ 0.11 & 0.923 $\pm$ 0.015 &  0.072 $\pm$ 0.010 &  0.112 $\pm$ 0.0034 \\ 
SNf20080514-002 & 0.0219 $\pm$ 0.000010 & 0.0216 $\pm$ 0.00050 & 0.071 & CfA & 54612.80 $\pm$ 0.00 & 1.360 $\pm$ 0.000 & -0.143 $\pm$ 0.000 &  0.027 $\pm$ 0.0014 \\ 
\hline
\end{tabular}
\end{center}
\tablecomments{
\\
$^a$ Heliocentric redshift from NED or the literature using $v_{\rm helio}$ from Table~\ref{TableDistanceMu1a}.\\
$^b$ Redshift corrected to the CMB frame and using the C15 local flow model or redshift-independent distance information from Table~\ref{TableDistanceMu2Special}. \\
$^c$ Uncertainty in the theoretical distance modulus because of the peculiar velocity, defined in Eq. (\ref{eq_PeculiarVelDefinition}).\\
$^d$ LC-data source. CfA: \citealt{woodvasey08,friedman15}, CSP: \citealt{krisciunas17}, Others: K04c: \citealt{krisciunas04c}; V03: \citealt{valentini03}; K03: \citealt{krisciunas03}; P08: \citealt{pignata08}; St07: \citealt{stanishev07}; L09: \citealt{leloudas09}; K07: \citealt{krisciunas07}. Also see Table 3 of \citealt{friedman15} for references. \\
$^e$ Determined by fitting the optical and NIR \lcs{} data with \snoopy{}.\\
$^f$ LC shape parameter: apparent-magnitude decline between B-band peak luminosity and 15 days after peak.\\
$^g$ Host-galaxy color excess, as measured by \snoopy{} fits to the optical and NIR \lcs{}. \\
$^h$ Milky-Way color excess, from the \cite{SchlaflyFinkbeiner2011} Milky Way dust maps. \\
}
\end{table*}
\renewcommand{\arraystretch}{1}
\renewcommand{\baselinestretch}{1}

\renewcommand{\arraystretch}{0.5}
\renewcommand{\baselinestretch}{0.8}
\begin{table*}[p!]
\begin{center}
\caption{\snIa{} Recession Velocities}
\label{TableDistanceMu1a}
\tiny
\begin{tabular}{lrr c cc rcc}
\hline \hline
\multicolumn{1}{l}{SN name$^{a}$}        &
\multicolumn{1}{c}{$RA$ (deg)}            &
\multicolumn{1}{c}{$DEC$ (deg)}         &
\multicolumn{1}{c}{Host Galaxy}     &
\multicolumn{1}{c}{$\vhel$}        &
\multicolumn{1}{c}{$\vcmb$}  &
\multicolumn{1}{c}{$\vcmbf$} &
\multicolumn{1}{c}{Ref(s).$^{g}$} &
\multicolumn{1}{c}{Code$^{h}$}  \\
\multicolumn{1}{l}{} &
\multicolumn{1}{r}{$\alpha(2000)^{b}$} &
\multicolumn{1}{r}{$\delta(2000)^{b}$}  &
\multicolumn{1}{c}{(or cluster)$^{c}$} &
\multicolumn{1}{c}{(km s$^{-1}$)$^{d}$}        &
\multicolumn{1}{c}{(km s$^{-1}$)$^{e}$}   &
\multicolumn{1}{c}{(km s$^{-1}$)$^{f}$} &
\multicolumn{1}{c}{} &
\multicolumn{1}{c}{} \\
\hline
SN1998bu & 161.69167 & 11.83528 & NGC 3368 & 888 $\pm$ 1 & 757 $\pm$ 70 & 242 $\pm$ 150 & NED;F01 & Cepheid \\ 
SN1999ee & 334.04167 & -36.84444 & IC 5179 & 3419 $\pm$ 3 & 3160 $\pm$ 3 & 3368 $\pm$ 150 & NED;C15 & Flow \\ 
SN1999ek & 84.13167 & 16.63833 & UGC 03329 & 5266 $\pm$ 2 & 5292 $\pm$ 2 & 5340 $\pm$ 150 & NED;C15 & Flow \\ 
SN2000E & 309.30750 & 66.09722 & NGC 6951 & 1424 $\pm$ 1 & 1267 $\pm$ 1 & 1685 $\pm$ 150 & NED;C15 & Flow \\ 
SN2000bh & 185.31292 & -21.99889 & ESO 573-G 014 & 6854 $\pm$ 8 & 7196 $\pm$ 8 & 7188 $\pm$ 150 & NED;C15 & Flow \\ 
SN2000ca & 203.84583 & -34.16028 & ESO 383-G 032 & 7080 $\pm$ 60 & 7351 $\pm$ 62 & 7167 $\pm$ 150 & NED;C15 & Flow \\ 
SN2001ba & 174.50750 & -32.33083 & MCG -05-28-001 & 8861 $\pm$ 10 & 9193 $\pm$ 10 & 9060 $\pm$ 150 & NED;C15 & Flow \\ 
SN2001bt & 288.44500 & -59.28972 & IC 4830 & 4388 $\pm$ 10 & 4331 $\pm$ 10 & 4260 $\pm$ 150 & NED;C15 & Flow \\ 
SN2001cn & 281.57417 & -65.76167 & IC 4758 & 4543 $\pm$ 38 & 4523 $\pm$ 38 & 4626 $\pm$ 150 & NED;C15 & Flow \\ 
SN2001cz & 191.87583 & -39.58000 & NGC 4679 & 4643 $\pm$ 8 & 4930 $\pm$ 8 & 5124 $\pm$ 150 & NED;C15 & Flow \\ 
SN2001el & 56.12750 & -44.63972 & NGC 1448 & 1168 $\pm$ 2 & 1340 $\pm$ 42 & 1568 $\pm$ 150 & NED;R16 & Cepheid \\ 
SN2002dj & 198.25125 & -19.51917 & NGC 5018 & 2816 $\pm$ 1 & 2479 $\pm$ 457 & 2023 $\pm$ 150 & NED;Co12 & SBF/TF \\ 
SN2003du & 218.64917 & 59.33444 & UGC 9391 & 1914 $\pm$ 4 & 2809 $\pm$ 105 & 3165 $\pm$ 150 & NED;R16 & Cepheid \\ 
SN2003hv & 46.03875 & -26.08556 & NGC 1201 & 1686 $\pm$ 11 & 1470 $\pm$ 101 & 1723 $\pm$ 150 & NED;Tu13 & SBF/TF \\ 
SN2004S & 101.43125 & -31.23111 & MCG -05-16-021 & 2788 $\pm$ 1 & 2937 $\pm$ 1 & 3213 $\pm$ 150 & NED;C15 & Flow \\ 
SN2004ef & 340.54175 & 19.99456 & UGC 12158 & 9289 $\pm$ 5 & 8931 $\pm$ 5 & 9015 $\pm$ 150 & NED;C15 & Flow \\ 
SN2004eo & 308.22579 & 9.92853 & NGC 6928 & 4684 $\pm$ 1 & 4398 $\pm$ 1 & 4560 $\pm$ 150 & NED;C15 & Flow \\ 
SN2004ey & 327.28254 & 0.44422 & UGC 11816 & 4749 $\pm$ 1 & 4405 $\pm$ 1 & 4617 $\pm$ 150 & NED;C15 & Flow \\ 
SN2004gs & 129.59658 & 17.62772 & MCG +03-22-020 & 8214 $\pm$ 2 & 8475 $\pm$ 2 & 8590 $\pm$ 150 & NED;C15 & Flow \\ 
SN2005bo & 192.42096 & -11.09647 & NGC 4708 & 4166 $\pm$ 8 & 4503 $\pm$ 9 & 4314 $\pm$ 150 & NED;C15 & Flow \\ 
SN2005cf & 230.38417 & -7.41306 & MCG -01-39-003 & 1929 $\pm$ 5 & 2077 $\pm$ 109 & 2034 $\pm$ 150 & NED;R16 & Cepheid \\ 
SN2005el & 77.95300 & 5.19428 & NGC 1819 & 4470 $\pm$ 5 & 4466 $\pm$ 5 & 4574 $\pm$ 150 & NED;C15 & Flow \\ 
SN2005iq & 359.63542 & -18.70917 & ESO 538- G 013 & 10206 $\pm$ 37 & 9880 $\pm$ 36 & 10058 $\pm$ 150 & NED;C15 & Flow \\ 
SN2005kc & 338.53058 & 5.56842 & NGC 7311 & 4524 $\pm$ 1 & 4159 $\pm$ 1 & 4343 $\pm$ 150 & NED;C15 & Flow \\ 
SN2005ki & 160.11758 & 9.20233 & NGC 3332 & 5833 $\pm$ 3 & 6185 $\pm$ 3 & 6080 $\pm$ 150 & NED;C15 & Flow \\ 
SN2005lu & 39.01546 & -17.26389 & ESO 545-G038 & 9596 $\pm$ 11 & 9389 $\pm$ 11 & 9515 $\pm$ 150 & NED;C15 & Flow \\ 
SN2005na & 105.40258 & 14.13325 & UGC 3634 & 7891 $\pm$ 25 & 8045 $\pm$ 25 & 8162 $\pm$ 150 & NED;C15 & Flow \\ 
SN2006D & 193.14142 & -9.77522 & MCG -01-33-034 & 2556 $\pm$ 5 & 2891 $\pm$ 6 & 2691 $\pm$ 150 & NED;C15 & Flow \\ 
SN2006N & 92.13000 & 64.72361 & CGCG 308-009 & 4280 $\pm$ 25 & 4278 $\pm$ 25 & 4354 $\pm$ 150 & NED;C15 & Flow \\ 
SN2006ac & 190.43708 & 35.08528 & NGC 4619 & 6923 $\pm$ 3 & 7175 $\pm$ 3 & 7113 $\pm$ 150 & NED;C15 & Flow \\ 
SN2006ax & 171.01442 & -12.29144 & NGC 3663 & 5014 $\pm$ 6 & 5382 $\pm$ 6 & 5386 $\pm$ 150 & NED;C15 & Flow \\ 
SN2006bh & 340.06708 & -66.48508 & NGC 7329 & 3252 $\pm$ 4 & 3148 $\pm$ 4 & 3222 $\pm$ 150 & NED;C15 & Flow \\ 
SN2006bt & 239.12721 & 20.04592 & CGCG 108-013 & 9618 $\pm$ 2 & 9714 $\pm$ 2 & 9211 $\pm$ 150 & NED;K17 & Flow \\ 
SN2006cp & 184.81208 & 22.42722 & UGC 7357 & 6682 $\pm$ 1 & 6990 $\pm$ 1 & 6673 $\pm$ 150 & NED;C15 & Flow \\ 
SN2006ej & 9.74904 & -9.01572 & NGC 191A & 6110 $\pm$ 2 & 5780 $\pm$ 2 & 6152 $\pm$ 150 & NED;C15 & Flow \\ 
SN2006kf & 55.46033 & 8.15694 & UGC 2829 & 6007 $\pm$ 3 & 5862 $\pm$ 3 & 5821 $\pm$ 150 & NED;C15 & Flow \\ 
SN2006lf & 69.62292 & 44.03361 & UGC 3108 & 3954 $\pm$ 5 & 3885 $\pm$ 5 & 3627 $\pm$ 150 & NED;C15 & Flow \\ 
SN2007A & 6.31942 & 12.88681 & NGC 105 & 5290 $\pm$ 26 & 4940 $\pm$ 24 & 5162 $\pm$ 150 & NED;C15 & Flow \\ 
SN2007af & 215.58763 & -0.39378 & NGC 5584 & 1638 $\pm$ 4 & 1667 $\pm$ 53 & 1418 $\pm$ 150 & NED;R16 & Cepheid \\ 
SN2007ai & 243.22392 & -21.63019 & MCG-04-38-004 & 9492 $\pm$ 41 & 9595 $\pm$ 41 & 9815 $\pm$ 150 & NED;C15 & Flow \\ 
SN2007as & 141.90004 & -80.17756 & ESO 018-G 018 & 5268 $\pm$ 138 & 5368 $\pm$ 141 & 5503 $\pm$ 150 & NED;C15 & Flow \\ 
SN2007bc & 169.81071 & 20.80903 & UGC 6332 & 6221 $\pm$ 2 & 6548 $\pm$ 2 & 6333 $\pm$ 150 & NED;C15 & Flow \\ 
SN2007bd & 127.88867 & -1.19944 & UGC 4455 & 9126 $\pm$ 30 & 9408 $\pm$ 31 & 9318 $\pm$ 150 & NED;C15 & Flow \\ 
SN2007ca & 202.77421 & -15.10183 & MCG -02-34-061 & 4217 $\pm$ 3 & 4520 $\pm$ 3 & 4339 $\pm$ 150 & NED;C15 & Flow \\ 
SN2007co & 275.76500 & 29.89722 & MCG +05-43-016 & 8083 $\pm$ 33 & 7963 $\pm$ 33 & 8229 $\pm$ 150 & NED;C15 & Flow \\ 
SN2007cq & 333.66833 & 5.08028 & 2MASX J22144070+0504435 & 7807 $\pm$ 24 & 7448 $\pm$ 23 & 7564 $\pm$ 150 & Ch13;C15 & Flow \\ 
SN2007jg & 52.46175 & 0.05683 & SDSS J032950.83+000316.0 & 11113 $\pm$ 4 & 10955 $\pm$ 4 & 11379 $\pm$ 150 & NED;C15 & Flow \\ 
SN2007le & 354.70171 & -6.52258 & NGC 7721 & 2017 $\pm$ 1 & 1660 $\pm$ 1 & 1939 $\pm$ 150 & NED;C15 & Flow \\ 
SN2007qe & 358.55417 & 27.40917 & NSF J235412.09+272432.3 & 7183 $\pm$ 15 & 6842 $\pm$ 14 & 7067 $\pm$ 150 & Ch13;C15 & Flow \\ 
SN2007sr & 180.47000 & -18.97269 & NGC 4038 & 1641 $\pm$ 9 & 1327 $\pm$ 75 & 611 $\pm$ 150 & NED;R16 & Cepheid \\ 
SN2007st & 27.17696 & -48.64939 & NGC 692 & 6350 $\pm$ 9 & 6195 $\pm$ 9 & 6330 $\pm$ 150 & NED;C15 & Flow \\ 
SN2008C & 104.29804 & 20.43714 & UGC 3611 & 4983 $\pm$ 4 & 5121 $\pm$ 4 & 5260 $\pm$ 150 & NED;C15 & Flow \\ 
SN2008af & 224.86875 & 16.65333 & UGC 9640 & 10020 $\pm$ 2 & 10199 $\pm$ 2 & 10195 $\pm$ 150 & NED;C15 & Flow \\ 
SN2008ar & 186.15800 & 10.83817 & IC 3284 & 7846 $\pm$ 3 & 8180 $\pm$ 3 & 8680 $\pm$ 150 & NED;C15 & Flow \\ 
SN2008bc & 144.63012 & -63.97378 & KK 1524 & 4523 $\pm$ 36 & 4711 $\pm$ 37 & 4677 $\pm$ 150 & NED;C15 & Flow \\ 
SN2008bf & 181.01208 & 20.24517 & ambiguous & 7045 $\pm$ 50 & 7365 $\pm$ 52 & 7608 $\pm$ 150 & K17;C15 & Flow \\ 
SN2008fl & 294.18683 & -37.55125 & NGC 6805 & 5960 $\pm$ 31 & 5815 $\pm$ 30 & 5980 $\pm$ 150 & NED;C15 & Flow \\ 
SN2008fr & 17.95475 & 14.64083 & SDSS J011149.19+143826.5 & 11692 $\pm$ 600 & 11373 $\pm$ 584 & 11503 $\pm$ 150 & NED;C15 & Flow \\ 
SN2008fw & 157.23321 & -44.66544 & NGC 3261 & 2563 $\pm$ 5 & 2851 $\pm$ 6 & 2587 $\pm$ 150 & NED;C15 & Flow \\ 
SN2008gb & 44.48792 & 46.86583 & UGC 2427 & 11092 $\pm$ 50 & 10921 $\pm$ 49 & 11428 $\pm$ 150 & NED;C15 & Flow \\ 
SN2008gg & 21.34600 & -18.17244 & NGC 539 & 9598 $\pm$ 7 & 9321 $\pm$ 7 & 9330 $\pm$ 150 & NED;C15 & Flow \\ 
SN2008gl & 20.22842 & 4.80531 & UGC 881 & 10198 $\pm$ 35 & 9885 $\pm$ 34 & 9954 $\pm$ 150 & NED;C15 & Flow \\ 
SN2008gp & 50.75304 & 1.36189 & MCG +00-9-74 & 9901 $\pm$ 21 & 9732 $\pm$ 21 & 10030 $\pm$ 150 & NED;C15 & Flow \\ 
SN2008hj & 1.00796 & -11.16875 & MCG -02-01-014 & 11360 $\pm$ 39 & 11018 $\pm$ 38 & 11140 $\pm$ 150 & NED;C15 & Flow \\ 
SN2008hm & 51.79542 & 46.94444 & 2MFGC 02845 & 5895 $\pm$ 23 & 5752 $\pm$ 22 & 6282 $\pm$ 150 & NED;C15 & Flow \\ 
SN2008hs & 36.37333 & 41.84306 & NGC 0910 (Abell 347) & 5207 $\pm$ 21 & 5655 $\pm$ 13 & 6186 $\pm$ 150 & NED;Dh18 & Cluster \\ 
SN2008hv & 136.89192 & 3.39225 & NGC 2765 & 3775 $\pm$ 2 & 4087 $\pm$ 2 & 4185 $\pm$ 150 & NED;C15 & Flow \\ 
SN2008ia & 132.64646 & -61.27794 & ESO 125-G 006 & 6578 $\pm$ 29 & 6761 $\pm$ 30 & 6754 $\pm$ 150 & NED;C15 & Flow \\ 
SN2009D & 58.59512 & -19.18172 & MCG -03-10-052 & 7497 $\pm$ 10 & 7397 $\pm$ 10 & 7275 $\pm$ 150 & NED;C15 & Flow \\ 
SN2009Y & 220.59938 & -17.24678 & NGC 5728 & 2793 $\pm$ 8 & 3019 $\pm$ 9 & 2827 $\pm$ 150 & NED;C15 & Flow \\ 
SN2009aa & 170.92617 & -22.27069 & ESO 570-G20 & 8187 $\pm$ 14 & 8543 $\pm$ 15 & 8597 $\pm$ 150 & NED;C15 & Flow \\ 
SN2009ab & 64.15162 & 2.76417 & UGC 2998 & 3349 $\pm$ 6 & 3260 $\pm$ 6 & 3090 $\pm$ 150 & NED;C15 & Flow \\ 
SN2009ad & 75.88908 & 6.65992 & UGC 3236 & 8514 $\pm$ 1 & 8496 $\pm$ 1 & 8602 $\pm$ 150 & NED;C15 & Flow \\ 
SN2009ag & 107.92004 & -26.68508 & ESO 492-2 & 2590 $\pm$ 2 & 2774 $\pm$ 2 & 3056 $\pm$ 150 & NED;C15 & Flow \\ 
SN2009al & 162.84196 & 8.57853 & NGC 3425 & 6627 $\pm$ 24 & 6982 $\pm$ 25 & 7007 $\pm$ 150 & NED;C15 & Flow \\ 
SN2009an & 185.69750 & 65.85111 & NGC 4332 & 2764 $\pm$ 2 & 2867 $\pm$ 2 & 3207 $\pm$ 150 & NED;C15 & Flow \\ 
SN2009bv & 196.83542 & 35.78444 & MCG +06-29-039 & 10966 $\pm$ 5 & 11201 $\pm$ 5 & 11539 $\pm$ 150 & NED;C15 & Flow \\ 
SN2009cz & 138.75008 & 29.73531 & NGC 2789 & 6344 $\pm$ 3 & 6601 $\pm$ 3 & 6548 $\pm$ 150 & NED;C15 & Flow \\ 
SN2009kk & 57.43458 & -3.26444 & 2MFGC 03182 & 3855 $\pm$ 45 & 3729 $\pm$ 44 & 3653 $\pm$ 150 & NED;C15 & Flow \\ 
SN2009kq & 129.06292 & 28.06722 & MCG +05-21-001 & 3507 $\pm$ 6 & 3739 $\pm$ 6 & 3766 $\pm$ 150 & NED;C15 & Flow \\ 
SN2010ai & 194.85000 & 27.99639 & SDSS J125925.04+275948.2 (Coma) & 5507 $\pm$ 37 & 7166 $\pm$ 54 & 7298 $\pm$ 150 & NED;P14 & Cluster \\ 
SN2010dw & 230.66792 & -5.92111 & 2MASX J15224062-0555214 & 11428 $\pm$ 45 & 11600 $\pm$ 46 & 11759 $\pm$ 150 & NED;C15 & Flow \\ 
SN2010iw & 131.31250 & 27.82278 & UGC 4570 & 6458 $\pm$ 2 & 6698 $\pm$ 2 & 6833 $\pm$ 150 & NED;C15 & Flow \\ 
SN2010kg & 70.03500 & 7.35000 & NGC 1633 & 4986 $\pm$ 2 & 4931 $\pm$ 2 & 5128 $\pm$ 150 & NED;C15 & Flow \\ 
SN2011B & 133.95208 & 78.21750 & NGC 2655 & 1400 $\pm$ 1 & 1419 $\pm$ 1 & 1670 $\pm$ 150 & NED;C15 & Flow \\ 
SN2011ao & 178.46250 & 33.36278 & IC 2973 & 3210 $\pm$ 1 & 3487 $\pm$ 1 & 3592 $\pm$ 150 & NED;C15 & Flow \\ 
SN2011by & 178.94000 & 55.32611 & NGC 3972 & 852 $\pm$ 1 & 1521 $\pm$ 61 & 1796 $\pm$ 150 & NED;R16 & Cepheid \\ 
SN2011df & 291.89000 & 54.38639 & NGC 6801 & 4361 $\pm$ 6 & 4205 $\pm$ 6 & 4500 $\pm$ 150 & NED;C15 & Flow \\ 
SNf20080514-002 & 202.30625 & 11.26889 & UGC 8472 & 6577 $\pm$ 3 & 6866 $\pm$ 3 & 6480 $\pm$ 150 & NED;C15 & Flow \\ 

\hline
\end{tabular}
\end{center}
\tablecomments{

$^a$ \snIa{} name. All \snIa{} in this work are spectroscopically normal (see \S\ref{sec:cuts}). \\ 
$^b$ Epoch J2000 equatorial coordinates in decimal degrees. \\
$^c$ Host galaxy or cluster from NASA/IPAC Extragalactic Database (NED) or the literature. See Ref(s) column. \\
$^d$ Heliocentric recession velocity from NED with smallest reported uncertainty (SDSS DR13 values are used even if earlier SDSS reported smaller uncertainties). When no uncertainty is reported we assume an error of 50 km/s.\\
$^e$ CMB frame recession velocity $\vcmb$ using $\vhel$, RA, DEC, and CMB dipole from (\citealt{planck2015viii}).\\
$^f$ $\vcmbf$ takes $\vcmb$, $RA$, $DEC$ as input and further corrects to the CMB frame via the local velocity flow model of \cite{carrick15} (hereafter C15), with assumed uncertainty of 150 km/s (see \S\ref{sec:z}).\\
$^g$ The first reference listed is for $v_{\rm helio}$ from NED or the literature. The second reference is for the effective $\vcmb$ derived using either the C15 local flow model or independent distance information for nearby galaxies with $\vhel \lesssim 3000$ km/s, special cases where host galaxy identification from NED is ambiguous, or some clusters which may have $v>3000$km/s (see \S\ref{sec:z} and Table~\ref{TableDistanceMu2Special}). Reference codes: C15: \citealt{carrick15}; Ch11: \citealt{childress11}; Ch13: \citealt{childress13a}; Co12: \citealt{courtois12}; Dh18: \citealt{dhawan18}; F01: \citealt{freedman01}; F15: \citealt{friedman15}; H12: \citealt{hicken12}; K17: \citealt{krisciunas17}; MO00: \citealt{mould00}; P14: \citealt{pimbblet14};  Pr07: \citealt{prieto07}; R14: \citealt{rest14}; R16: \citealt{riess16}; Tu13: \citealt{tully13}; Tu16: \citealt{tully16}.\\
$^h$ If Code=Flow, we use $\vcmbf$ from C15 in our Hubble diagram. If Code $\ne$ Flow, we use $\vcmb$. Other codes include Cepheid: HST Cepheid distances from SHOES \citealt{riess16} or HST Key Project \citealt{freedman01}; Cluster: Mean redshift of galaxy cluster from NED (e.g. Virgo, Coma, Abell*); SBF/TF: Surface Brightness Fluctuations (SBF) or Tully-Fisher relation (TF) (e.g. \citealt{courtois12,tully13,tully16}).
}
\end{table*}
\renewcommand{\arraystretch}{1}
\renewcommand{\baselinestretch}{1}

\renewcommand{\arraystretch}{0.001}
\renewcommand{\baselinestretch}{0.8}
\begin{table*}
\begin{center}
\caption{\snIa{} With Redshift-Independent Distance Moduli}
\begin{tabular}{lcccclc}
\hline \hline
\multicolumn{1}{l}{SN name}               & \multicolumn{1}{c}{Host Galaxy}    & \multicolumn{1}{c}{$\mu^{\prime}$} & \multicolumn{1}{c}{$\Ho^{\prime}$} &  \multicolumn{1}{c}{$\mu_{\rm eff}$} & \multicolumn{1}{c}{Ref.$^{c}$}  &
\multicolumn{1}{c}{Code$^{d}$}  \\
\multicolumn{1}{c}{}        &
\multicolumn{1}{c}{(or cluster)}   & \multicolumn{1}{c}{(mag)$^{a}$}    &
\multicolumn{1}{c}{(\Hounits)}     & \multicolumn{1}{c}{(mag)$^{b}$}    &  \multicolumn{1}{c}{}     &
\multicolumn{1}{c}{} \\
\hline
SN1998bu & NGC 3368 & 30.110 $\pm$ 0.200 & 72.00 $\pm$ 8.00 & 30.073 $\pm$ 0.200 & F01 & Cepheid \\ 
SN2001el & NGC 1448 & 31.311 $\pm$ 0.045 & 73.24 $\pm$ 1.74 & 31.311 $\pm$ 0.045 & R16 & Cepheid \\ 
SN2002dj & NGC 5018 & 32.570 $\pm$ 0.400 & 75.90 $\pm$ 3.80 & 32.647 $\pm$ 0.400 & Co12 & SBF/TF \\ 
SN2003du & UGC 9391 & 32.919 $\pm$ 0.063 & 73.24 $\pm$ 1.74 & 32.919 $\pm$ 0.063 & R16 & Cepheid \\ 
SN2003hv & NGC 1201 & 31.120 $\pm$ 0.250 & 74.40 $\pm$ 3.00 & 31.154 $\pm$ 0.250 & Tu13 & SBF/TF \\ 
SN2005cf & MCG -01-39-003 & 32.263 $\pm$ 0.102 & 73.24 $\pm$ 1.74 & 32.263 $\pm$ 0.102 & R16 & Cepheid \\ 
SN2007af & NGC 5584 & 31.786 $\pm$ 0.046 & 73.24 $\pm$ 1.74 & 31.786 $\pm$ 0.046 & R16 & Cepheid \\ 
SN2007sr & NGC 4038 & 31.290 $\pm$ 0.112 & 73.24 $\pm$ 1.74 & 31.290 $\pm$ 0.112 & R16 & Cepheid \\ 
SN2011by & NGC 3972 & 31.587 $\pm$ 0.070 & 73.24 $\pm$ 1.74 & 31.587 $\pm$ 0.070 & R16 & Cepheid \\ 

\hline
\end{tabular}
\label{TableDistanceMu2Special}
\end{center}
\tablecomments{
\\
$^a$ Reported distance modulus $\mu^{\prime}$ on Hubble scale $\Ho^{\prime}$. \\
$^b$ This is converted to a distance modulus $\mu_{\rm eff}$ on the Hubble scale of $\Ho=73.24$ \Hounits{} via Eq.~(\ref{mufromHo}). For \snIa{} with Cepheid distances from \citealt{riess16}, since $\Ho^{\prime}=\Ho$ and $\mu^{\prime}=\mu_{\rm eff}$, effective distance moduli $\mu_{\rm eff}$ are already on the Hubble scale used for this paper. We compute the effective CMB frame recession velocity $v_{\rm eff}$ in Table~\ref{TableDistanceMu1a} via Eqs.~\ref{ceph2vel}-\ref{ceph2velerr} using $\mu^{\prime}$ and $\Ho^{\prime}$ (or equivalently $\mu_{\rm eff}$ and $\Ho$). This is then used to construct an effective redshift or recession velocity for use in Hubble diagrams. \\
$^c$ Reference codes: Co12: \citealt{courtois12}; F01: \citealt{freedman01}; R16: \citealt{riess16}; Tu13: \citealt{tully13}; Tu16: \citealt{tully16} \\
$^d$ Same as in Table~\ref{TableDistanceMu1a}.
}
\end{table*}
\renewcommand{\arraystretch}{1}
\renewcommand{\baselinestretch}{1}

\subsection{Host Galaxy Redshifts}
\label{sec:z}

Heliocentric galaxy recession velocities and CMB frame redshifts are shown in Tables~\ref{Table_LC_params} and \ref{TableDistanceMu1a}. We obtained heliocentric host galaxy recession velocities using the NASA/IPAC Extragalactic Database (NED), using measurements with the smallest reported uncertainty.\footnote{However, even if earlier SDSS data releases report a smaller redshift error, we use the SDSS DR13 (2016) reported heliocentric redshift from NED where available (\citealt{albareti17}; \url{http://www.sdss.org/dr13/data\_access/bulk/}).} If the host galaxy was anonymous or had no reported NED redshift, we used redshifts reported in the literature. When no uncertainties are available, we assume a recession velocity uncertainty of 50 km/s.

To further correct the CMB frame redshifts for local velocity flows and to estimate uncertainties, we used the model of \citealt{carrick15}.\footnote{\url{http://cosmicflows.iap.fr/table\_query.html}. We used defaults of $\Omega_{M}=0.3$ (implicitly $\Omega_{\Lambda}=0.7$ for a flat universe), $\Ho=73.24$ km s$^{-1}$Mpc$^{-1}$ (\citealt{riess16}), $\beta=0.43$, and bulk flows of $(V_x,V_y,V_z)=(89,-131,17)$ km/s (\citealt{carrick15}).} Such corrections are most important for \snIa{} with $z<0.01$ ($v < 3000$ km/s), but we also use them for \snIa{} further into the Hubble flow.

In special cases, we did not use the \citealt{carrick15} flow model and instead used independent information for individual objects. For several \snIa{} that have $\vhel > 3000$ km/s, but are members of known galaxy clusters, to avoid large peculiar velocities from the cluster velocity dispersion, following \citealt{dhawan18}, we used the mean recession velocity of the cluster based on the cluster redshift from NED to estimate the CMB frame recession velocity for the \snIa{} host galaxy. For \sn{}2008hs in Abell 347, we used $\vcmb=5655 \pm 13$ km/s. For \sn{}2010ai in the Coma cluster, we used $\vcmb=7166 \pm 54$ km/s (\citealt{pimbblet14}).

To further avoid peculiar velocity systematics for \snIa{} with $\vhel < 3000$ km/s, where available, we also used redshift-independent distance information from Cepheid variable stars, surface brightness fluctuations (SBF), or the Tully-Fisher method (TF) to estimate an effective CMB frame redshift (see Tables~\ref{TableDistanceMu1a}-\ref{TableDistanceMu2Special} for references).

Of the 19 \snIa{} with Cepheid distances $\mu_{\rm Ceph}$ and uncertainties $\sigma_{\mu_{\rm Ceph}}$ in the HST SHOES program (Table~5 of \citealt{riess16}), \nCepheid{} with NIR data are included in our Table~\ref{TableDistanceMu1a} (\sn{}2001el, \sn{}2003du, \sn{}2005cf, \sn{}2007af, \sn{}2007sr, and \sn{}2011by). One other \snIa{} (\sn{}1998bu) also has Cepheid distance from the HST Key Project (Table~4 of \citealt{freedman01}).\footnote{We use the metallicity corrected values $\mu_Z$ and $\sigma_Z$ from Table~4 of \citealt{freedman01}.} Lastly, \nSBFTF{} more \snIa{} with NIR data (\sn{}2002dj, \sn{}2003hv) had redshift-independent host galaxy distance information from TF and/or SBF (\citealt{courtois12,tully13,tully16}).

For all of these methods, we convert the reported distance modulus $\mu^{\prime}$ on a given Hubble scale $\Ho^{\prime}$ to the Hubble scale of $\Ho=73.24$ km s$^{-1}$Mpc$^{-1}$ as measured by \citealt{riess16} and use this value of $\Ho$ throughout the rest of the paper. More specifically, for Hubble constants in units of km s$^{-1}$Mpc$^{-1}$, the distance modulus $\mu_{\rm eff}$ on our fiducial Hubble scale $\Ho$ is given by
\begin{equation}\label{mufromHo}
\mu_{\rm eff} = 5 \log_{10} \Big(\frac{\Ho^{\prime}}{\Ho} \Big) + \mu^{\prime}
\end{equation}
See Table~\ref{TableDistanceMu2Special}.

For these objects, we convert the redshift independent distance modulus $\muind$ to an effective CMB frame recession velocity with Hubble's law in the linear regime:
\begin{eqnarray}
v_{\rm eff} = c \ z_{\rm eff} & = & \Ho \ d_L(\muind) = \Ho \times 1 {\rm Mpc} \times 10^{\frac{\muind-25}{5}}
\label{ceph2vel}
\end{eqnarray}
with an uncertainty given by\footnote{We do not propagate the uncertainty on $\Ho$ in Eq.~\ref{ceph2vel} because we have fixed the Hubble scale for this work.}
\begin{equation}\label{ceph2velerr}
\sigma_{v_{\rm eff}} = c \ \sigma_{z_{\rm eff}} = c \ z_{\rm eff} \Big( \frac{ \ln 10 }{5} \dmuind \Big)
\end{equation}
For \snIa{} with Cepheid distances, we assume that the only contribution to the recession velocity uncertainty comes from Eq.~\ref{ceph2velerr} and therefore adopt a peculiar velocity uncertainty of $0$ km/s for these objects.

For objects without Cepheid or other redshift-independent distances, we assume a peculiar velocity uncertainty of $\sigmaVpec = 150$ km/s, following \citealt{radburnsmith04}.\footnote{Estimates in the literature range from $\sigmaVpec=150-400$ km/s: $\sigmaVpec=150$ km/s (\citealt{radburnsmith04}), $300$ km/s (\citealt{davis11}), $\sigmaVpec=360$ km/s (\citealt{kessler09}), or $\sigmaVpec = 400$ km/s (\citealt{woodvasey07}).} As shown in Section \ref{sec_disc}, the value of $\sigmaVpec = 150$ km/s yields a more conservative determination of the Hubble diagram intrinsic scatter compared with larger values of $\sigmaVpec$ that tend to produce a misleadingly small value. However, statistics like the RMS, which we also use to compare various methods, are relatively insensitive to the assumed value of $\sigmaVpec$.

\section{NIR \lc{} Templates}
\label{SecTemplate}

\begin{figure*}
\centering
\includegraphics[width=7.1in]{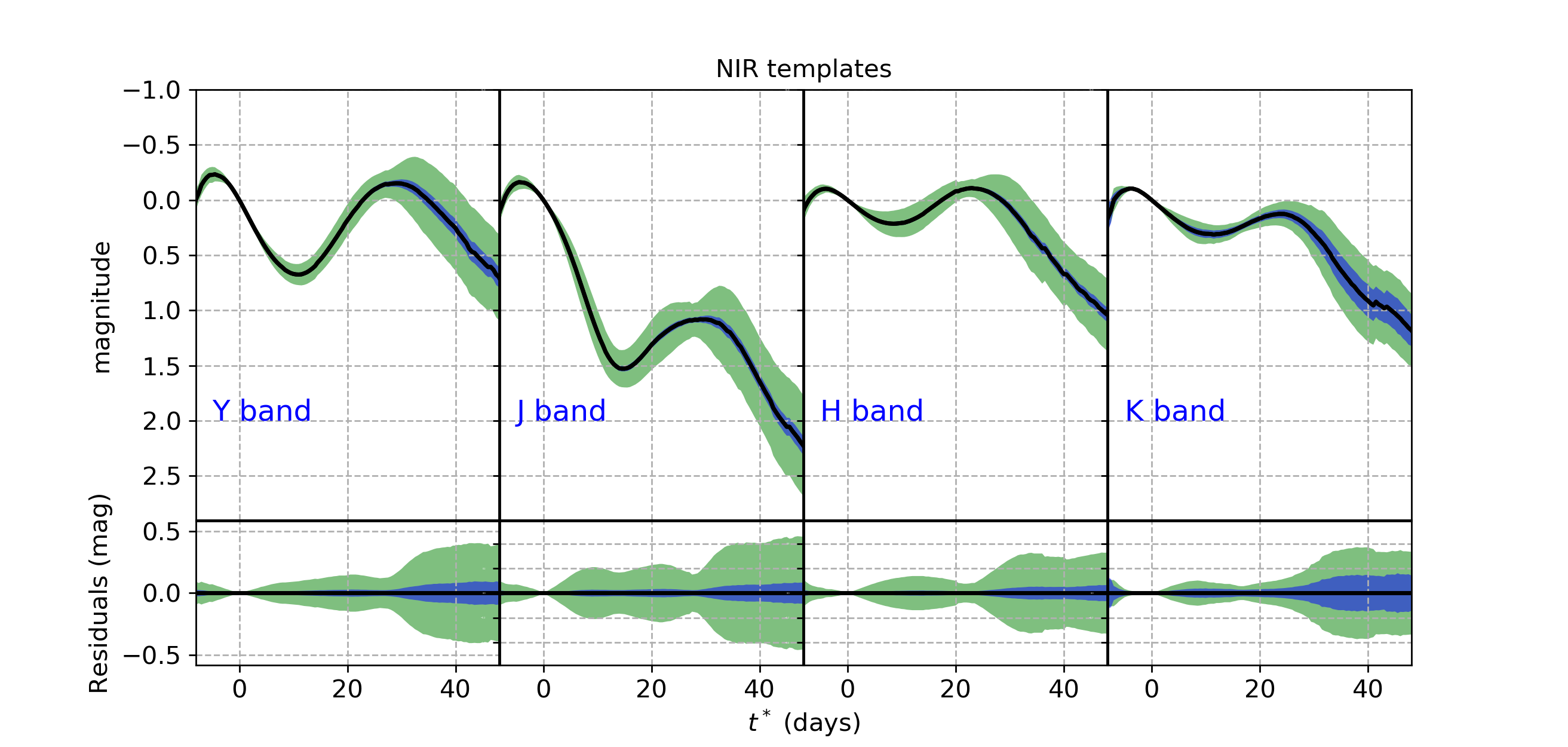}
\caption{
Upper and lower panels show the normalized mean \yjhk{} templates and residual plots, respectively. By construction, we normalize the templates so that they have magnitude zero at $t^* = 0$, with reference to the time of $B$-band maximum light. The numerical values of these templates are tabulated with 1-day sampling in Table~\ref{tab_NIR_templates}. The black curves show the normalized mean magnitude $\hypermeanHBM(t^*)$ vs. rest-frame phase $t^*$, while the green and blue bands correspond to the population standard deviation, $\hyperStdDevHBM(t^*)$, and the uncertainty in $\hypermeanHBM$, respectively, determined using the hierarchical Bayesian model and Gaussian process method described in \S\ref{SecTemplate}. We use 28, 67, 68 and 25 \snIa{} that we can determine $\meanNormaLCVec$ as described in \S\ref{sec_gaussProcess} to build the $Y, J, H$ and $K_s$ templates respectively.}
\label{Fig_templates}
\end{figure*}

\renewcommand{\tabcolsep}{2pt}
\begin{table*}
\begin{center}
\caption{Normalized $\yjhk$ \lc{} Templates}
\scriptsize
\begin{tabular}{c rr rr rr rr}
\hline \hline
 $t^*$  &
 \multicolumn{1}{c}{$\hypermeanHBM^{(Y)}$}  & \multicolumn{1}{c}{$\hyperStdDevHBM^{(Y)}$} & \multicolumn{1}{c}{$\hypermeanHBM^{(J)}$}  & \multicolumn{1}{c}{$\hyperStdDevHBM^{(J)}$} & \multicolumn{1}{c}{$\hypermeanHBM^{(H)}$}  & \multicolumn{1}{c}{$\hyperStdDevHBM^{(H)}$} & \multicolumn{1}{c}{$\hypermeanHBM^{(K)}$}  & \multicolumn{1}{c}{$\hyperStdDevHBM^{(K)}$} \\
\multicolumn{1}{c}{(days)} &  \multicolumn{1}{c}{(mag)}  & \multicolumn{1}{c}{(mag)} &  \multicolumn{1}{c}{(mag)}  & \multicolumn{1}{c}{(mag)} &  \multicolumn{1}{c}{(mag)}  & \multicolumn{1}{c}{(mag)} &  \multicolumn{1}{c}{(mag)}  & \multicolumn{1}{c}{(mag)} \\
\hline
-10  & 0.303 $\pm$ 0.067 & 0.168 & 0.428 $\pm$ 0.037 & 0.136 & 0.367 $\pm$ 0.047 & 0.184 & 0.439 $\pm$ 0.290 & 0.178 \\
-9  & 0.152 $\pm$ 0.034 & 0.108 & 0.261 $\pm$ 0.025 & 0.106 & 0.213 $\pm$ 0.032 & 0.136 & 0.289 $\pm$ 0.335 & 0.194 \\
-8  & -0.007 $\pm$ 0.026 & 0.092 & 0.092 $\pm$ 0.020 & 0.094 & 0.075 $\pm$ 0.023 & 0.104 & 0.150 $\pm$ 0.125 & 0.112 \\
-7  & -0.135 $\pm$ 0.023 & 0.092 & -0.038 $\pm$ 0.015 & 0.076 & -0.010 $\pm$ 0.016 & 0.070 & 0.001 $\pm$ 0.058 & 0.106 \\
-6  & -0.204 $\pm$ 0.018 & 0.078 & -0.115 $\pm$ 0.013 & 0.070 & -0.065 $\pm$ 0.012 & 0.054 & -0.058 $\pm$ 0.034 & 0.066 \\
-5  & -0.228 $\pm$ 0.016 & 0.072 & -0.153 $\pm$ 0.012 & 0.070 & -0.093 $\pm$ 0.010 & 0.046 & -0.088 $\pm$ 0.021 & 0.034 \\
-4  & -0.224 $\pm$ 0.012 & 0.056 & -0.159 $\pm$ 0.010 & 0.060 & -0.102 $\pm$ 0.007 & 0.034 & -0.103 $\pm$ 0.016 & 0.020 \\
-3  & -0.200 $\pm$ 0.008 & 0.042 & -0.148 $\pm$ 0.008 & 0.050 & -0.092 $\pm$ 0.006 & 0.032 & -0.096 $\pm$ 0.011 & 0.014 \\
-2  & -0.151 $\pm$ 0.005 & 0.028 & -0.116 $\pm$ 0.005 & 0.038 & -0.069 $\pm$ 0.004 & 0.024 & -0.073 $\pm$ 0.008 & 0.010 \\
-1  & -0.082 $\pm$ 0.003 & 0.014 & -0.065 $\pm$ 0.003 & 0.022 & -0.037 $\pm$ 0.002 & 0.014 & -0.040 $\pm$ 0.004 & 0.006 \\
0  & 0.000 $\pm$ 0.000 & 0.000 & 0.000 $\pm$ 0.000 & 0.000 & 0.000 $\pm$ 0.000 & 0.000 & 0.000 $\pm$ 0.000 & 0.000 \\
1  & 0.090 $\pm$ 0.002 & 0.012 & 0.075 $\pm$ 0.003 & 0.024 & 0.039 $\pm$ 0.002 & 0.016 & 0.041 $\pm$ 0.005 & 0.014 \\
2  & 0.184 $\pm$ 0.005 & 0.024 & 0.162 $\pm$ 0.007 & 0.050 & 0.079 $\pm$ 0.005 & 0.032 & 0.083 $\pm$ 0.011 & 0.030 \\
3  & 0.276 $\pm$ 0.007 & 0.036 & 0.259 $\pm$ 0.010 & 0.080 & 0.115 $\pm$ 0.007 & 0.048 & 0.125 $\pm$ 0.017 & 0.046 \\
4  & 0.363 $\pm$ 0.009 & 0.050 & 0.369 $\pm$ 0.014 & 0.108 & 0.147 $\pm$ 0.009 & 0.064 & 0.163 $\pm$ 0.022 & 0.062 \\
5  & 0.441 $\pm$ 0.011 & 0.060 & 0.492 $\pm$ 0.018 & 0.140 & 0.173 $\pm$ 0.012 & 0.078 & 0.200 $\pm$ 0.026 & 0.074 \\
6  & 0.510 $\pm$ 0.013 & 0.072 & 0.628 $\pm$ 0.022 & 0.168 & 0.194 $\pm$ 0.013 & 0.092 & 0.233 $\pm$ 0.030 & 0.086 \\
7  & 0.568 $\pm$ 0.015 & 0.080 & 0.774 $\pm$ 0.025 & 0.188 & 0.207 $\pm$ 0.015 & 0.104 & 0.262 $\pm$ 0.034 & 0.096 \\
8  & 0.612 $\pm$ 0.016 & 0.086 & 0.921 $\pm$ 0.026 & 0.202 & 0.214 $\pm$ 0.017 & 0.114 & 0.284 $\pm$ 0.037 & 0.100 \\
9  & 0.650 $\pm$ 0.016 & 0.088 & 1.066 $\pm$ 0.028 & 0.210 & 0.214 $\pm$ 0.018 & 0.122 & 0.298 $\pm$ 0.037 & 0.100 \\
10  & 0.670 $\pm$ 0.017 & 0.092 & 1.199 $\pm$ 0.027 & 0.208 & 0.208 $\pm$ 0.018 & 0.128 & 0.306 $\pm$ 0.038 & 0.094 \\
11  & 0.676 $\pm$ 0.017 & 0.096 & 1.316 $\pm$ 0.027 & 0.200 & 0.198 $\pm$ 0.019 & 0.134 & 0.309 $\pm$ 0.035 & 0.086 \\
12  & 0.666 $\pm$ 0.019 & 0.104 & 1.419 $\pm$ 0.026 & 0.186 & 0.179 $\pm$ 0.020 & 0.138 & 0.309 $\pm$ 0.035 & 0.080 \\
13  & 0.640 $\pm$ 0.020 & 0.108 & 1.487 $\pm$ 0.024 & 0.170 & 0.156 $\pm$ 0.020 & 0.138 & 0.305 $\pm$ 0.034 & 0.078 \\
14  & 0.600 $\pm$ 0.022 & 0.116 & 1.525 $\pm$ 0.024 & 0.166 & 0.127 $\pm$ 0.021 & 0.138 & 0.295 $\pm$ 0.033 & 0.074 \\
15  & 0.540 $\pm$ 0.023 & 0.118 & 1.529 $\pm$ 0.024 & 0.170 & 0.089 $\pm$ 0.019 & 0.134 & 0.278 $\pm$ 0.032 & 0.068 \\
16  & 0.477 $\pm$ 0.024 & 0.126 & 1.513 $\pm$ 0.026 & 0.182 & 0.053 $\pm$ 0.019 & 0.128 & 0.257 $\pm$ 0.028 & 0.058 \\
17  & 0.409 $\pm$ 0.025 & 0.132 & 1.478 $\pm$ 0.028 & 0.195 & 0.017 $\pm$ 0.019 & 0.122 & 0.234 $\pm$ 0.027 & 0.056 \\
18  & 0.336 $\pm$ 0.026 & 0.138 & 1.429 $\pm$ 0.029 & 0.208 & -0.018 $\pm$ 0.018 & 0.116 & 0.211 $\pm$ 0.028 & 0.064 \\
19  & 0.261 $\pm$ 0.026 & 0.142 & 1.373 $\pm$ 0.030 & 0.216 & -0.049 $\pm$ 0.017 & 0.108 & 0.188 $\pm$ 0.030 & 0.074 \\
20  & 0.183 $\pm$ 0.027 & 0.144 & 1.312 $\pm$ 0.032 & 0.226 & -0.079 $\pm$ 0.016 & 0.100 & 0.169 $\pm$ 0.031 & 0.082 \\
21  & 0.114 $\pm$ 0.029 & 0.150 & 1.263 $\pm$ 0.033 & 0.232 & -0.091 $\pm$ 0.014 & 0.080 & 0.149 $\pm$ 0.033 & 0.088 \\
22  & 0.054 $\pm$ 0.028 & 0.148 & 1.221 $\pm$ 0.034 & 0.236 & -0.102 $\pm$ 0.014 & 0.076 & 0.138 $\pm$ 0.034 & 0.094 \\
23  & -0.007 $\pm$ 0.027 & 0.142 & 1.183 $\pm$ 0.032 & 0.228 & -0.108 $\pm$ 0.014 & 0.082 & 0.130 $\pm$ 0.036 & 0.100 \\
24  & -0.056 $\pm$ 0.026 & 0.134 & 1.151 $\pm$ 0.032 & 0.218 & -0.102 $\pm$ 0.016 & 0.094 & 0.126 $\pm$ 0.039 & 0.110 \\
25  & -0.095 $\pm$ 0.025 & 0.126 & 1.124 $\pm$ 0.029 & 0.198 & -0.092 $\pm$ 0.019 & 0.120 & 0.132 $\pm$ 0.043 & 0.122 \\
26  & -0.123 $\pm$ 0.024 & 0.120 & 1.106 $\pm$ 0.026 & 0.180 & -0.075 $\pm$ 0.023 & 0.152 & 0.147 $\pm$ 0.047 & 0.134 \\
27  & -0.144 $\pm$ 0.024 & 0.124 & 1.091 $\pm$ 0.025 & 0.170 & -0.050 $\pm$ 0.027 & 0.182 & 0.172 $\pm$ 0.054 & 0.156 \\
28  & -0.147 $\pm$ 0.026 & 0.134 & 1.085 $\pm$ 0.024 & 0.154 & -0.018 $\pm$ 0.031 & 0.214 & 0.205 $\pm$ 0.060 & 0.174 \\
29  & -0.151 $\pm$ 0.031 & 0.162 & 1.081 $\pm$ 0.028 & 0.178 & 0.021 $\pm$ 0.036 & 0.244 & 0.248 $\pm$ 0.068 & 0.200 \\
30  & -0.149 $\pm$ 0.038 & 0.196 & 1.082 $\pm$ 0.034 & 0.224 & 0.067 $\pm$ 0.041 & 0.272 & 0.302 $\pm$ 0.084 & 0.238 \\
31  & -0.137 $\pm$ 0.046 & 0.238 & 1.090 $\pm$ 0.040 & 0.274 & 0.124 $\pm$ 0.043 & 0.294 & 0.356 $\pm$ 0.095 & 0.262 \\
32  & -0.115 $\pm$ 0.053 & 0.274 & 1.112 $\pm$ 0.048 & 0.322 & 0.183 $\pm$ 0.046 & 0.312 & 0.414 $\pm$ 0.111 & 0.306 \\
33  & -0.083 $\pm$ 0.058 & 0.304 & 1.134 $\pm$ 0.054 & 0.354 & 0.248 $\pm$ 0.049 & 0.316 & 0.484 $\pm$ 0.118 & 0.322 \\
34  & -0.038 $\pm$ 0.065 & 0.332 & 1.186 $\pm$ 0.059 & 0.382 & 0.324 $\pm$ 0.052 & 0.326 & 0.566 $\pm$ 0.133 & 0.342 \\
35  & 0.007 $\pm$ 0.070 & 0.340 & 1.232 $\pm$ 0.063 & 0.396 & 0.375 $\pm$ 0.052 & 0.320 & 0.636 $\pm$ 0.138 & 0.350 \\
36  & 0.054 $\pm$ 0.075 & 0.356 & 1.305 $\pm$ 0.065 & 0.408 & 0.438 $\pm$ 0.053 & 0.320 & 0.699 $\pm$ 0.144 & 0.358 \\
37  & 0.105 $\pm$ 0.078 & 0.364 & 1.378 $\pm$ 0.065 & 0.396 & 0.472 $\pm$ 0.050 & 0.298 & 0.762 $\pm$ 0.146 & 0.362 \\
38  & 0.156 $\pm$ 0.077 & 0.370 & 1.464 $\pm$ 0.068 & 0.408 & 0.545 $\pm$ 0.050 & 0.294 & 0.816 $\pm$ 0.148 & 0.372 \\
39  & 0.208 $\pm$ 0.077 & 0.372 & 1.555 $\pm$ 0.069 & 0.408 & 0.605 $\pm$ 0.050 & 0.294 & 0.868 $\pm$ 0.141 & 0.368 \\
40  & 0.256 $\pm$ 0.082 & 0.384 & 1.647 $\pm$ 0.068 & 0.402 & 0.672 $\pm$ 0.050 & 0.292 & 0.912 $\pm$ 0.148 & 0.370 \\
41  & 0.325 $\pm$ 0.084 & 0.388 & 1.736 $\pm$ 0.069 & 0.406 & 0.703 $\pm$ 0.049 & 0.274 & 0.955 $\pm$ 0.142 & 0.356 \\
42  & 0.390 $\pm$ 0.086 & 0.396 & 1.820 $\pm$ 0.071 & 0.418 & 0.759 $\pm$ 0.050 & 0.282 & 0.947 $\pm$ 0.140 & 0.334 \\
43  & 0.469 $\pm$ 0.090 & 0.404 & 1.922 $\pm$ 0.076 & 0.436 & 0.818 $\pm$ 0.053 & 0.294 & 0.981 $\pm$ 0.133 & 0.332 \\
44  & 0.507 $\pm$ 0.094 & 0.404 & 1.992 $\pm$ 0.077 & 0.440 & 0.852 $\pm$ 0.055 & 0.302 & 0.987 $\pm$ 0.149 & 0.335 \\
45  & 0.557 $\pm$ 0.091 & 0.400 & 2.055 $\pm$ 0.081 & 0.454 & 0.908 $\pm$ 0.060 & 0.312 & 1.028 $\pm$ 0.149 & 0.338 \\
\hline
\end{tabular}
\label{tab_NIR_templates}
\end{center}
\tablecomments{
Mean NIR \lc{} templates in the $\yjhk$-bands using the hierarchical Bayesian model and Gaussian process method described in \S\ref{SecTemplate}. These are referenced to the time of $B$-band maximum light, such that $t^*=0$ at $t_{B {\rm max}}$. See Fig.~\ref{Fig_templates}.
}
\end{table*}
\renewcommand{\tabcolsep}{6pt}
\renewcommand{\arraystretch}{1}

We determine the normalized mean $\yjhk$ \lc{} templates, as shown in Figure~\ref{Fig_templates} and Table~\ref{tab_NIR_templates}, using the \snIa{} in Table~\ref{Table_LC_params} as follows. In each band, we convert the photometry from the observer-frame apparent magnitude to the rest-frame absolute magnitude. We further apply $K$-corrections to the rest-frame and correct for Milky Way dust extinction. These steps are detailed in \S\ref{sec:datacorr}. We then use a \textit{Gaussian process} method, as described in \S\ref{sec_gaussProcess}, to fit the \lc{} in each NIR band. Finally, in \S\ref{sec_HBM}, using a hierarchical Bayesian model we average all the \lcs{} in a given NIR band to determine the normalized mean \lc{} template, its uncertainty, and the population standard deviation.

\subsection{Rest-Frame Absolute Magnitudes}
\label{sec:datacorr}

For a given supernova $s$ observed through filter $O$, we convert the apparent magnitude $m_s$ datum observed at the modified Julian day (MJD) $t_{\text{MJD}}$ to the absolute magnitude $M_s$ at rest-frame phase $t$, via
\begin{equation}\label{eq_absmagDef}
    M_s(t) = m_s(t) - \muLCDM(z_{s}) - K^s_{O Q} - A^s_{O},
\end{equation}
where $z_s$ is the spectroscopic redshift of the supernova $s$ with respect to the CMB, including any local flow models (see Table~\ref{Table_LC_params}). The phase ${t} \equiv (t_{\text{MJD}} - \TBmax)/(1+z_{{\rm helio},s})$ is the rest-frame observation time, corrected for cosmological time dilation,  $z_{{\rm helio},s}$ is the heliocentric redshift, and $\TBmaxx$ is the time of $B$-band maximum light. The term $K_{O Q}$ is the $K$-correction from the observed band $O$ to the rest-frame band $Q$, and $A_O$ is the Milky Way foreground extinction defined as $A_O = R_O  \ebvmw$, where $R_O$ is the total-to-selective extinction ratio in filter $O$ and $\ebvmw$ is the Milky Way color excess. We use the \citet{SchlaflyFinkbeiner2011} dust reddening map for $\ebvmw$, and the CCM+O \citep{ODonnell1994} reddening laws to determine $R_O$ for the NIR and optical bands respectively. We assume a $V$-band total-to-selective extinction ratio for the Milky Way of $R_V = 3.1$.

We determine $\TBmaxx$ and compute the $K$-correction $K^s_{OQ}$ terms using a module in the \snoopy{} \lc{} package  (\citealt{BurnsEtal2011Snoopy}), which uses the normal \snIa{} spectroscopic template of \citealt{hsiao07} that is ``mangled'' to match the actual colors derived from the data.

The theoretical distance modulus is defined as
\begin{equation}\label{eq:dmodth}
\muLCDM(z_{s}) = 5 \log_{10}\Bigg[ \frac{d_L(z_{s})}{1 \text{Mpc} } \Bigg] + 25
\end{equation}
We assume the luminosity distance $d_L(z)$ for a spatially flat $\Lambda$CDM Universe, ignoring radiation, is approximately given by
\begin{equation}\label{eq:dl}
d_L(z) = \Big(\frac{c}{\Ho}\Big)(1+z) \int_{0}^{z} \frac{dz}{E(z)}
\end{equation}
where $E(z) = \sqrt{\Omega_{\rm m}(1+z)^3 + \Omega_\Lambda}$ and $c$ is the speed of light. We assume fiducial values for the matter and energy density fractions of $\Omega_{\rm m} = 0.28$ and $\Omega_\Lambda = 0.72$ and a Hubble constant of $\Ho = 73.24$ km s$^{-1}$ Mpc$^{-1}$ (\citealt{riess16}).

Every value of $M_s$ has an error variance
\begin{equation}\label{eq_err_AbsMag}
    \sigma^2_M = \sigma^2_m  + \smupecNoS^2 + \sigma^2_A + \sigma^2_{\rm Kcorr},
\end{equation}
where $\sigma_m$ is the measurement uncertainty of the apparent magnitude $m_s$, $\smupecNoS$ is the uncertainty in the distance modulus $\muLCDM(z_{s})$ due to the peculiar velocity and redshift uncertainties, given as
\begin{equation} \label{eq_PeculiarVelDefinition}
\smupec^2(z_s)  = \left( \frac{5}{z_s\ln(10)} \right)^2 \left[ \left( \frac{\sigmaVpec}{c} \right)^2 + \sigma^2_{z, s} \right].
\end{equation}
For the SN in Table \ref{TableDistanceMu2Special} with independent distance estimates, we use those corresponding distance modulus uncertainties. The term $\sigma_A$ in Eq. (\ref{eq_err_AbsMag}) is the uncertainty in the Milky Way dust extinction $A_O$ computed as, $\sigma_A = R_O \, \sigma_{\text{EBV}}$, where  $\sigma_{\text{EBV}}$ is the uncertainty in the Milky Way color excess $\ebvmw$, and $\sigma_{\rm Kcorr}$ is the K-correction uncertainty estimated using Monte Carlo simulations of the full optical and NIR dataset $\{ m_s\}$ for a given SN. In this Monte Carlo approach, for each photometric datum at a given MJD time and band, $m_s(T_{\rm MJD})$, we simulate a realization of this datum by drawing a random value from a Gaussian distribution with mean and standard deviation equal to the measured values $m_s$ and $\sigma_m$. For each simulated full optical+NIR dataset, we compute the K-corrections and then determine the mean and standard deviation of the distribution of the K-correction values for each photometric datum at a given MJD time and band. We use this standard deviation as an estimation of the uncertainty of the K-correction value for that datum.

\subsection{\lc{} Fitting: Gaussian process}
\label{sec_gaussProcess}

The Gaussian process technique is a non-parametric Bayesian method that we use to fit the NIR \lcs{} for each \snIa{} in Table~\ref{Table_LC_params}. A Gaussian process defines a prior over functions.  Realizations from a GP, evaluated on a discrete set of times, are random vectors drawn from a joint multivariate Gaussian distribution, $\GaussDistr$, of dimension equal to the number of components in the vector. Given a dataset, the GP formalism allows us to coherently determine the posterior mean function that fits the dataset along with its posterior covariance. The GP methodology is especially helpful in accounting for missing data (in our cases, phases with no observations), and when the data are correlated as in the case of the \snIa{} \lcs{}. \citet{Rasmussen_Williams_GP} provide an introduction to GPs for machine learning.

The following description applies to a \lc{} of a single supernova in a given NIR band. We model the absolute magnitude $M$ at phase $t$ as a noisy measurement of the latent (true) absolute magnitude $\latM$ at that phase, given by $M(t) = \latM(t) + \epsilon$, where $\epsilon \sim \GaussDistr(0, \sigma^2_M)$. In vector notation we express the collection of absolute magnitude data of a given \lc{} as $\mathbf{M} \equiv \left[ M(t_1), M(t_2), ..., M(t_{\nLCs}) \right]^\top$, measured at phases $\mathbf{t} \equiv \left[t_1, t_2, ..., t_{\nLCs} \right]^\top$, where $\nLCs$ is the number of data in the \lc{}, and $\top$ means the transpose.

Using GP, we estimate the posterior mean, $\postGPmeanVec$, and the posterior covariance, $\postGPcovMatrix$, of the latent absolute magnitudes $\latMVec^* \equiv \left[  \latM(t^*_1), \latM(t^*_2), ..., \latM(t^*_{\nGPgrid}) \right]^\top$ on a regular grid of phases $\mathbf{t}^* \equiv [t^*_1, t^*_2, ..., t^*_{\nGPgrid}]^\top$, where $\nGPgrid$ is the number of times in the grid determined from a sequence of phases between ${t}_{{\rm min}, s}$ and ${t}_{\text{max}, s} $ in steps of 0.5 days, where ${t}_{\text{min}, s}$ and ${t}_{\text{max}, s} $ are the minimum and maximum phases in $\mathbf{t}^*$. Thus the number of times in the regular grid is $\nGPgrid = ({t}_{\text{max},s} - {t}_{\text{min},s} )/0.5$.
In Appendix \ref{Sec_GaussianProc}, we provide the mathematical details to determine $\postGPmeanVec$ and $\postGPcovMatrix$.

\subsubsection{Normalization of the GP light curves}
\label{Sec_NormalizationGP}

Our goal with the GP fitting is to determine the \textit{shape} of the \lc{} to be used later in Section \ref{sec_HBM} to construct NIR templates to fit the data and estimate distance moduli. So once we determine the posterior mean and covariance of the latent absolute magnitude \lc{} for a given supernova $s$ using GP, we \textit{normalize} the \lcs{} to extract the information about their shape regardless of their absolute magnitudes.  The normalized LC $L(t)$ is the function, over phase, of the difference in magnitudes relative to the peak phase, so that $L(t_{Bmax}) = 0$.

To estimate the distance moduli, we choose to use the phase of $B$-band maximum light, $\tbmaxx$, as the reference time to derive the distances. In Section \ref{sec_dmod_gp}, we also implement the estimation of distance moduli using the time of NIR-band maximum light instead of $\tbmaxx$ as the reference time.

We define the vector $\normaLCVec$, corresponding to the normalized \lc{} derived from the latent absolute magnitude \lc{} $\latMVec^*$, evaluated on the phase grid $\mathbf{t}^*$, as
\begin{equation}\label{eq_defNormaLC}
    \normaLCVec \equiv \latMVec^* - \latM_{0} \Ivector
\end{equation}
where $\latM_{0}$ is the latent absolute magnitude at $\tbmaxx$ and $\Ivector$ is a vector of dimension $\nGPgrid$ with all its elements equal to one. Since this is a linear transformation of $\latMVec^*$ into $\mathbf{L}$, and $\latMVec^*$ is Gaussian, therefore $\normaLCVec$ is also Gaussian and described completely by its mean $\meanNormaLCVec$ and covariance $\covNormaLCMatrix$. See Appendix \ref{Sec_NormalizationGP_Appendix} for details. In the next section we use $(\meanNormaLCVec,\covNormaLCMatrix)$ to construct the NIR light curve templates.

\subsection{Hierarchical Bayesian Model for the Normalized Magnitudes}
\label{sec_HBM}

In this section, we describe how we construct NIR \lc{} templates for the $Y$, $J$, $H$, and $K_s$ bands that correspond to the mean \textit{shape} of \snIa{} \lcs{} in each of these bands. To do so, we combine the normalized \lcs{} described by $( \meanNormaLCVec, \covNormaLCMatrix )$, from all the supernovae at a given phase $t^*$ using a hierarchical Bayesian model to determine the mean normalized magnitude. Then we repeat the procedure described below over all the phases in $\mathbf{t}^*$ to construct the final NIR \lc{} templates.

First, we assume the normalized magnitude at  phase $t^*$, $ \meanNormaLC_s $,  for the supernova $s$ is drawn from a Gaussian distribution with true value $\AbsMagTilde$ and standard deviation $\AbsMagTildeSigma$,
\begin{equation}\label{EqGaussianParams1}
\meanNormaLC_s \sim \GaussDistr (\AbsMagTilde, \AbsMagTildeSigmaSq)
\end{equation}
where the value of $\AbsMagTildeSigmaSq$ is given by the $(t^*_s, t^*_s)$  element in the diagonal of the covariance matrix $\covNormaLCMatrix_s$ [see Eq. (\ref{eq_covNormaLCMatrix_1})]. Next, we assume that the set of values $\{ \AbsMagTilde \}$ from all the $\nSNTast$ supernovae at phase $t^*$, are independent draws from a Gaussian population distribution with population mean $\hypermeanHBM$ and variance $\hyperStdDevHBM^2$,
\begin{equation}\label{EqGaussianHyperpars1}
p \left( \{ \AbsMagTilde \} |  \hypermeanHBM, \hyperStdDevHBM^2  \right) = \prod^{\nSNTast}_{s=1} \GaussDistr \left( \AbsMagTilde  | \hypermeanHBM, \hyperStdDevHBM^2  \right)
\end{equation}

In Appendix \ref{Section_HierarBayes}, we write the expression for the joint posterior distribution of the hierarchical model and describe additional decompositions in order to make the computations more tractable~to determine the posterior inference\footnote{We use the {\it median} of the posterior probability distribution as the best estimated value.} of ($\{ \AbsMagTilde\}, \hypermeanHBM, \hyperStdDevHBM$) given the data $\{ (\meanNormaLC_s, \AbsMagTildeSigma) \}$ at phase $t^*$.

We repeat the above procedure for all phases in the range $t^* = (-10, 45)$ days, every 0.5 days, to determine $(\hypermeanHBM, \hyperStdDevHBM)$ for all $t^*$ in this range. Figure{}~\ref{Fig_templates} shows the $\yjhk$ templates constructed with this procedure and Table~\ref{tab_NIR_templates} reports the numerical values of the templates. The posterior estimates of the population mean and variance of the normalized LC, $(\hypermeanHBM, \hyperStdDevHBM^2)$, and the uncertainty in the determination of $\hypermeanHBM$, are shown in Figure{}~\ref{Fig_templates} as black curves, green bands, and blue bands, respectively.

\section{Hubble diagram}
\label{SecHubbleDiagram}
We implement two different methods to derive the distance modulus for each supernova from the NIR \lcs. We call them the \textit{template method} and the \textit{Gaussian-process method} (GP). The GP method requires data near the NIR maximum for all NIR bands being used, while the template method works for arbitrarily sampled data, even if the \lc{} is sparse near maximum. For this reason, we have more objects in the template method Hubble diagrams. We describe these methods in more detail in the following sections.

Any of these NIR-only approaches approximately treat the information in each of the $\yjhk$ bands as independent. However, this simple approach does not take maximal advantage of the cross-band correlations between each of the NIR and optical bands, as is done using a more sophisticated hierarchical Bayesian model (e.g. \bayesn{}: \citealt{mandel09,mandel11,mandel14a}). Nor does this approach use the fact that there is only one true distance to the supernova.

To alleviate this problem, we also derive the distance modulus for each supernova from the \textit{combined} distance moduli in each NIR band. However, instead of computing a simple average distance modulus from the individual distance moduli, we instead estimate the covariance matrix of the $\yjhk$ distance moduli (and submatrices of it) and then derive the \textit{weighted} average distance modulus. The advantage of this procedure is that it takes into account the correlations among the magnitudes in the NIR bands and then derives more realistic mean distance moduli and their uncertainties. More details are in Section \ref{sec_dmod_AnyNIR}.

For our NIR-only Hubble diagrams, only NIR \lcs{} are used to directly construct distance moduli. However, auxiliary optical data is used to estimate $\tbmaxx$, \dm{}, and mangled K-corrections, and is employed in the input data selection cuts described in \S\ref{sec:cuts}.

\subsection{Distance Modulus: Template method}
\label{sec_dmod_template}

To determine the photometric distance modulus $\mu_s$ of the supernova $s$ in a given NIR band, we use the normalized mean template, $\hypermeanHBM$, computed in Section \ref{SecTemplate}, to determine the apparent magnitude at phase zero, $m_{0,s} \equiv m_s(t = 0)$, by fitting the template to the sometimes sparse photometric \lc{} data $\{ m_s(t) \}$.
We define the difference
\begin{equation}
\Delta m_s({t}) \equiv m_s({t}) - \hypermeanHBM(t) - m_{0,s}
\end{equation}
where $m_s(t)$ and $\hypermeanHBM(t)$ are the apparent magnitude and the magnitude of the normalized template at phase ${t}$, respectively. We can express this difference for all the $\nLCs$ phases in a given \lc{} as the vector,
\begin{equation}
\Delta \boldsymbol{m}_s \equiv
\begin{pmatrix}
\Delta m_s({t}_1) \\
\Delta m_s({t}_2) \\
\vdots \\
\Delta m_s({t}_{\nLCs})
\end{pmatrix}.
\end{equation}
Then, to determine $m_{0,s}$ we minimize the negative of the log likelihood function $\likelihood(m_{0,s})$ defined as
\begin{equation}\label{Eq_NegLogLikeMu}
- 2 \ln \likelihood(m_{0,s}) = \Delta \boldsymbol{m}_s^{\top} \cdot \mathbf{C}^{-1} \cdot \Delta \boldsymbol{m}_s + {\rm constant},
\end{equation}
where $\mathbf{C}$ is the $\nLCs$-dimensional covariance matrix where the $(t_i, t_j)$ component is given by:
\begin{align}
C_{ij} \equiv  & \text{Cov} \left( \Delta m_s(t_i), \Delta m_s(t_j) \right) \\
= & \hyperStdDevHBM(t_i) \hyperStdDevHBM(t_j)  \exp \left[ - \frac{(t_i -t_j)^2}{2 l^2} \right] + \nonumber \\
& \hat{\sigma}_{m,s}^2(t_i) \delta_{ij},
\end{align}
where $\hyperStdDevHBM({t})$ is the population standard deviation of the sample distribution of magnitudes at phase ${t}$, determined from Eq.~(\ref{EqJointPosterior2}) during the training process used to construct the mean \lc{} template, $\hat{\sigma}_{m,s}^2(t_i)$ is the photometric error of the datum $m_s(t_i)$, and $l$ is the hyperparameter of GP kernel determined from Eq. (\ref{EqGlobalLike}) and with values shown in Table \ref{tab_GP_hyperpars}.

From Eq.~(\ref{Eq_NegLogLikeMu}), we can calculate an analytic expression for the maximum likelihood estimator (MLE) of the apparent magnitude at $B$-band maximum light, $\hat{m}_{0,s}$, given by:
\begin{multline}
\hat{m}_{0,s} = \left[ \sum^{\nLCs}_{i,j} (C^{-1})_{ij} \right]^{-1} \times  \\
\sum^{\nLCs}_i \left[ \Bigl( m_s (t_i) - \hypermeanHBM(t_i) \Bigr) \sum^{\nLCs}_j (C^{-1})_{ij}  \right],
\label{eq:mus}
\end{multline}
with the MLE of the uncertainty of  $\hat{m}_{0,s}$ given as
\begin{equation}\label{err_fit}
\sigmafithat = \left[ \sum_{i,j}^{\nLCs} (C^{-1})_{ij} \right]^{-1/2},
\end{equation}
which corresponds to the fitting error of the light curve.  This error incorporates the photometric measurement error and the sparsity of the actual data points.

\begin{figure}
\centering
\includegraphics[width=8cm]{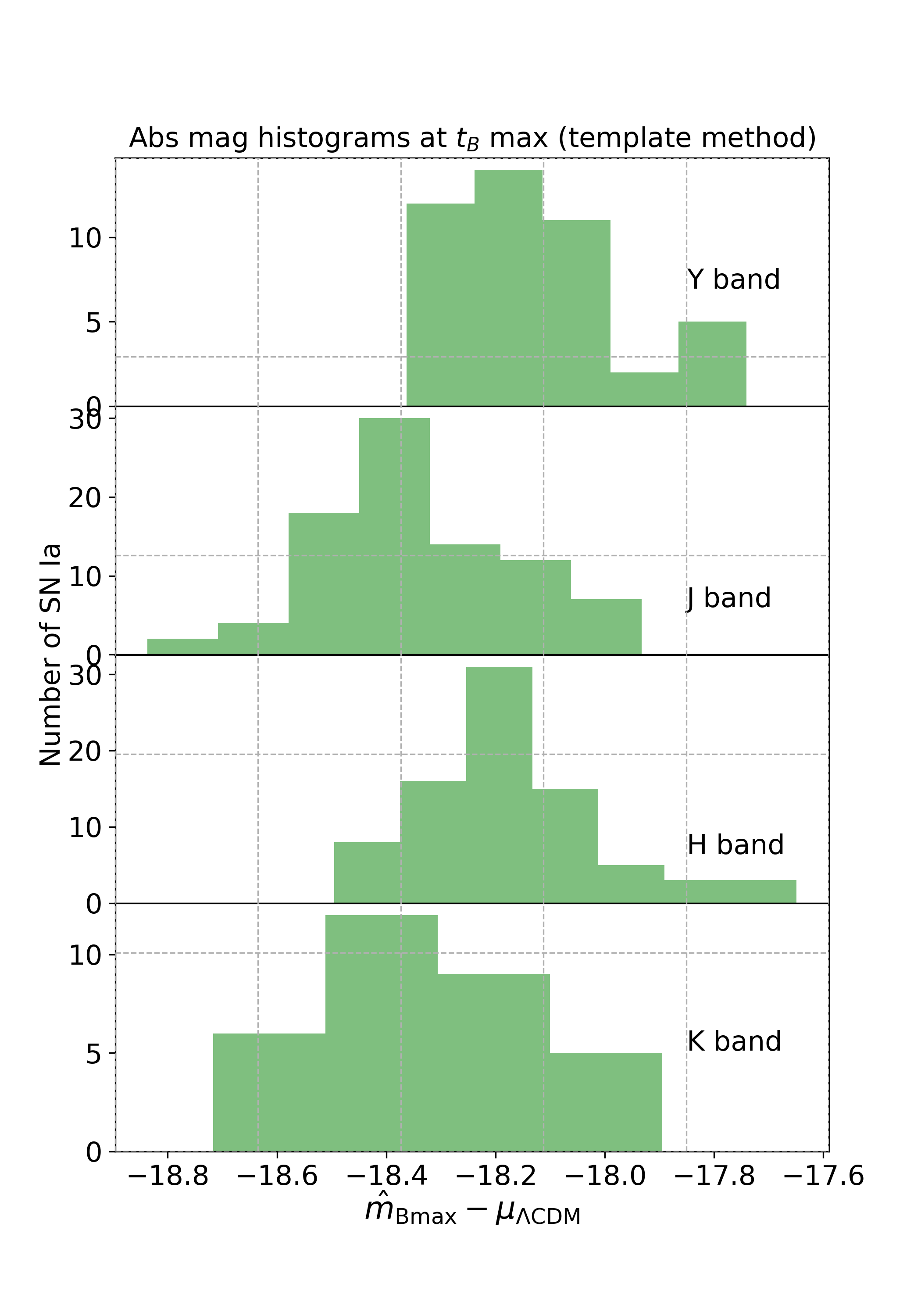}
\caption{
 Histograms of the absolute magnitudes at phase zero ($t^{*}=\tbmaxx$), defined as $M_{0,s} \equiv \hat{m}_{0,s} - \muLCDM(z_s)$ for the \snIa{} sample using the template method. The sample mean, standard deviation, and the number of supernovae in each histogram are shown in Table \ref{Table_AbsMags}.
}
\label{Fig_histo_AbsMag_Template}
\end{figure}

Now, from the distribution of absolute magnitudes at phase zero estimated as $M_{0,s} \equiv  \hat{m}_{0,s} - \mu_{\Lambda{\rm CDM}}(z_s)$ (see Fig.~\ref{Fig_histo_AbsMag_Template}), we compute the sample mean absolute magnitude, $\langle M_0\rangle$, and the sample standard deviation of the distribution, obtaining the values reported in Table~\ref{Table_AbsMags}.  The sample standard deviation describes the total scatter of the absolute magnitude estimates.  Below, we decompose this into the contributions from peculiar velocity-distance errors, measurement/fitting errors, and intrinsic dispersion.

\begin{table}
\begin{center}
\caption{Mean $\yjhk$ absolute magnitudes at $t_{B{\rm max}}$ or $t_{\rm NIR}$ max.}
\begin{tabular}{l c c c}
\hline \hline \\[-0.3cm]
Band & $\nSNHD$ & $\langle M \rangle$ & Std. deviation \\
 &  & (mag) & (mag) \\[0.1cm]
\multicolumn{4}{c}{Template method} \\
\hline \\[-0.3cm]
$Y$ & 44 & $-18.12$ & $0.15$ \\
$J$ & 87 & $-18.34$ & $0.17$ \\
$H$ & 81 & $-18.18$ & $0.17$ \\
$K_s$ & 32 & $-18.35$ & $0.21$ \\
\hline \\
\multicolumn{4}{c}{Gaussian-process method at NIR max} \\
\hline \\[-0.3cm]
$Y$ & 29 & $-18.39$ & $0.11$ \\
$J$ & 52 & $-18.52$ & $0.14$ \\
$H$ & 44 & $-18.30$ & $0.11$ \\
$K_s$ & 14 & $-18.37$ & $0.18$ \\
\hline \\
\multicolumn{4}{c}{Gaussian-process method at $B$ max} \\
\hline \\[-0.3cm]
$Y$ & 29 & $-18.16$ & $0.12$ \\
$J$ & 52 & $-18.34$ & $0.15$ \\
$H$ & 44 & $-18.19$ & $0.12$ \\
$K_s$ & 14 & $-18.28$ & $0.17$ \\
\hline \\
\end{tabular}
\label{Table_AbsMags}
\end{center}
\tablecomments{
We use the sample mean values of the absolute magnitudes, $\langle M \rangle$, in each band to determine the distance modulus in the template and GP methods using Eqs. (\ref{DistMu_EachBand}) and (\ref{DistMu_EachBand_GP}) respectively. For the template method $\langle M \rangle \equiv \langle M_0 \rangle$
and for the GP method $\langle M \rangle \equiv \langle M_{\rm NIRmax} \rangle$, $\langle M_{\rm Bmax} \rangle$. Figs.~\ref{Fig_histo_AbsMag_Template}, \ref{Fig_histo_AbsMag_GP} and \ref{Fig_histo_AbsMag_GP_Bmax} show the histograms of $M_{0,s}$ and $ M_{{\rm NIRmax},s}$, respectively. In this table, we also present the sample standard deviation of the absolute magnitude sample distribution just as a reference, we do not use those values in any part of the computations.
}
\end{table}

\begin{figure}
\centering
\includegraphics[width=8cm]{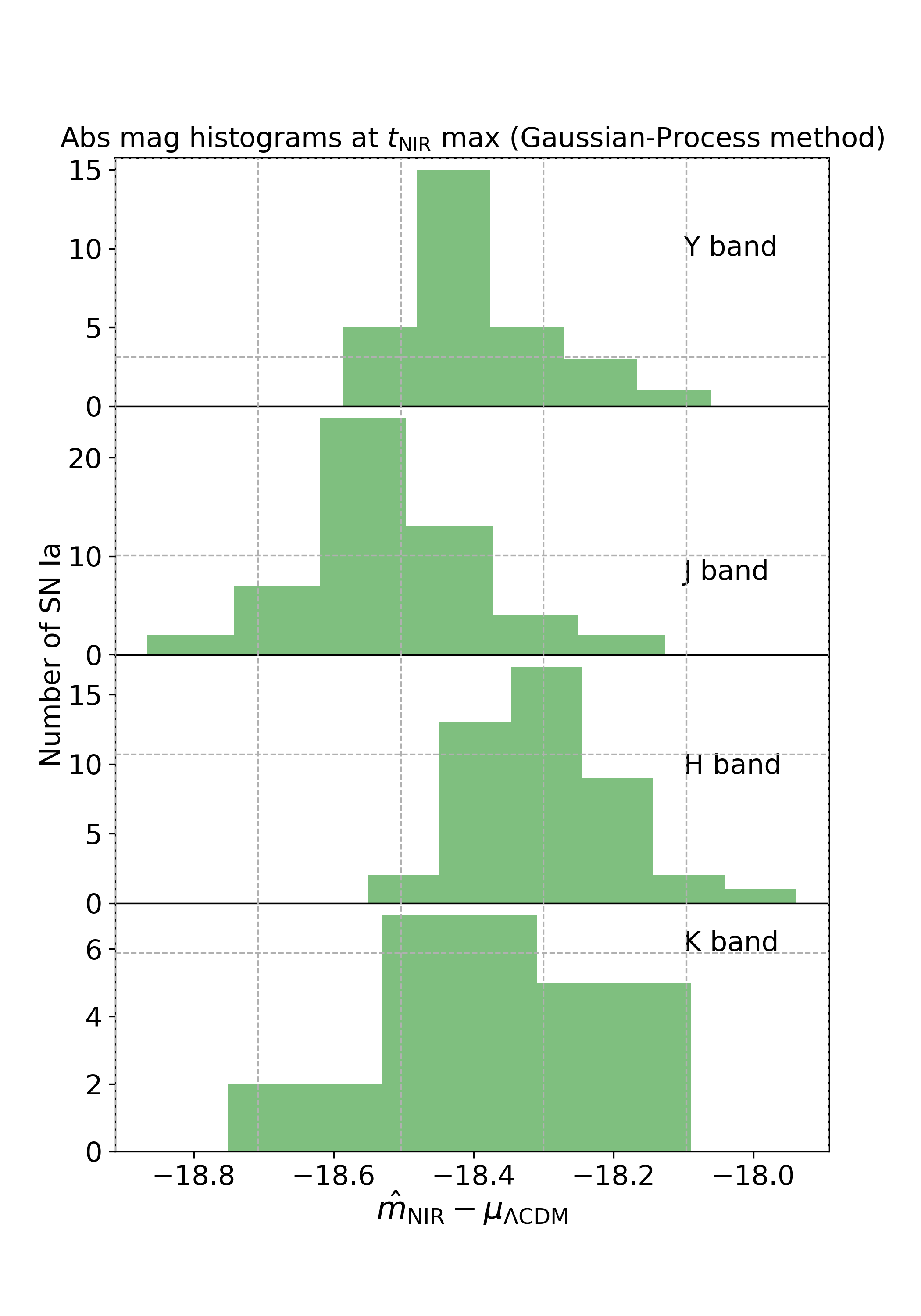}
\caption{
 Histograms of the absolute magnitudes at phase $= {\rm NIR}_{\rm max}$, defined as $M_{{\rm NIR}_{\rm max},s} \equiv \appmagNIRmaxHat - \muLCDM(z_s)$ for the \snIa{} sample  in the GP method at NIR max. The sample mean, standard deviation, and the number of supernovae in each histogram are shown in Table \ref{Table_AbsMags}.
}
\label{Fig_histo_AbsMag_GP}
\end{figure}

\begin{figure}
\centering
\includegraphics[width=8cm]{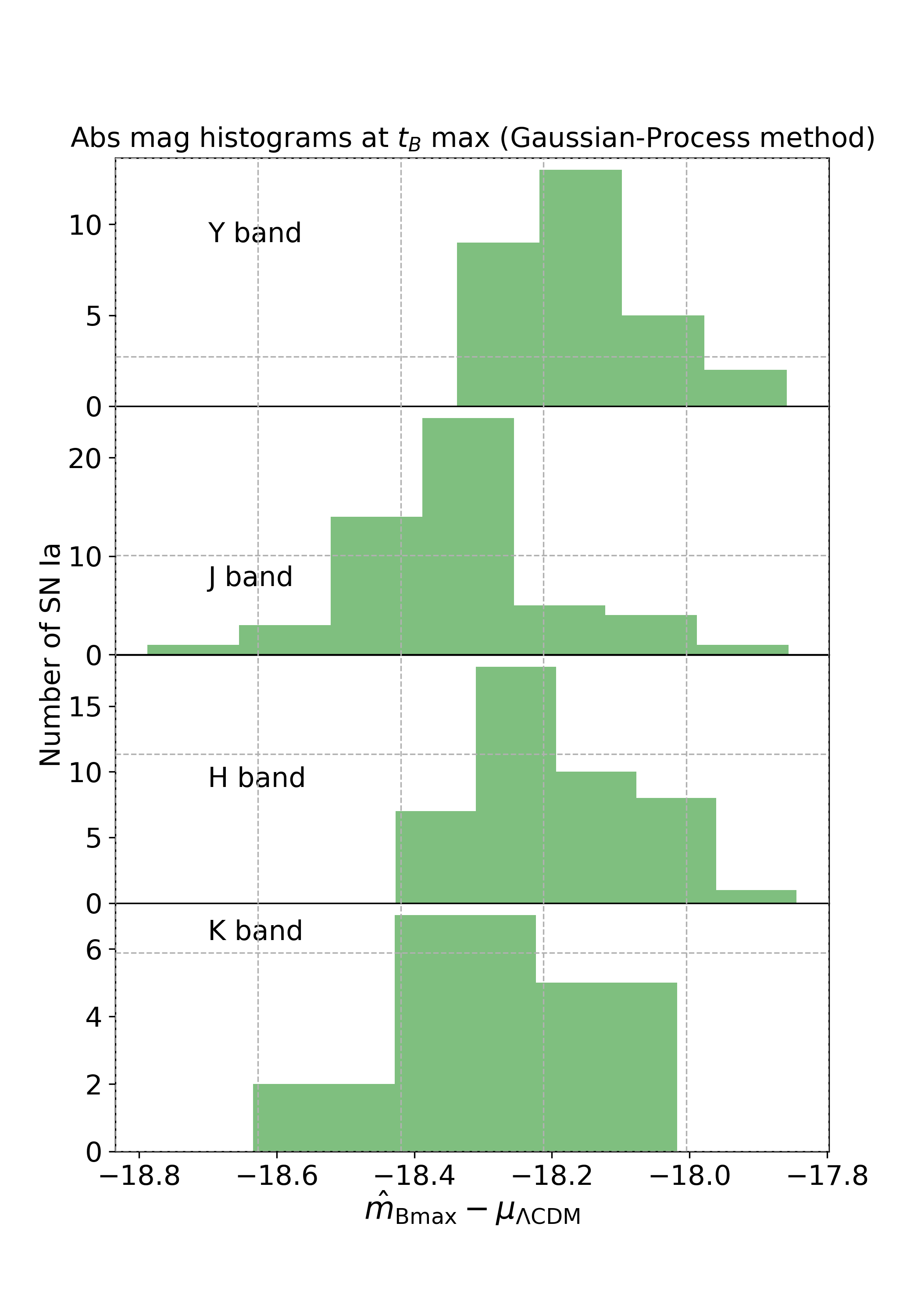}
\caption{
 Histograms of the absolute magnitudes at phase $= {B}_{\rm max}$, defined as $M_{{B}_{\rm max},s} \equiv \appmagBmaxHat - \muLCDM(z_s)$ for the \snIa{} sample in the GP method at $B$ max. The sample mean, standard deviation, and the number of supernovae in each histogram are shown in Table \ref{Table_AbsMags}.
}
\label{Fig_histo_AbsMag_GP_Bmax}
\end{figure}

Finally, we estimate the photometric distance modulus for supernova $s$ in a given NIR band as
\begin{equation}\label{DistMu_EachBand}
    \hat{\mu}_s = \hat{m}_{0,s} - \langle M_{0} \rangle\,  .
\end{equation}

The uncertainty on $\hat{\mu}_s$ is composed of two sources of errors: the fitting uncertainty $\sigmafithat$ estimated in Eq. (\ref{err_fit}) for each individual supernova, and the \textit{intrinsic scatter}, $\sigmaint$, which primarily comes from the intrinsic variation of \snIa{} absolute magnitudes and is estimated by fitting an entire sample of \snIa{} on the Hubble diagram  (see Appendix \ref{sec_WRMS_IntDisp} for more details).
So the variance of the photometric distance modulus is given as
\begin{equation}\label{DistMuError_EachBand_2}
\sigmamuphotohat^2 = \sigmafithat^2 + \sigmainthat^2.
\end{equation}

The Hubble residual for supernova $s$ is defined as
\begin{equation}\label{eq_HubbleResidual1}
\Delta \mu_s \equiv  \hat{\mu}_s - \muLCDM(z_s).
\end{equation}

The uncertainty on $\muLCDM(z_s)$ is given by Eq. (\ref{eq_PeculiarVelDefinition}).
The variance on the Hubble residual for supernova $s$, $\sigmaResidual^2$, comes from the propagation of uncertainties on $\hat{\mu}_s$  and $\muLCDM(z_s)$, it is,
\begin{equation}\label{eq_sigmas_residuals}
\sigmaResidual^2 = \sigmafithat^2 + \sigmainthat^2 + \smupec^2.
\end{equation}

In addition to $\sigmaint$, to quantify the dispersion in the Hubble residuals, we also compute both the RMS and the inverse-variance weighted root-mean-square (wRMS, see Appendix \ref{sec_WRMS_IntDisp}).  The RMS and wRMS are measures of the total scatter in the Hubble Diagram. The wRMS is relatively insensitive to the assumed value of the peculiar velocity uncertainty, and the formula for the RMS does not depend on the asssumed value of $\sigmaVpec$ at all and is therefore more straightforward to compare with other works.

For the template method, the values of $\hat{\mu}_s$, $\sigmaint$ and wRMS in the Hubble diagram residual for a given NIR band depend on the phase range of the NIR \lc{} template used to determine the distance modulus. We found that phase range of $t^* = (-8,30)$ days in each of the $\yjhk$ bands minimized the scatter in the Hubble residual, as measured by $\sigmaint$ or wRMS.

Table~\ref{tab_distanceMu_template} reports the distance moduli $\hat{\mu}_s$ and their fitting uncertainty $\sigmafithat$ we obtain with this procedure for each supernova in each band, and Fig.~\ref{fig_hubbles_Template_Individuals} shows the Hubble diagram and residuals.

\subsection{Distance Modulus: Gaussian-process Method}
\label{sec_dmod_gp}

The nearby low-$z$ NIR sample now contains a sufficient number of \snIa{} well-sampled around maximum light in the $\yjhk$-bands, that we can explore referencing various distance estimation approaches to the times of these NIR maxima, rather than $B$-max, for which there has long been sufficiently well sampled optical photometry.

An alternative approach that we implement to derive distance moduli is by determining the apparent magnitude at the time of NIR maximum light, $\tnirmaxx$, and $B$ maximum light, $\tbmaxx$, using the GP technique to interpolate the \lc{} data. The method follows the same procedure as the one described in Section \ref{sec_gaussProcess}, but instead of GP fitting the absolute magnitude \lcs{}, $\{M_s(t) \}$, we directly GP fit the \textit{apparent} magnitude \lcs{}, $\{m_s(t) \}$. By doing this, we do not include $\smupecNoS$ in the error budget for each $m_s(t)$ because we do not subtract $\muLCDM$($z$).

To determine the posterior mean of the apparent magnitude \lc{}, $\{ \bar{m}_s(t^*) \}$, and the posterior covariance of a GP fit to $\{m_s(t) \}$ we use the Eqs. (\ref{EqMeanGP}) and (\ref{EqCovarianceGP}) where we set $\smupec^2 = 0$. For each \lc{}, we use the average of the apparent magnitude data as the GP prior mean, and use the same values for the hyperparameters of the GP kernel shown in Table \ref{tab_GP_hyperpars}, given that the shape and dispersion of the apparent magnitude \lc{} data is very similar to the absolute magnitude \lcs{} we fitted with GP in Section \ref{sec_gaussProcess} for each supernova. We verified that the GP fits to the \lcs{} are insensitive to these choices.

We only consider \lcs{} that have data either around $\tnirmaxx$ or $\tbmaxx$ so that we can determine the GP fit at those references phases.  By construction, $\tbmaxx$ corresponds to the phase $= 0$ days. For the case of $\tnirmaxx$ we limit the search for the maximum to the phase range $-8.5 < \tnirmaxx < -2.5$ days to remove cases where maximum of the posterior mean happens \textit{after} $\tbmaxx$, which we found can be artifacts of the GP fit when there are too few data points before $\tbmaxx$. For the rest of this section we denote the subscripts ``NIRmax'' and ``$B$max'' simply as ``max''.

From each set $\{ \bar{m}_s(t^*) \}$, we estimate, $\appmagmaxHat$, the GP interpolated apparent magnitude at $\tmax$. Then we estimate the distance modulus as
\begin{equation}\label{DistMu_EachBand_GP}
    \hat{\mu}_s = \appmagmaxHat - \langle M_{\rm max} \rangle
\end{equation}
where $\langle M_{\rm max} \rangle$ is the mean absolute magnitude at $\tmax$ from all the supernovae in a given NIR band (see Fig.~\ref{Fig_histo_AbsMag_GP}), with $ M_{{\rm max},s} \equiv \appmagmaxHat - \muLCDM(z_s)$. The uncertainty on the photometric distance modulus $\hat{\mu}_s$ in this case is $\sigmafithat$, which is equal to the uncertainty in the apparent magnitude at $\tmax$ inferred from the GP fit to the LC.

Figure~\ref{fig_hubbles_GP_Individuals} shows Hubble diagrams constructed from the distance moduli inferred from the GP method for each of the \yjhk{} bands, with numerical values reported in Table~\ref{tab_distanceMu_GaussianProc_NIRmax}.

\subsection{Distance modulus from the combined NIR bands}
\label{sec_dmod_AnyNIR}

From the estimated distance moduli $(\hat{\mu}^Y_s, \hat{\mu}^J_s, \hat{\mu}^H_s, \hat{\mu}^K_s)$ for a given supernova $s$ determined from each NIR band using either of the three methods described above, we estimate the weighted average of the distance modulus ${\mu}_s$ from each method. First we define the vector of residuals
\begin{equation}\label{VectorDistMu1}
\delta \boldsymbol{\mu}_s \equiv
\begin{pmatrix}
\hat{\mu}^Y_s - {\mu}_s  \\
\hat{\mu}^J_s - {\mu}_s \\
\hat{\mu}^H_s - {\mu}_s \\
\hat{\mu}^K_s - {\mu}_s
\end{pmatrix}.
\end{equation}
where $\hat{\mu}^Y_s, \hat{\mu}^J_s, \hat{\mu}^H_s, \hat{\mu}^K_s$ are determined by either Eqs. (\ref{DistMu_EachBand}) or (\ref{DistMu_EachBand_GP}), for the template or GP methods, respectively. Then, to estimate ${\mu}_s$, we minimize the negative of the likelihood function $\likelihood({\mu}_s)$ defined as
\begin{equation}\label{Eq_LogLike_TotalMu}
-2 \ln \likelihood({\mu}_s) = \delta \boldsymbol{\mu}_s^{\top} \cdot C^{-1}_{\mu} \cdot \delta \boldsymbol{\mu}_s + {\rm constant},
\end{equation}
where $C_{\mu}$ is the sample covariance matrix computed from the Hubble residuals (see Eq. (\ref{eq_HubbleResidual1})) $\{\Delta {\mu}^Y_s, \Delta{\mu}^J_s, \Delta{\mu}^H_s, \Delta{\mu}^K_s \}$, the collection of distance-modulus residuals from all \snIa{} with observations in the four $\yjhk$ bands. For supernovae with observations in only three, two, or one bands, we construct the respective covariance matrices based on those supernova subsamples, and the vector defined in Eq. (\ref{VectorDistMu1}) becomes three, two, or one dimensional, respectively. In Appendix \ref{sec_covmatrix}, we provide numerical values of the covariance matrix $C_\mu$ for these different subcases.

We derive an analytic expression for the minimization of Eq. (\ref{Eq_LogLike_TotalMu}) with respect to $\mu_s$ and obtain the maximum likelihood estimate for the combined distance modulus given by,
\begin{equation}
    \hat{\mu}_s = \sum_{b} w_b \, \hat{\mu}^b_s
\end{equation}
where $\hat{\mu}^b_s \in \{\hat{\mu}^Y_s, \hat{\mu}^J_s, \hat{\mu}^H_s, \hat{\mu}^K_s$\} (the index $b$ stands for \textit{band}), and
\begin{equation}
    w_b =  \left[ \sum^{}_{b'} (C^{-1})_{bb'} \right] \times \left[ \sum^{}_{b',b''} (C^{-1})_{b'b''} \right]^{-1}.
\end{equation}

Now, assuming that the uncertainties in the distance modulus estimated from each individual \yjhk{} band, $\hat{\sigma}_{{\rm fit}, s, Y}, \, \hat{\sigma}_{{\rm fit}, s, J}, \, \hat{\sigma}_{{\rm fit}, s, H}, \, \hat{\sigma}_{{\rm fit}, s, K}$, are independent between bands $b$ and also independent of the intrinsic scatter $\sigmaint$, then we can propagate the uncertainty in the combined distance modulus due to the fitting only as:
\begin{equation}
    \sigmafithat = \sqrt{\sum_{b} w^2_b \, \hat{\sigma}^2_{s, b}}
\end{equation}
where $\hat{\sigma}_{s, b} \equiv \hat{\sigma}_{{\rm fit}, s, Y}, \, \hat{\sigma}_{{\rm fit}, s, J}, \, \hat{\sigma}_{{\rm fit}, s, H}, \, \hat{\sigma}_{{\rm fit}, s, K}$.

The last column in Tables~\ref{tab_distanceMu_template}-\ref{tab_distanceMu_GaussianProc_Bmax} show the combined distance moduli we obtain with this procedure for the template and GP methods respectively. The reported uncertainties correspond to $\sigmafithat$ in all cases.

\subsection{Distance modulus from optical bands}
\label{sec_dmod_optical}

We wish to assess how well the \snIa{} observed in NIR bands perform as standard candles, specifically when using $\tnirmaxx$ as opposed to $\tbmaxx$, as the time reference to estimate their distance. To do so, we determine the distance moduli using only \textit{optical} $BVR$-bands \lcs{} for exactly the same \nAnyYJHKgp{} supernovae in the ``any \yjhk'' Hubble diagram  set that was used for the GP method (see left panel in Fig.~\ref{fig_hubbles_GP_AnyYJHK} and the SN listed in Table~\ref{tab_distanceMu_GaussianProc_NIRmax}). Then we can compare the intrinsic scatter and RMS or wRMS in the Hubble-diagram residuals between the optical-only and NIR-only Hubble diagrams. A smaller intrinsic scatter, wRMS, or RMS, including the uncertainties, would indicate evidence that \snIa{} are better standard candles using that data and Hubble diagram construction method.

\subsubsection{SALT2 distance modulus}
\label{sec_dmod_salt}

We use the optical photometric data compiled in the public SNANA (\citealt{Kessler_SNANA}) database\footnote{http://snana.uchicago.edu. Version Oct 18, 2017.} but replace the CMB redshift values in the SNANA photometric files with the $\zcmb$ values in Table~\ref{Table_LC_params}. Using the latest SALT2 model (SALT2.JLA-B14) (\citealt{guy07}) already trained on the JLA sample (\citealt{BetouleEtal2014_JLA}), we fit the optical data and determine the SALT2 light-curve fit parameters for each supernova. For the CSP data, we added an additional 0.01 mag in quadrature to the photometric errors to have a more conservative uncertainties on those values when fitting the data in SALT2.
We use the SALT2 outputs including the apparent magnitude $m_B$ at $B$-band maximum light, the stretch parameter $x_1$, and the color term $c$, as well as their correlations.

We convert the SALT2-fit parameters to distance moduli for each supernova using the Tripp formula (\citealt{Tripp1998}),
\begin{equation}\label{eq_Tripp}
    \mu_s =  m_{B,s} - M_B + \alpha x_{1,s} - \beta c_{s},
\end{equation}
where $M_B$ is the expected absolute magnitude at $B$-band maximum light for a \snIa{} with $x_1=0, c=0$, while $\alpha$ and $\beta$ are coefficients parametrizing correlations between luminosity and stretch or luminosity and color, respectively.

For the global parameters $M, \alpha, \beta$ we use the values reported by \citet{scolnic18}; $\alpha = 0.147$, $\beta = 3.00$, and assume the fiducial values of $\Ho = 73.24$ \Hounits{} and $M_B = -19.36$ mag. We then adjust the latter to $M_B = \MBfidSALT$ mag so that the weighted-average Hubble residual is zero.

The standard deviation of the measurement error $\sigmafit$ from the SALT2 fitting comes from propagating the uncertainties on Eq. (\ref{eq_Tripp}), including their correlations. Interestingly we found that for the supernovae with high Milky Way color excess $E(B-V)>0.2$ the uncertainty on $m_{B,s}$ is larger than the propagated uncertainty on the SALT2 distance modulus, $\mu_s$, derived from optical bands. This evidence further emphasizes how \snIa{} are more negatively affected by dust when deriving distances using optical data, as compared to NIR observations.

The variance  of  the  photometric  distance  modulus  is  given  by
\begin{equation}
    \sigmamuphoto^2 = \sigmafit^2 + \sigmaint^2.
\end{equation}
Using SALT2 in this way, we obtain an intrinsic scatter in the Hubble residuals of $\sigmaint = \sigmaIntSALT $, an inverse-variance weighted RMS of wRMS=$\wrmsSALT $ mag, and a simple RMS $=\rmsSALT $ mag. The third column of Table~\ref{tab_distanceMu_Optical} and the left panel of Fig.~\ref{fig_hubbles_SALT2_Snoopy} show the distance moduli derived from the SALT2 fits, along with the Hubble diagram and residuals, respectively. The uncertainties shown in Table \ref{tab_distanceMu_Optical} and Fig.~\ref{fig_hubbles_SALT2_Snoopy} are the values of $\sigmafit $.

Note that we are not applying the usual SALT2 cuts to this subsample of SN because we are interested in comparing the scatter in the Hubble residuals using exactly the same \nAnyYJHKgp{} \snIa{} used in the ``any $\yjhk$" Hubble diagram for the GP method. We find that when applying the SALT2 cut on color, $-0.3 < c < 0.3$, there is only 1 \snIa{} in the subsample that does not pass this cut.
All \snIa{} in the sample pass these SALT2 cuts: $-3 < x_1 < 3$, uncertainty in $x_1 < 1$, and uncertainty in $t_{B {\rm max}} < 2$ days. However, 21 \snIa{} fail to pass the SALT2 cut requiring that the probability that the data are represented by the model, given the $\chi^2$ per degree of freedom of the fit, is larger than $0.001$ (a.k.a, FITPROB $> 0.001$). However, a low fit probability does not necessarily indicate a poor \snIa{} light curve fit and may instead be an indication that the photometric uncertainties or the model uncertainties are unrealistically small. We visually inspected the light curve fits of these 21 \snIa{}, finding that they are reasonably well-fit by the model and can therefore be used to yield accurate distance measurements.

\subsubsection{\snoopy{} distance modulus}
\label{sec_dmod_snoopy}

As a second cross check of the scatter in the optical-only Hubble diagram, we also fit the $BVR$-bands \lcs{} using the \snoopy{} \lc{} fitting package's \verb+EBV_model+ (\citealt{BurnsEtal2011Snoopy}), where every observed apparent magnitude $m_s$ in band $O \equiv B,V,R$ is compared to the model
\begin{multline}
    m_s(t) = \mu_s + T_Q( t, \Delta m_{15, s}) + M_{Q}(\Delta m_{15, s})\\
    + R_{O} {\ebvmw}_{,s} + R_{Q} {\ebvhost}_{,s} +\\
    K^{s}_{OQ} \Bigl(z_s,t, {\ebvmw}_{,s}, {\ebvhost}_{,s} \Bigr),
\end{multline}
where $ T_Q( t, \Delta m_{15, s})$ is a light-curve template for the rest-frame band $Q$ that depend on $t$ and $\Delta m_{15, s}$, and $M_{Q,s}(\Delta m_{15, s})$ is the absolute magnitude band $Q$.
In this model, the free parameters that \snoopy{} estimates (along with their uncertainties) are $\mu_s$, $\Delta m_{15,s}$, ${\ebvhost}_{,s}$ and $\TBmaxx$.
We consider the estimated uncertainty on $\mu_s$ output by \snoopy{} as the $\sigmafit $ in our analysis. We refer the reader to \citet{BurnsEtal2011Snoopy} for details on how \snoopy{} estimates the uncertainty on $\mu_s$.

We obtain an intrinsic scatter in the Hubble residuals of $\sigmaint = \sigmaIntSnoopy$, a wRMS$= \wrmsSnoopy$ mag, and a RMS $= \rmsSnoopy $ mag. The fourth column of Table~\ref{tab_distanceMu_Optical} and right panel of Fig.~\ref{fig_hubbles_SALT2_Snoopy} show the distance moduli derived from the \snoopy{} fits, along with the Hubble diagram and residuals, respectively.

\begin{table*}
\begin{center}
\caption{Hubble Residual Intrinsic Scatter, $\sigmaint$, and RMS.}
\footnotesize
\begin{tabular}{c c c c c c c}
\hline \hline \\[-0.3cm]
Band  & Method  &  $N_{\rm SN}$ & $\sigmaint$ [mag] & $\sigmaint$ [mag] &  wRMS [mag]  & RMS [mag]\\
   &  &  &  ($\sigmaVpec = 150$ km/s) &  ($\sigmaVpec = 250$ km/s) &  ($\sigmaVpec = 150$ km/s)  &  \\
\hline \\[-0.3cm]
Optical $BVR$ & SALT2    & 56 & $0.133 \pm 0.022$   & $0.107 \pm 0.025$   & $0.174 \pm 0.020$  & $0.179 \pm 0.018$  \\
Optical $BVR$ & SNooPy   & 56 & $0.128 \pm 0.018$   & $0.111 \pm 0.020$   & $0.159 \pm 0.019$  & $0.174 \pm 0.021$  \\
any $YJHK_s$ & Template  & 56 & $0.112 \pm 0.016$   & $0.096 \pm 0.019$   & $0.140 \pm 0.016$  & $0.138 \pm 0.014$  \\
any $YJHK_s$     & GP (NIR max) & 56 & $0.047 \pm 0.018$   & $0.000 \pm 0.000^{*}$   & $0.100 \pm 0.013$  & $0.117 \pm 0.014$  \\
any $YJHK_s$     & GP ($B$ max) & 56 & $0.066 \pm 0.016$   & $0.044 \pm 0.023$   & $0.106 \pm 0.010$  & $0.115 \pm 0.011$  \\[0.1cm]
\hline \hline
\end{tabular}
\label{Tab_scatter_GPSubsample}
\end{center}
\tablecomments{
We compare the Hubble residual scatter for the optical and NIR bands using exactly the same \nAnyYJHKgp{} supernovae for several methods. We compute the intrinsic scatter, $\sigmaint$ (see Appendix \ref{sec_WRMS_IntDisp}), the inverse-variance weighted root-mean square (wRMS), and the simple RMS, using two standard \lc{} fitters: SALT2 (\citealt{guy07}) and \snoopy{} (\citealt{BurnsEtal2011Snoopy}) to fit the optical $BVR$-band \lc{} data, as well as the three NIR methods we implement in this work: NIR \lc{} templates at $B$-max [Template], and GP regression at NIR-max [GP (NIR max) or $B$-band maximum [GP ($B$ max)]. We are limited to \nAnyYJHKgp{} \snIa{} because these are all the supernovae that we can fit using the GP method. Columns 4 and 5 show $\sigmaint$, assuming $\sigmaVpec = 150$ and $250$ km/s respectively. The estimated intrinsic scatter $\sigmaint$ decreases as the assumed peculiar velocity uncertainty $\sigmaVpec$ increases from commonly assumed values of 150 km/s (\citealt{radburnsmith04}) to 250 km/s (\citealt{scolnic18}), making $\sigmaint$ somewhat model dependent. By contrast, the wRMS value only changes by thousandths of a magnitude for $\sigmaVpec$ in the same range. Column 6 shows the wRMS assuming $\sigmaVpec = 150$ km/s. Column 7 shows the simple RMS, which makes no assumptions about error weighting and does not depend on $\sigmaVpec$. Both optical methods apply \lc{} shape and dust corrections but still yield a larger scatter than the NIR methods quantified with any of $\sigmaint$, wRMS or RMS (see also Table \ref{Tab_IntDispersion_PecVel}). Figs.~\ref{fig_hubbles_GP_AnyYJHK}-\ref{fig_hubbles_SALT2_Snoopy} show Hubble diagrams and residuals for this subsample.\\
$^*$ For $\sigmaVpec=250$ km/s, the estimated value of $\sigmaint$ is consistent with 0. See the paragraph below Eq. (B.7) in \citet{BlondinEtal2011} for the explicit approximation we use to estimate the uncertainty, which breaks down at $\sigmaint = 0$.
}
\end{table*}

\begin{table*}
\begin{center}
\caption{Hubble Residual Intrinsic Scatter, $\sigmaint$, and RMS.}
\footnotesize
\begin{tabular}{c c c c c c c}
\hline \hline \\[-0.3cm]
Band  & Method  &  $N_{\rm SN}$ & $\sigmaint$ [mag] & $\sigmaint$ [mag] &  wRMS [mag]  & RMS [mag]\\
   &  &  &  ($\sigmaVpec = 150$ km/s) &  ($\sigmaVpec = 250$ km/s) &  ($\sigmaVpec = 150$ km/s)  &  \\
\hline \\[-0.3cm]

$Y$     & Template     & 44 & $0.105 \pm 0.018$   & $0.093 \pm 0.021$   & $0.139 \pm 0.013$  & $0.152 \pm 0.016$  \\
$Y$     & GP (NIR max) & 29 & $0.066 \pm 0.020$   & $0.037 \pm 0.032$   & $0.102 \pm 0.015$  & $0.111 \pm 0.018$  \\
$Y$     & GP ($B$ max) & 29 & $0.080 \pm 0.019$   & $0.062 \pm 0.024$   & $0.110 \pm 0.013$  & $0.118 \pm 0.017$  \\
$J$     & Template     & 87 & $0.136 \pm 0.016$   & $0.122 \pm 0.018$   & $0.170 \pm 0.013$  & $0.175 \pm 0.013$  \\
$J$     & GP (NIR max) & 52 & $0.107 \pm 0.017$   & $0.090 \pm 0.021$   & $0.136 \pm 0.017$  & $0.139 \pm 0.016$  \\
$J$     & GP ($B$ max) & 52 & $0.124 \pm 0.019$   & $0.110 \pm 0.021$   & $0.153 \pm 0.020$  & $0.151 \pm 0.021$  \\
$H$     & Template     & 81 & $0.126 \pm 0.016$   & $0.112 \pm 0.018$   & $0.162 \pm 0.015$  & $0.166 \pm 0.016$  \\
$H$     & GP (NIR max) & 44 & $0.032 \pm 0.027$   & $0.000\pm 0.000^{*}$  & $0.095 \pm 0.010$  & $0.114 \pm 0.015$  \\
$H$     & GP ($B$ max) & 44 & $0.063 \pm 0.020$   & $0.037 \pm 0.033$   & $0.111 \pm 0.011$  & $0.120 \pm 0.013$  \\
$K$     & Template     & 32 & $0.175 \pm 0.032$   & $0.163 \pm 0.035$   & $0.211 \pm 0.023$  & $0.207 \pm 0.020$  \\
$K$     & GP (NIR max) & 14 & $0.093 \pm 0.054$   & $0.077 \pm 0.060$   & $0.163 \pm 0.033$  & $0.179 \pm 0.029$  \\
$K$     & GP ($B$ max) & 14 & $0.094 \pm 0.054$   & $0.059 \pm 0.069$   & $0.162 \pm 0.035$  & $0.170 \pm 0.027$  \\
any $YJHK_s$     & Template     & 89 & $0.123 \pm 0.014$   & $0.107 \pm 0.016$   & $0.154 \pm 0.013$  & $0.161 \pm 0.013$  \\
$JH$      & Template     & 81 & $0.127 \pm 0.015$   & $0.112 \pm 0.017$   & $0.158 \pm 0.015$  & $0.164 \pm 0.015$  \\
$JH$      & GP (NIR max) & 42 & $0.039 \pm 0.024$   & $0.000\pm 0.000^{*}$  & $0.096 \pm 0.011$  & $0.114 \pm 0.016$  \\
$JH$      & GP ($B$ max) & 42 & $0.069 \pm 0.019$   & $0.046 \pm 0.028$   & $0.112 \pm 0.014$  & $0.118 \pm 0.015$  \\
$YJH$     & Template     & 40 & $0.093 \pm 0.018$   & $0.080 \pm 0.022$   & $0.121 \pm 0.013$  & $0.137 \pm 0.018$  \\
$YJH$     & GP (NIR max) & 21 & $0.044 \pm 0.028$   & $0.000\pm 0.000^{*}$  & $0.088 \pm 0.014$  & $0.087 \pm 0.013$  \\
$YJH$     & GP ($B$ max) & 21 & $0.068 \pm 0.023$   & $0.056 \pm 0.031$   & $0.097 \pm 0.014$  & $0.098 \pm 0.014$  \\

\hline \hline
\end{tabular}
\label{Tab_IntDispersion_PecVel}
\end{center}
\tablecomments{
Scatter in the Hubble residuals for different NIR band subsets, quantified by the intrinsic scatter, $\sigmaint$, the wRMS, and the simple RMS, using three methods: NIR \lc{} templates at $B$-max [Template], and GP regression at NIR-max [GP (NIR max) or $B$-band maximum [GP ($B$ max)]. Column 3 shows the number of \snIa{} in each Hubble diagram. Also see Table \ref{Tab_scatter_GPSubsample}. Columns 4 and 5 show $\sigmaint$, assuming $\sigmaVpec = 150$ and $250$ km/s respectively. Note that by increasing the value of $\sigmaVpec$, the $\sigmaint$ decreases even to zero in some cases with uncertainty denoted by $\infty$. For the GP method, we use exactly the same supernovae at $B$-max or NIR-max.  For all NIR band subsets, the GP (NIR max) method produces the smallest scatter, followed by the GP ($B$-max) method, while the template method always yields the largest scatter and wRMS. Figs.~\ref{fig_hubbles_Template_AnyYJHK}-\ref{fig_hubbles_GP_AnyYJHK_Bmax} and \ref{fig_hubbles_Template_Individuals}-\ref{fig_hubbles_GP_Individuals_Bmax} show Hubble diagrams and residuals for most of the NIR subsets listed in this table.\\
$^*$ For $\sigmaVpec= 250$ km/s, the estimated value of $\sigmaint$ in these cases is consistent with zero. See the paragraph below Eq. (B.7) in \citet{BlondinEtal2011} for the explicit approximation we use to estimate the uncertainty, which breaks down at $\sigmaint = 0$.
}
\end{table*}

\begin{table*}
\begin{center}
\caption{Optical - NIR intrinsic scatter}
\begin{tabular}{c c c c c c c}
\hline \hline \\[-0.3cm]
Optical $BVR$ Method - NIR band(s) & $\Delta\sigmaint$ & $n$-$\sigma$ & $\Delta$wRMS & $n$-$\sigma$ & $\Delta$RMS & $n$-$\sigma$\\
\hline
\hline
SALT2 - $Y$ & $0.067 \pm 0.029$ & 2.3 & $0.073 \pm 0.025$ & 2.9 & $0.068 \pm 0.026$ & 2.6 \\
\snoopy{} - $Y$ & $0.062 \pm 0.027$ & 2.3 & $0.057 \pm 0.024$ & 2.4 & $0.063 \pm 0.028$ & 2.2 \\
\hline
SALT2 - $J$ & $0.027 \pm 0.028$ & 1.0 & $0.038 \pm 0.026$ & 1.5 & $0.040 \pm 0.024$ & 1.6 \\
\snoopy{} - $J$ & $0.021 \pm 0.025$ & 0.8 & $0.023 \pm 0.025$ & 0.9 & $0.035 \pm 0.027$ & 1.3 \\
\hline
SALT2 - $H$ & $0.101 \pm 0.035$ & 2.9 & $0.079 \pm 0.022$ & 3.6 & $0.065 \pm 0.023$ & 2.8 \\
\snoopy{} - $H$ & $0.095 \pm 0.033$ & 2.9 & $0.063 \pm 0.021$ & 3.0 & $0.060 \pm 0.026$ & 2.3 \\
\hline
SALT2 - $K_s$ & $0.040 \pm 0.058$ & 0.7 & $0.011 \pm 0.039$ & 0.3 & $0.000 \pm 0.035$ & 0.0 \\
\snoopy{} - $K_s$ & $0.034 \pm 0.057$ & 0.6 & $-0.005 \pm 0.038$ & -0.1 & $-0.005 \pm 0.036$ & -0.1 \\
\hline
SALT2 - any $YJHK_s$ & $0.086 \pm 0.028$ & 3.0 & $0.074 \pm 0.024$ & 3.2 & $0.062 \pm 0.023$ & 2.7 \\
\snoopy{} - any $YJHK_s$ & $0.080 \pm 0.026$ & 3.1 & $0.059 \pm 0.023$ & 2.6 & $0.057 \pm 0.025$ & 2.3 \\
\hline
SALT2 - $JH$  & $0.095 \pm 0.032$ & 2.9 & $0.078 \pm 0.023$ & 3.5 & $0.065 \pm 0.024$ & 2.7 \\
\snoopy{} - $JH$  & $0.089 \pm 0.030$ & 2.9 & $0.062 \pm 0.022$ & 2.8 & $0.060 \pm 0.026$ & 2.3 \\
\hline
SALT2 - $YJH$ & $0.089 \pm 0.036$ & 2.5 & $0.086 \pm 0.024$ & 3.5 & $0.092 \pm 0.022$ & 4.1 \\
\snoopy{} - $YJH$ & $0.084 \pm 0.034$ & 2.5 & $0.070 \pm 0.024$ & 3.0 & $0.087 \pm 0.025$ & 3.5 \\
\hline

\hline
\end{tabular}
\label{Tab_IntDispersion_PecVel1}
\end{center}
\tablecomments{
We show $\Delta\sigmaint$, $\Delta$wRMS, and $\Delta$RMS, where the first is defined as the difference in Hubble residuals intrinsic scatter between the optical $BVR$ data, fit using SALT2 or \snoopy{}, and the indicated subset of NIR data using the Gaussian process method at NIR max. The quantities $\Delta$wRMS and $\Delta$RMS are defined in a similar way to $\Delta\sigmaint$ but using wRMS and RMS instead of the intrinsic scatter, respectively. The uncertainties are given by the quadrature sum of the $\sigmaint$, wRMS, or RMS, uncertainties from columns 4, 6, or 7, respectively of Tables~\ref{Tab_scatter_GPSubsample} and \ref{Tab_IntDispersion_PecVel} for $\sigmaVpec = 150$ km/s. 
Columns 3, 5, and 6, show $n$-$\sigma$ defined as the number $n$ of standard deviations $\sigma$ by which the NIR data yields {\it smaller} $\sigmaint$, wRMS, or RMS, than the optical data using these methods, respectively. Excluding the $K_s$-band on its own, where our \lc{} compilation contains much less data than the $YJH$ bands, in general, NIR data subsets yield smaller RMS than the optical data at the $\sim \optminusnirRMSnsSmallest$-$\optminusnirRMSnsLargest \sigma$ level.
In the best case, the $JH$, $YJH$, and $\yjhk$-bands perform $\sim \optminusnirRMSnsSmallestCombined$-$\optminusnirRMSnsLargestCombined \sigma$ better than either SALT2 or \snoopy{} fits to the $BVR$ data in terms of the RMS, while in the worst case, $J$-band, still performs $\optminusnirsigmaJWorstRMS \sigma$ better than optical data.  For simplicity, the stated uncertainties on the difference in dispersion estimates between any two methods ignores the fact that the actual peculiar velocity-distance errors are exactly the same between the optical and NIR samples, since they contain exactly the same SN.  The effect of accounting for this correlation is to decrease the uncertainty of the difference, and increase the significance (\S \ref{sec_disc}).
}
\end{table*}

\begin{table*}
\begin{center}
\caption{GP Method intrinsic scatter for $B$ max vs. NIR max}
\begin{tabular}{c c c c c c c}
\hline \hline \\[-0.3cm]
NIR band(s) & $\Delta\sigmaint$ & $n$-$\sigma$ & $\Delta {\rm wRMS}$ & $n$-$\sigma$ & $\Delta {\rm RMS}$ & $n$-$\sigma$\\
\hline

$Y$ & $0.014 \pm 0.028$ & 0.49 & $0.009 \pm 0.020$ & 0.42 & $0.007 \pm 0.025$ & 0.26 \\
$J$ & $0.018 \pm 0.025$ & 0.70 & $0.017 \pm 0.026$ & 0.65 & $0.012 \pm 0.026$ & 0.46 \\
$H$ & $0.031 \pm 0.034$ & 0.92 & $0.016 \pm 0.014$ & 1.12 & $0.006 \pm 0.020$ & 0.32 \\
$K_s$ & $0.001 \pm 0.076$ & 0.01 & $-0.001 \pm 0.048$ & -0.03 & $-0.009 \pm 0.040$ & -0.23 \\
any $YJHK_s$ & $0.019 \pm 0.024$ & 0.77 & $0.006 \pm 0.016$ & 0.38 & $-0.002 \pm 0.018$ & -0.10 \\
$JH$  & $0.030 \pm 0.031$ & 0.99 & $0.016 \pm 0.018$ & 0.89 & $0.004 \pm 0.021$ & 0.17 \\
$YJH$ & $0.024 \pm 0.037$ & 0.66 & $0.008 \pm 0.020$ & 0.41 & $0.011 \pm 0.019$ & 0.58 \\

\hline \hline
\end{tabular}
\label{Tab_NIRmax_vs_Bmax}
\end{center}
\tablecomments{
Similar to Table \ref{Tab_IntDispersion_PecVel1}, we show $\Delta\sigmaint$, $\Delta$wRMS, and $\Delta$RMS, defined here as the difference in Hubble residuals scatter between the Gaussian process method referenced to $B$ max or NIR max. As in Table~\ref{Tab_IntDispersion_PecVel1}, the uncertainties are given by the quadrature sum of the $\sigmaint$ or wRMS uncertainties from columns 4 or 6 of Tables~\ref{Tab_scatter_GPSubsample} and \ref{Tab_IntDispersion_PecVel} for $\sigmaVpec = 150$ km/s. Columns 3, 5, and 7, show $n$-$\sigma$ defined as the number $n$ of standard deviations $\sigma$ by which the NIR data referenced to NIR max yields {\it smaller} intrinsic scatter, wRMS, or RMS, than when referenced to $B$-max, respectively.
For every individual band and subset of NIR bands, the GP method yields smaller estimated intrinsic scatter when referencing to NIR max instead of $B$ max, where the largest difference is $n$-$\sigma = \BmaxminusNIRmaxLargestNS \sigma$ for \BmaxminusNIRmaxLargestNSBand{} band. 
This trend is also observed when comparing the wRMS values, again, excluding, $K_s$ band, where our sample lacks enough data to draw meaningful conclusions.
}
\end{table*}

\begin{figure*}
\centering
\begin{center}
\includegraphics[width=18cm]{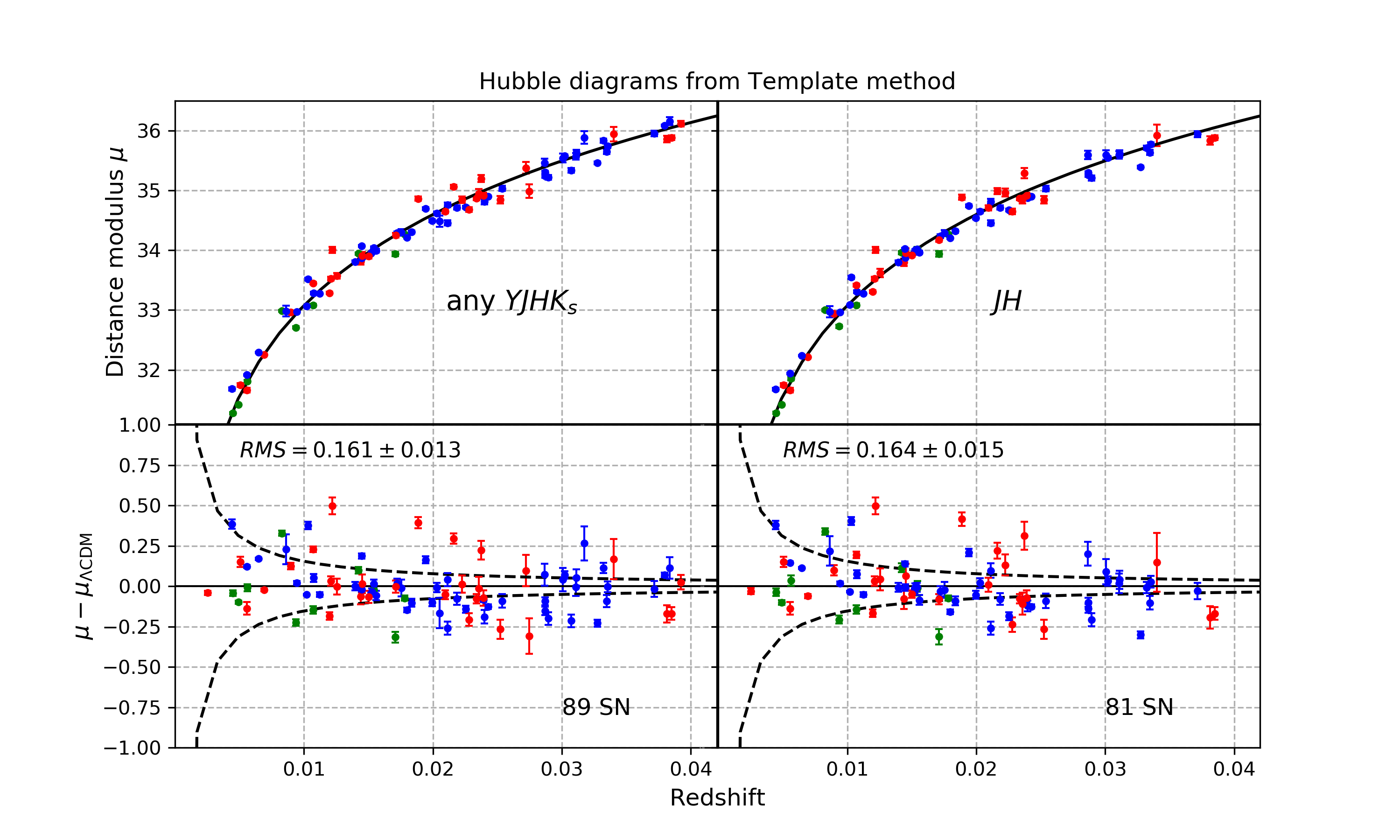}
\caption{$\yjhk$ Hubble diagrams (top row) and residuals (bottom row) using the template method on the \nAnyYJHKtemp{} supernovae that passed our cuts. The error bars plotted for each supernova correspond to the fitting uncertainties $\sigmafithat$. The left panel corresponds to the case when we determine a single distance modulus by combining any of the available $1$, $2$, $3$, or $4$ $\yjhk$ distance moduli for a given \snIa. The right panel shows the case when we require only \snIa{} with $J$ and $H$-band data, which allows us to include the majority of data from the CfA and CSP samples. Points are color coded by NIR photometric data source, including the CfA (red; \citealt{woodvasey08,friedman15}), the CSP (blue; \citealt{krisciunas17}), and other data from the literature (green; see Table~\ref{Table_LC_params} and references therein). Note that only the CSP used a $Y$-band filter. Table~\ref{Tab_IntDispersion_PecVel} summarizes the intrinsic scatter in the Hubble diagrams, while Table~\ref{tab_distanceMu_template} reports the numerical values of the distance moduli from this figure.
}
\label{fig_hubbles_Template_AnyYJHK}
\end{center}
\end{figure*}

\begin{figure*}
\begin{center}
\includegraphics[width=18cm]{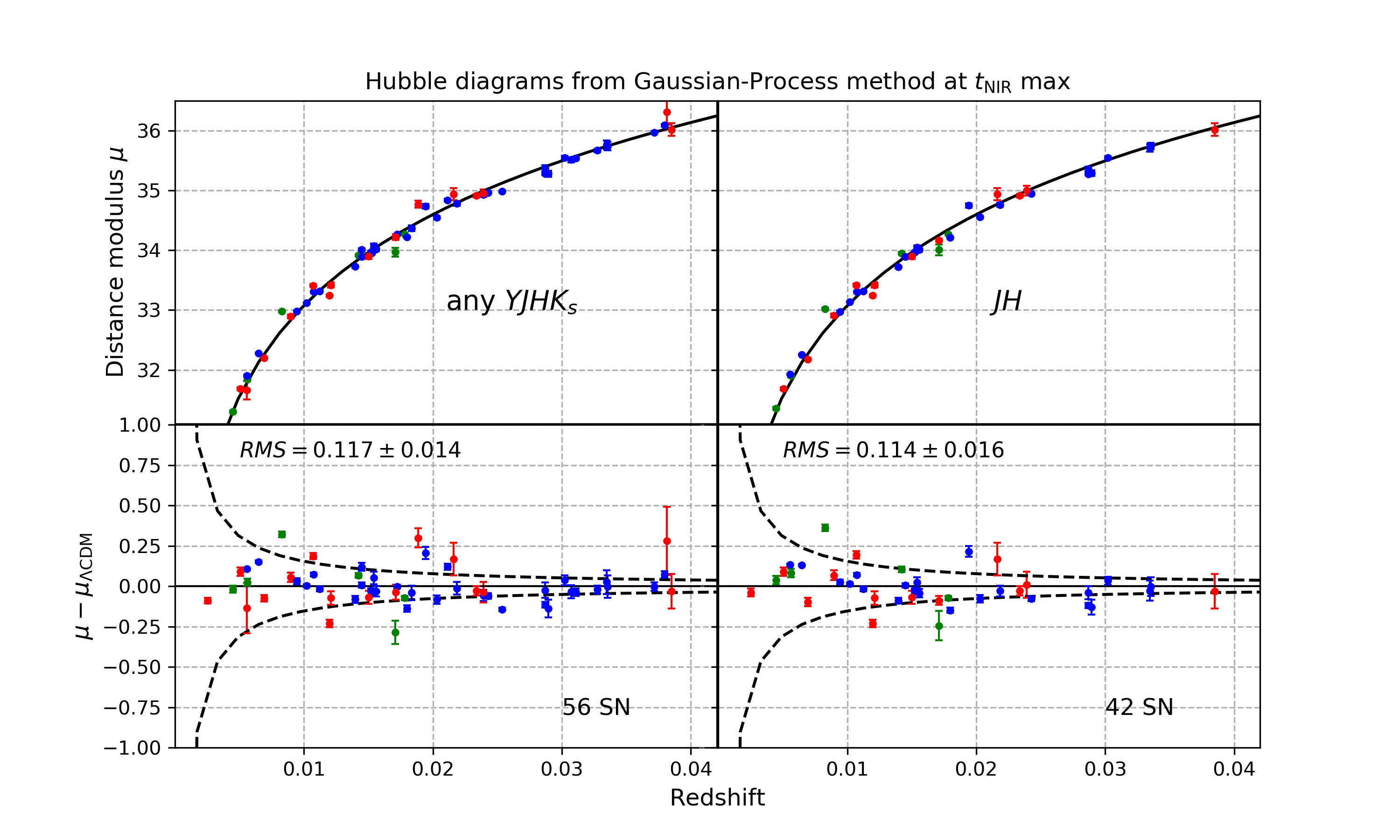}
\caption{Similar to Fig.~\ref{fig_hubbles_Template_AnyYJHK}, but for $\yjhk$ Hubble diagrams (top row) and residuals (bottom row) using the GP method at NIR max. Again, Tables~\ref{Tab_scatter_GPSubsample} and \ref{Tab_IntDispersion_PecVel} summarize the intrinsic scatter in the Hubble diagrams, while
Table~\ref{tab_distanceMu_GaussianProc_NIRmax} lists numerical values of the distance moduli from this figure. }
\label{fig_hubbles_GP_AnyYJHK}
\end{center}
\end{figure*}

\begin{figure*}
\begin{center}
\includegraphics[width=18cm]{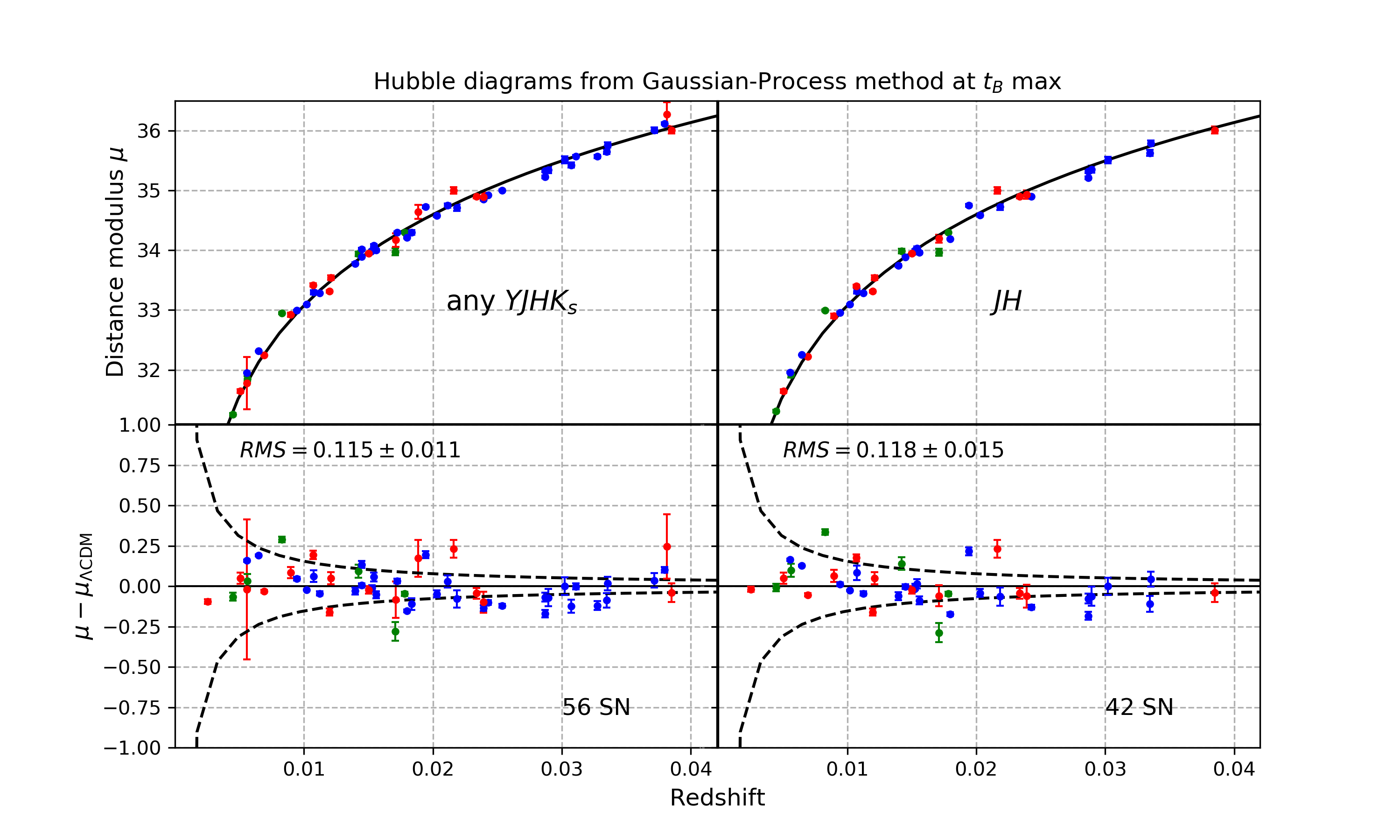}
\caption{Similar to Fig.~\ref{fig_hubbles_GP_AnyYJHK}, but for $\yjhk$ Hubble diagrams (top row) and residuals (bottom row) using the Gaussian-process method at $B$ max. Again, Tables~\ref{Tab_scatter_GPSubsample} and \ref{Tab_IntDispersion_PecVel} summarizes the intrinsic scatter in the Hubble diagram while Table~\ref{tab_distanceMu_GaussianProc_Bmax} shows numerical values of the distance moduli from this figure. }
\label{fig_hubbles_GP_AnyYJHK_Bmax}
\end{center}
\end{figure*}

\begin{figure*}
\begin{center}
\includegraphics[width=18cm]{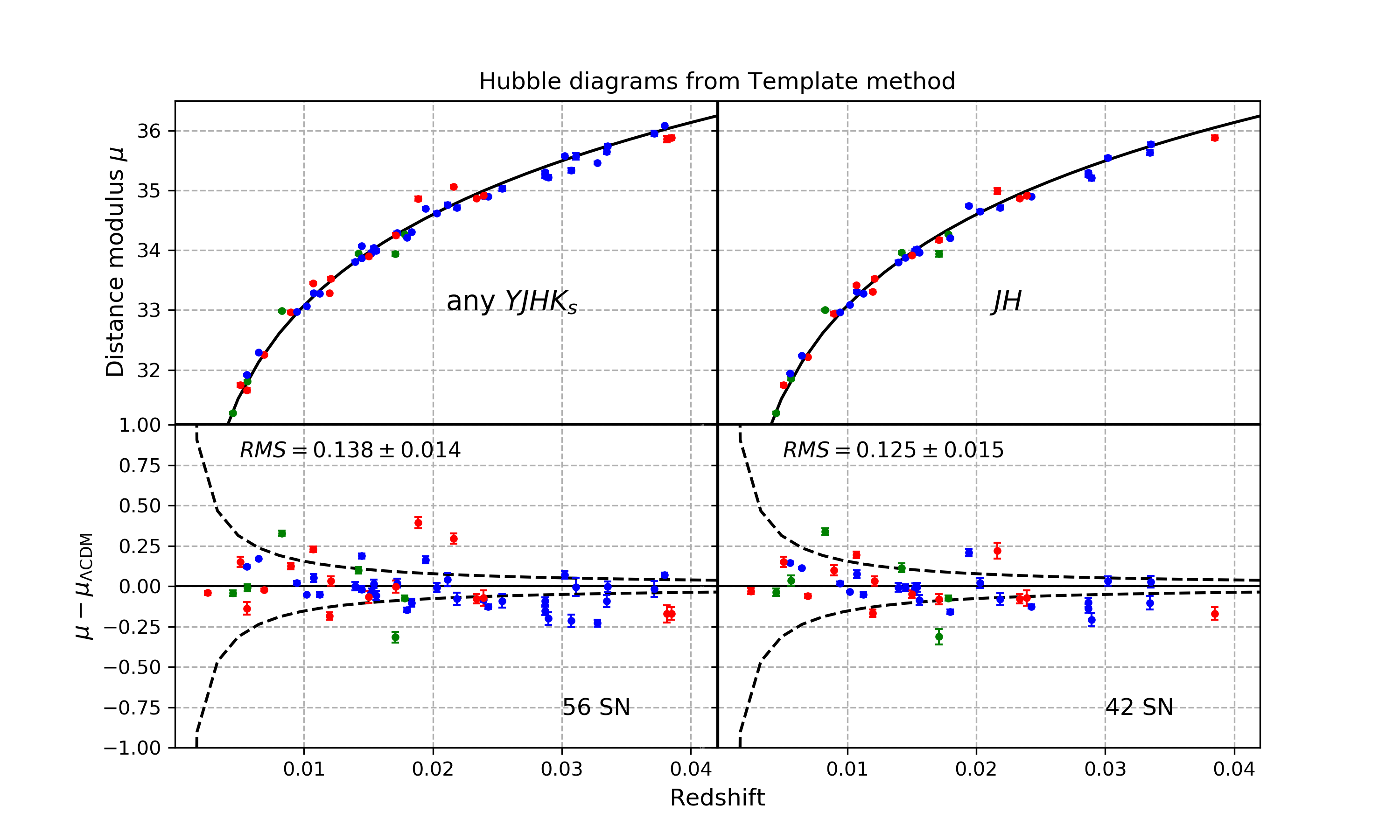}
\caption{Similar to Fig.~\ref{fig_hubbles_Template_AnyYJHK} but applying the template method to exactly same \nAnyYJHKgp{} supernovae shown in Fig.~\ref{fig_hubbles_GP_AnyYJHK} and \ref{fig_hubbles_GP_AnyYJHK_Bmax}. Again, Table~\ref{Tab_scatter_GPSubsample} summarizes the intrinsic scatter in the Hubble diagram while Table~\ref{tab_distanceMu_template} shows numerical values of the distance moduli from this figure. }
\label{fig_hubbles_Template_AnyYJHK_GPNIRsample}
\end{center}
\end{figure*}

\begin{figure*}
\begin{center}
\includegraphics[width=16cm]{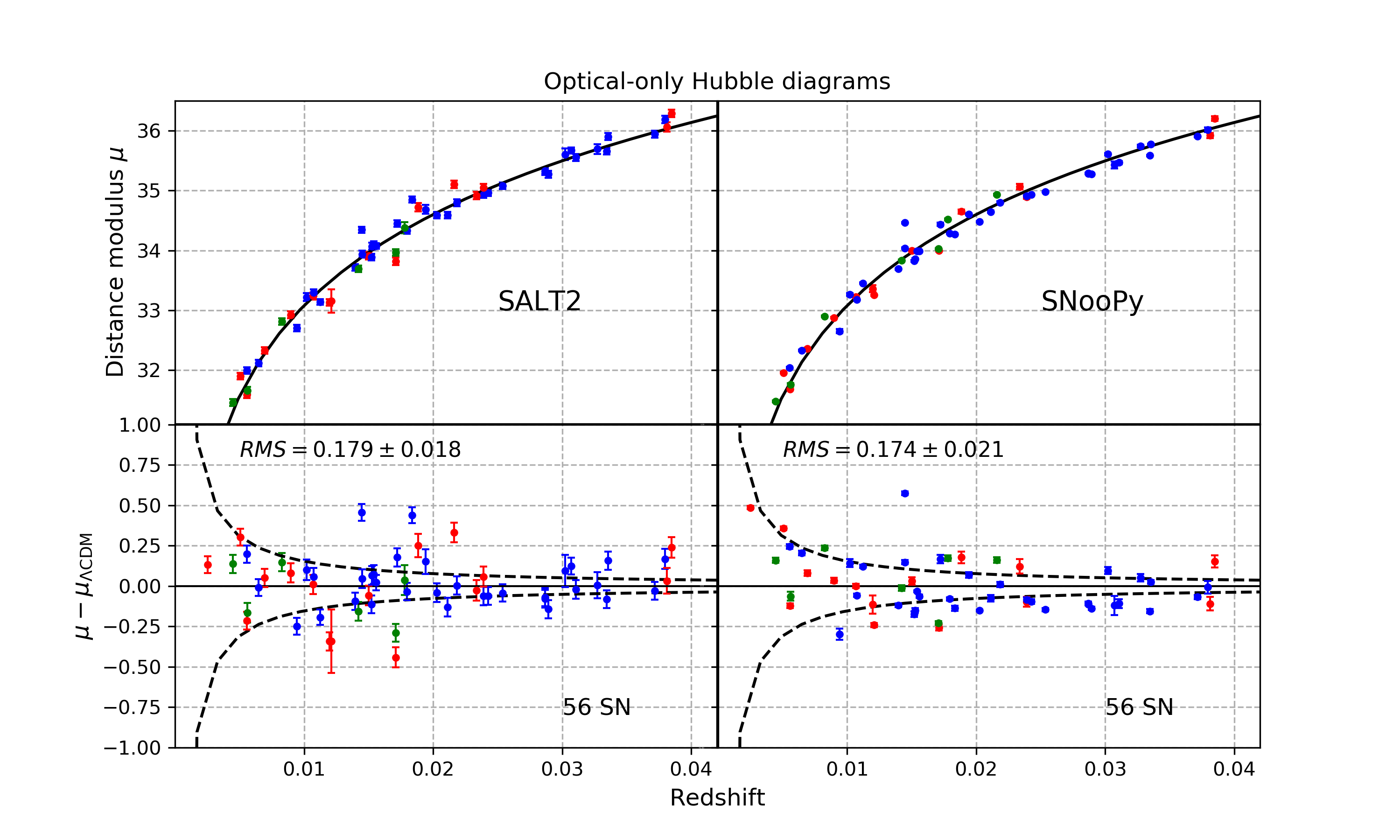}
\caption{Hubble diagram (top row) and residuals (bottom row) using SALT2 and \snoopy{} to fit only the {\it optical} $BVR$-band \lcs{} for exactly same sample of \nAnyYJHKgp{} \snIa{} used for the ``any \yjhk{} '' GP (NIR max) Hubble diagram shown in the left panel of Fig.~\ref{fig_hubbles_GP_AnyYJHK} and listed in Table~\ref{tab_distanceMu_GaussianProc_NIRmax}. As emphasized in Tables~\ref{Tab_scatter_GPSubsample}-\ref{Tab_IntDispersion_PecVel1}, the intrinsic scatter is clearly larger in these optical only Hubble diagrams compared with the GP NIR max ones constructed for the same \nAnyYJHKgp{} \snIa{}. Table~\ref{tab_distanceMu_Optical}
shows numerical values of the distance moduli from this figure.
}
\label{fig_hubbles_SALT2_Snoopy}
\end{center}
\end{figure*}

\renewcommand{\arraystretch}{0.001}
\renewcommand{\tabcolsep}{3pt}
\begin{table*}
\begin{center}
\caption{\snIa{} $\yjhk$ Distance Moduli from Template method}
\scriptsize
\begin{tabular}{l c cccc c}
\hline \hline
 & & $\hat{\mu}^{\rm Y}$ & $\hat{\mu}^{\rm J}$  & $\hat{\mu}^{\rm H}$ & $\hat{\mu}^{\rm K}$ & $\hat{\mu}$  \\
SN name & Source & (mag)   & (mag)  & (mag)  & (mag) & (mag) \\
\hline
SN1998bu     & CfA & ... & $30.11 \pm 0.03$ & $30.02 \pm 0.02$ & $29.97 \pm 0.01$ & $30.03 \pm 0.013$ \\ 
SN1999ee     & CSP & ... & $33.30 \pm 0.02$ & $33.27 \pm 0.02$ & ... & $33.28 \pm 0.016$ \\ 
SN1999ek     & Others & ... & $34.27 \pm 0.02$ & $34.27 \pm 0.02$ & ... & $34.27 \pm 0.016$ \\ 
SN2000bh     & CSP & $34.83 \pm 0.07$ & $35.00 \pm 0.06$ & $34.83 \pm 0.04$ & ... & $34.81 \pm 0.041$ \\ 
SN2000ca     & CSP & ... & $34.87 \pm 0.02$ & $34.92 \pm 0.02$ & ... & $34.91 \pm 0.016$ \\ 
SN2000E      & Others & ... & $31.72 \pm 0.04$ & $31.91 \pm 0.04$ & $31.78 \pm 0.04$ & $31.81 \pm 0.022$ \\ 
SN2001ba     & CSP & ... & $35.56 \pm 0.04$ & $35.54 \pm 0.04$ & $35.66 \pm 0.04$ & $35.58 \pm 0.023$ \\ 
SN2001bt     & Others & ... & $33.94 \pm 0.04$ & $33.97 \pm 0.03$ & $33.92 \pm 0.03$ & $33.94 \pm 0.021$ \\ 
SN2001cn     & Others & ... & $34.06 \pm 0.06$ & $34.00 \pm 0.06$ & $33.95 \pm 0.09$ & $34.00 \pm 0.039$ \\ 
SN2001cz     & Others & ... & $33.88 \pm 0.04$ & $33.96 \pm 0.06$ & $33.96 \pm 0.07$ & $33.94 \pm 0.034$ \\ 
SN2001el     & Others & ... & $31.32 \pm 0.03$ & $31.27 \pm 0.03$ & $31.24 \pm 0.03$ & $31.28 \pm 0.017$ \\ 
SN2002dj     & Others & ... & $32.99 \pm 0.03$ & $33.00 \pm 0.03$ & $32.97 \pm 0.03$ & $32.99 \pm 0.015$ \\ 
SN2003du     & Others & ... & $32.64 \pm 0.04$ & $32.76 \pm 0.03$ & $32.72 \pm 0.03$ & $32.71 \pm 0.020$ \\ 
SN2003hv     & Others & ... & $31.38 \pm 0.02$ & $31.43 \pm 0.02$ & $31.45 \pm 0.02$ & $31.42 \pm 0.011$ \\ 
SN2004ef     & CSP & $35.54 \pm 0.07$ & $35.66 \pm 0.12$ & $35.56 \pm 0.10$ & ... & $35.54 \pm 0.073$ \\ 
SN2004eo     & CSP & $33.91 \pm 0.01$ & $33.95 \pm 0.01$ & $33.98 \pm 0.02$ & ... & $33.95 \pm 0.013$ \\ 
SN2004ey     & CSP & $34.01 \pm 0.02$ & $33.95 \pm 0.03$ & $34.04 \pm 0.04$ & ... & $34.04 \pm 0.026$ \\ 
SN2004gs     & CSP & $35.36 \pm 0.07$ & $35.65 \pm 0.10$ & $35.57 \pm 0.09$ & ... & $35.47 \pm 0.068$ \\ 
SN2004S      & Others & ... & $33.05 \pm 0.04$ & $33.10 \pm 0.03$ & $33.10 \pm 0.05$ & $33.08 \pm 0.023$ \\ 
SN2005bo     & CfA & ... & $33.77 \pm 0.11$ & $33.81 \pm 0.08$ & $33.87 \pm 0.06$ & $33.81 \pm 0.050$ \\ 
SN2005cf     & CfA & ... & $32.22 \pm 0.02$ & $32.21 \pm 0.02$ & $32.34 \pm 0.01$ & $32.25 \pm 0.009$ \\ 
SN2005el     & CSP & $33.95 \pm 0.02$ & $34.02 \pm 0.03$ & $33.99 \pm 0.03$ & ... & $33.97 \pm 0.020$ \\ 
SN2005iq     & CSP & $35.69 \pm 0.02$ & $35.76 \pm 0.03$ & $35.78 \pm 0.05$ & ... & $35.74 \pm 0.033$ \\ 
SN2005kc     & CSP & $33.86 \pm 0.02$ & $33.88 \pm 0.03$ & $33.88 \pm 0.02$ & ... & $33.87 \pm 0.019$ \\ 
SN2005ki     & CSP & $34.58 \pm 0.02$ & $34.64 \pm 0.03$ & $34.65 \pm 0.04$ & ... & $34.62 \pm 0.029$ \\ 
SN2005lu     & CSP & $35.89 \pm 0.11$ & ... & ... & ... & $35.89 \pm 0.106$ \\ 
SN2005na     & CfA & ... & $35.31 \pm 0.13$ & ... & $35.48 \pm 0.15$ & $35.38 \pm 0.097$ \\ 
SN2006ac     & CfA & ... & $35.20 \pm 0.08$ & $35.32 \pm 0.12$ & $35.03 \pm 0.09$ & $35.20 \pm 0.059$ \\ 
SN2006ax     & CSP & $34.18 \pm 0.01$ & $34.15 \pm 0.02$ & $34.23 \pm 0.02$ & ... & $34.22 \pm 0.014$ \\ 
SN2006bh     & CSP & $33.27 \pm 0.03$ & $33.32 \pm 0.04$ & $33.31 \pm 0.03$ & ... & $33.29 \pm 0.025$ \\ 
SN2006bt     & CSP & ... & $35.33 \pm 0.04$ & ... & ... & $35.33 \pm 0.039$ \\ 
SN2006cp     & CfA & ... & $35.12 \pm 0.11$ & $34.90 \pm 0.08$ & $34.47 \pm 0.10$ & $34.85 \pm 0.054$ \\ 
SN2006D      & CfA & ... & $32.93 \pm 0.03$ & $32.94 \pm 0.04$ & $33.03 \pm 0.04$ & $32.97 \pm 0.021$ \\ 
SN2006ej     & CSP & $34.42 \pm 0.14$ & $34.60 \pm 0.09$ & ... & ... & $34.49 \pm 0.094$ \\ 
SN2006kf     & CSP & $34.65 \pm 0.02$ & $34.75 \pm 0.02$ & $34.74 \pm 0.03$ & ... & $34.70 \pm 0.022$ \\ 
SN2006lf     & CfA & ... & $33.48 \pm 0.03$ & $33.54 \pm 0.04$ & ... & $33.53 \pm 0.030$ \\ 
SN2006N      & CfA & ... & $34.05 \pm 0.11$ & $33.92 \pm 0.09$ & $33.74 \pm 0.09$ & $33.91 \pm 0.058$ \\ 
SN2007A      & CSP & $34.30 \pm 0.02$ & $34.18 \pm 0.05$ & $34.26 \pm 0.04$ & ... & $34.29 \pm 0.029$ \\ 
SN2007af     & CSP & $31.91 \pm 0.01$ & $31.96 \pm 0.01$ & $31.93 \pm 0.01$ & ... & $31.92 \pm 0.009$ \\ 
SN2007ai     & CSP & $35.57 \pm 0.02$ & $35.42 \pm 0.03$ & $35.38 \pm 0.03$ & ... & $35.46 \pm 0.022$ \\ 
SN2007as     & CSP & $34.26 \pm 0.02$ & $34.27 \pm 0.02$ & $34.34 \pm 0.04$ & ... & $34.31 \pm 0.025$ \\ 
SN2007bc     & CSP & $34.68 \pm 0.03$ & $34.80 \pm 0.04$ & $34.82 \pm 0.06$ & ... & $34.76 \pm 0.042$ \\ 
SN2007bd     & CSP & $35.54 \pm 0.03$ & $35.58 \pm 0.04$ & $35.60 \pm 0.08$ & ... & $35.57 \pm 0.057$ \\ 
SN2007ca     & CSP & $34.17 \pm 0.01$ & $34.07 \pm 0.01$ & $34.01 \pm 0.02$ & ... & $34.07 \pm 0.016$ \\ 
SN2007co     & CfA & ... & ... & ... & $34.99 \pm 0.11$ & $34.99 \pm 0.110$ \\ 
SN2007cq     & CfA & ... & $34.87 \pm 0.03$ & $34.84 \pm 0.08$ & ... & $34.85 \pm 0.060$ \\ 
SN2007jg     & CSP & $36.05 \pm 0.02$ & $36.16 \pm 0.02$ & ... & ... & $36.09 \pm 0.014$ \\ 
SN2007le     & CSP & $32.36 \pm 0.01$ & $32.24 \pm 0.01$ & $32.24 \pm 0.01$ & ... & $32.29 \pm 0.007$ \\ 
SN2007qe     & CfA & ... & $34.70 \pm 0.17$ & $34.91 \pm 0.07$ & $35.26 \pm 0.15$ & $34.95 \pm 0.075$ \\ 
SN2007sr     & CSP & $31.68 \pm 0.05$ & $31.67 \pm 0.06$ & $31.68 \pm 0.03$ & ... & $31.68 \pm 0.029$ \\ 
SN2007st     & CSP & ... & $34.22 \pm 0.09$ & $34.55 \pm 0.04$ & ... & $34.46 \pm 0.041$ \\ 
SN2008af     & CfA & ... & $35.98 \pm 0.19$ & $35.90 \pm 0.24$ & $35.96 \pm 0.18$ & $35.94 \pm 0.124$ \\ 
SN2008ar     & CSP & $35.30 \pm 0.02$ & $35.30 \pm 0.04$ & $35.17 \pm 0.06$ & ... & $35.22 \pm 0.038$ \\ 
SN2008bc     & CSP & $34.07 \pm 0.02$ & $34.02 \pm 0.04$ & $33.95 \pm 0.04$ & ... & $34.00 \pm 0.029$ \\ 
SN2008bf     & CSP & $34.96 \pm 0.01$ & $34.95 \pm 0.01$ & $35.07 \pm 0.06$ & ... & $35.03 \pm 0.042$ \\ 
SN2008C      & CSP & $34.27 \pm 0.08$ & $34.24 \pm 0.09$ & $34.31 \pm 0.06$ & ... & $34.30 \pm 0.053$ \\ 
SN2008fl     & CSP & $34.42 \pm 0.03$ & $34.52 \pm 0.05$ & $34.55 \pm 0.03$ & ... & $34.49 \pm 0.024$ \\ 
SN2008fr     & CSP & $36.11 \pm 0.06$ & $36.23 \pm 0.14$ & ... & ... & $36.16 \pm 0.067$ \\ 
SN2008fw     & CSP & $33.07 \pm 0.11$ & $33.06 \pm 0.14$ & $32.94 \pm 0.12$ & ... & $32.98 \pm 0.091$ \\ 
SN2008gb     & CfA & ... & $35.98 \pm 0.08$ & $35.78 \pm 0.09$ & $35.83 \pm 0.11$ & $35.86 \pm 0.053$ \\ 
SN2008gg     & CSP & $35.63 \pm 0.05$ & $35.60 \pm 0.10$ & $35.63 \pm 0.07$ & ... & $35.63 \pm 0.051$ \\ 
SN2008gl     & CSP & $35.97 \pm 0.02$ & $35.70 \pm 0.03$ & $35.72 \pm 0.05$ & ... & $35.83 \pm 0.033$ \\ 
SN2008gp     & CSP & $35.55 \pm 0.02$ & $35.50 \pm 0.03$ & $35.69 \pm 0.06$ & ... & $35.65 \pm 0.038$ \\ 
SN2008hj     & CSP & $36.03 \pm 0.05$ & $36.02 \pm 0.07$ & $35.91 \pm 0.07$ & ... & $35.95 \pm 0.049$ \\ 
SN2008hm     & CfA & ... & $34.59 \pm 0.02$ & $34.76 \pm 0.06$ & $34.57 \pm 0.04$ & $34.65 \pm 0.027$ \\ 
SN2008hs     & CfA & ... & $34.86 \pm 0.06$ & $34.90 \pm 0.06$ & $34.82 \pm 0.07$ & $34.86 \pm 0.035$ \\ 
SN2008hv     & CSP & $33.80 \pm 0.02$ & $33.78 \pm 0.02$ & $33.81 \pm 0.04$ & ... & $33.81 \pm 0.026$ \\ 
SN2008ia     & CSP & $34.80 \pm 0.02$ & $34.72 \pm 0.03$ & $34.66 \pm 0.03$ & ... & $34.72 \pm 0.022$ \\ 
SN2009aa     & CSP & $35.23 \pm 0.03$ & $35.27 \pm 0.04$ & $35.25 \pm 0.03$ & ... & $35.24 \pm 0.026$ \\ 

SN2009ab     & CSP & $33.46 \pm 0.02$ & $33.51 \pm 0.03$ & $33.56 \pm 0.03$ & ... & $33.52 \pm 0.023$ \\
SN2009ad     & CSP & $35.24 \pm 0.01$ & $35.21 \pm 0.02$ & $35.33 \pm 0.04$ & ... & $35.30 \pm 0.025$ \\
SN2009ag     & CSP & $33.06 \pm 0.00$ & $33.11 \pm 0.01$ & $33.08 \pm 0.01$ & ... & $33.07 \pm 0.005$ \\
SN2009al     & CfA & ... & $34.92 \pm 0.05$ & $34.84 \pm 0.03$ & ... & $34.87 \pm 0.028$ \\
SN2009an     & CfA & ... & $33.46 \pm 0.03$ & $33.40 \pm 0.03$ & $33.51 \pm 0.04$ & $33.45 \pm 0.017$ \\
SN2009bv     & CfA & ... & $36.03 \pm 0.05$ & $35.82 \pm 0.05$ & ... & $35.88 \pm 0.040$ \\
SN2009cz     & CSP & $34.69 \pm 0.05$ & $34.68 \pm 0.06$ & $34.73 \pm 0.04$ & ... & $34.71 \pm 0.037$ \\
SN2009D      & CSP & $34.90 \pm 0.02$ & $34.90 \pm 0.01$ & $34.90 \pm 0.02$ & ... & $34.90 \pm 0.014$ \\
SN2009kk     & CfA & ... & $33.92 \pm 0.05$ & $34.04 \pm 0.07$ & ... & $34.01 \pm 0.051$ \\
SN2009kq     & CfA & ... & $33.53 \pm 0.09$ & $33.65 \pm 0.09$ & $33.52 \pm 0.06$ & $33.57 \pm 0.048$ \\
SN2009Y      & CSP & $32.97 \pm 0.01$ & $32.96 \pm 0.02$ & $32.97 \pm 0.01$ & ... & $32.97 \pm 0.011$ \\
SN2010ai     & CfA & ... & $35.04 \pm 0.03$ & $34.87 \pm 0.06$ & ... & $34.92 \pm 0.047$ \\
SN2010dw     & CfA & ... & $36.12 \pm 0.04$ & ... & ... & $36.12 \pm 0.045$ \\
SN2010iw     & CfA & ... & $34.70 \pm 0.04$ & $34.63 \pm 0.06$ & $34.73 \pm 0.10$ & $34.68 \pm 0.039$ \\
SN2010kg     & CfA & ... & $34.24 \pm 0.04$ & $34.14 \pm 0.04$ & $34.40 \pm 0.11$ & $34.25 \pm 0.037$ \\
SN2011ao     & CfA & ... & $33.35 \pm 0.03$ & $33.29 \pm 0.03$ & $33.22 \pm 0.06$ & $33.29 \pm 0.023$ \\
SN2011B      & CfA & ... & $31.62 \pm 0.07$ & $31.68 \pm 0.05$ & ... & $31.66 \pm 0.038$ \\
SN2011by     & CfA & ... & $31.76 \pm 0.06$ & $31.74 \pm 0.04$ & ... & $31.75 \pm 0.032$ \\
SN2011df     & CfA & ... & $33.97 \pm 0.01$ & $33.90 \pm 0.03$ & $33.83 \pm 0.12$ & $33.90 \pm 0.037$ \\
SNf20080514-002 & CfA & ... & $35.03 \pm 0.04$ & $34.97 \pm 0.07$ & $35.23 \pm 0.05$ & $35.07 \pm 0.032$ \\
\hline
\end{tabular}
\label{tab_distanceMu_template}
\end{center}
\tablecomments{
Distance moduli and their fitting uncertainties $\sigmafithat$, estimated from the different NIR bands, either alone (see columns 3-6) or combined (column 7), using the template method.
Corresponding Hubble diagrams are shown in Figs.~\ref{fig_hubbles_Template_AnyYJHK} and \ref{fig_hubbles_Template_Individuals}.
}
\end{table*}
\renewcommand{\tabcolsep}{6pt}
\renewcommand{\arraystretch}{1}

\renewcommand{\arraystretch}{0.001}
\renewcommand{\tabcolsep}{3pt}
\begin{table*}
\begin{center}
\caption{\snIa{} $\yjhk$ Distance Moduli from Gaussian-process Method at NIR max}
\scriptsize
\begin{tabular}{l c cccc c}
\hline \hline
 & & $\hat{\mu}^{\rm Y}$ & $\hat{\mu}^{\rm J}$  & $\hat{\mu}^{\rm H}$ & $\hat{\mu}^{\rm K}$ & $\hat{\mu}$  \\
SN name & Source & (mag)   & (mag)  & (mag)  & (mag) & (mag) \\
\hline
SN1998bu     & CfA & ... & $30.09 \pm 0.03$ & $30.03 \pm 0.03$ & $29.87 \pm 0.02$ & $29.99 \pm 0.018$ \\
SN1999ee     & CSP & ... & $33.30 \pm 0.02$ & $33.32 \pm 0.02$ & ... & $33.32 \pm 0.016$ \\
SN1999ek     & Others & ... & $34.24 \pm 0.02$ & $34.28 \pm 0.02$ & ... & $34.27 \pm 0.014$ \\
SN2000ca     & CSP & ... & $34.94 \pm 0.03$ & ... & ... & $34.94 \pm 0.034$ \\
SN2000E      & Others & ... & $31.72 \pm 0.02$ & $31.93 \pm 0.03$ & $31.71 \pm 0.04$ & $31.84 \pm 0.022$ \\
SN2001ba     & CSP & ... & $35.55 \pm 0.02$ & $35.55 \pm 0.03$ & $35.56 \pm 0.08$ & $35.55 \pm 0.031$ \\
SN2001bt     & Others & ... & $33.91 \pm 0.03$ & $33.96 \pm 0.02$ & $33.83 \pm 0.03$ & $33.91 \pm 0.014$ \\
SN2001cz     & Others & ... & $33.90 \pm 0.05$ & $34.03 \pm 0.11$ & $33.87 \pm 0.11$ & $33.97 \pm 0.072$ \\
SN2001el     & Others & ... & $31.45 \pm 0.02$ & $31.34 \pm 0.03$ & $31.17 \pm 0.04$ & $31.30 \pm 0.022$ \\
SN2002dj     & Others & ... & $33.01 \pm 0.03$ & $33.03 \pm 0.02$ & $32.89 \pm 0.03$ & $32.98 \pm 0.017$ \\
SN2004eo     & CSP & $33.97 \pm 0.01$ & $33.97 \pm 0.01$ & $33.97 \pm 0.01$ & ... & $33.97 \pm 0.011$ \\
SN2004ey     & CSP & $34.03 \pm 0.01$ & $33.96 \pm 0.01$ & $34.06 \pm 0.04$ & ... & $34.08 \pm 0.038$ \\
SN2005cf     & CfA & ... & $32.21 \pm 0.06$ & $32.17 \pm 0.03$ & $32.25 \pm 0.03$ & $32.20 \pm 0.020$ \\
SN2005el     & CSP & $33.88 \pm 0.01$ & $33.95 \pm 0.01$ & $33.97 \pm 0.01$ & ... & $33.96 \pm 0.013$ \\
SN2005iq     & CSP & $35.76 \pm 0.03$ & $35.75 \pm 0.04$ & $35.74 \pm 0.07$ & ... & $35.74 \pm 0.069$ \\
SN2005kc     & CSP & $33.92 \pm 0.01$ & $33.90 \pm 0.01$ & $33.89 \pm 0.02$ & ... & $33.90 \pm 0.015$ \\
SN2005ki     & CSP & $34.58 \pm 0.01$ & $34.59 \pm 0.01$ & $34.55 \pm 0.03$ & ... & $34.55 \pm 0.027$ \\
SN2006ax     & CSP & $34.17 \pm 0.01$ & $34.16 \pm 0.01$ & $34.22 \pm 0.02$ & ... & $34.22 \pm 0.019$ \\
SN2006bh     & CSP & $33.31 \pm 0.01$ & $33.30 \pm 0.01$ & $33.31 \pm 0.01$ & ... & $33.31 \pm 0.012$ \\
SN2006bt     & CSP & ... & $35.51 \pm 0.04$ & ... & ... & $35.51 \pm 0.041$ \\
SN2006D      & CfA & ... & $32.85 \pm 0.02$ & $32.92 \pm 0.04$ & $32.86 \pm 0.06$ & $32.90 \pm 0.029$ \\
SN2006kf     & CSP & $34.73 \pm 0.01$ & $34.77 \pm 0.01$ & $34.75 \pm 0.04$ & ... & $34.74 \pm 0.039$ \\
SN2006lf     & CfA & ... & $33.40 \pm 0.03$ & $33.42 \pm 0.05$ & ... & $33.42 \pm 0.043$ \\
SN2007A      & CSP & $34.26 \pm 0.01$ & $34.17 \pm 0.02$ & ... & ... & $34.27 \pm 0.012$ \\
SN2007af     & CSP & $31.92 \pm 0.01$ & $32.00 \pm 0.00$ & $31.92 \pm 0.01$ & ... & $31.90 \pm 0.007$ \\
SN2007ai     & CSP & $35.67 \pm 0.02$ & ... & ... & ... & $35.67 \pm 0.025$ \\
SN2007as     & CSP & ... & ... & $34.37 \pm 0.04$ & ... & $34.37 \pm 0.044$ \\
SN2007bc     & CSP & ... & $34.84 \pm 0.02$ & ... & ... & $34.84 \pm 0.019$ \\
SN2007bd     & CSP & $35.54 \pm 0.02$ & $35.56 \pm 0.04$ & ... & ... & $35.54 \pm 0.023$ \\
SN2007ca     & CSP & ... & ... & $34.01 \pm 0.03$ & ... & $34.01 \pm 0.025$ \\
SN2007jg     & CSP & $36.09 \pm 0.02$ & $36.11 \pm 0.02$ & ... & ... & $36.09 \pm 0.021$ \\
SN2007le     & CSP & $32.33 \pm 0.01$ & $32.23 \pm 0.01$ & $32.26 \pm 0.01$ & ... & $32.28 \pm 0.006$ \\
SN2008ar     & CSP & $35.33 \pm 0.01$ & $35.34 \pm 0.03$ & $35.28 \pm 0.05$ & ... & $35.28 \pm 0.054$ \\
SN2008bc     & CSP & $34.02 \pm 0.01$ & $33.98 \pm 0.02$ & $34.01 \pm 0.03$ & ... & $34.02 \pm 0.031$ \\
SN2008bf     & CSP & $34.98 \pm 0.01$ & $34.87 \pm 0.01$ & ... & ... & $34.98 \pm 0.010$ \\
SN2008gb     & CfA & ... & ... & ... & $36.31 \pm 0.21$ & $36.31 \pm 0.211$ \\
SN2008gp     & CSP & $35.63 \pm 0.03$ & $35.53 \pm 0.05$ & $35.75 \pm 0.07$ & ... & $35.76 \pm 0.074$ \\
SN2008hj     & CSP & $35.98 \pm 0.02$ & $36.08 \pm 0.04$ & ... & ... & $35.97 \pm 0.026$ \\
SN2008hs     & CfA & ... & $34.86 \pm 0.05$ & ... & $34.67 \pm 0.11$ & $34.77 \pm 0.059$ \\
SN2008hv     & CSP & $33.77 \pm 0.00$ & $33.72 \pm 0.01$ & $33.72 \pm 0.02$ & ... & $33.73 \pm 0.022$ \\
SN2009aa     & CSP & $35.31 \pm 0.01$ & $35.28 \pm 0.01$ & $35.28 \pm 0.02$ & ... & $35.28 \pm 0.018$ \\
SN2009ad     & CSP & $35.27 \pm 0.01$ & $35.27 \pm 0.02$ & $35.37 \pm 0.05$ & ... & $35.37 \pm 0.048$ \\
SN2009ag     & CSP & $33.20 \pm 0.01$ & $33.20 \pm 0.01$ & $33.12 \pm 0.01$ & ... & $33.12 \pm 0.009$ \\
SN2009al     & CfA & ... & $35.07 \pm 0.02$ & $34.88 \pm 0.03$ & ... & $34.91 \pm 0.029$ \\
SN2009an     & CfA & ... & $33.40 \pm 0.02$ & $33.42 \pm 0.03$ & $33.39 \pm 0.03$ & $33.41 \pm 0.019$ \\
SN2009bv     & CfA & ... & $36.07 \pm 0.04$ & $36.01 \pm 0.13$ & ... & $36.02 \pm 0.108$ \\
SN2009cz     & CSP & $34.77 \pm 0.01$ & $34.72 \pm 0.02$ & $34.77 \pm 0.04$ & ... & $34.78 \pm 0.038$ \\
SN2009D      & CSP & $34.98 \pm 0.01$ & $34.91 \pm 0.01$ & $34.95 \pm 0.02$ & ... & $34.97 \pm 0.018$ \\
SN2009Y      & CSP & $33.02 \pm 0.01$ & $32.98 \pm 0.01$ & $32.97 \pm 0.02$ & ... & $32.98 \pm 0.018$ \\
SN2010ai     & CfA & ... & $35.05 \pm 0.05$ & $34.99 \pm 0.10$ & $34.85 \pm 0.09$ & $34.95 \pm 0.063$ \\
SN2010kg     & CfA & ... & $34.22 \pm 0.03$ & $34.16 \pm 0.03$ & $34.33 \pm 0.14$ & $34.22 \pm 0.047$ \\
SN2011ao     & CfA & ... & $33.33 \pm 0.05$ & $33.23 \pm 0.03$ & ... & $33.24 \pm 0.023$ \\
SN2011B      & CfA & ... & $31.66 \pm 0.16$ & ... & ... & $31.66 \pm 0.156$ \\
SN2011by     & CfA & ... & $31.71 \pm 0.05$ & $31.68 \pm 0.03$ & ... & $31.69 \pm 0.026$ \\
SN2011df     & CfA & ... & $33.94 \pm 0.03$ & $33.89 \pm 0.05$ & ... & $33.90 \pm 0.041$ \\
SNf20080514-002 & CfA & ... & $35.03 \pm 0.14$ & $34.92 \pm 0.12$ & ... & $34.94 \pm 0.101$ \\
\hline
\end{tabular}
\label{tab_distanceMu_GaussianProc_NIRmax}
\end{center}
\tablecomments{
Distance moduli and their fitting uncertainties $\sigmafithat$, estimated from the different NIR bands, either alone (see columns 3-6) or combined (column 7) using the Gaussian-process method at NIR max.
The Hubble diagrams from these data are shown in Figs.~\ref{fig_hubbles_GP_AnyYJHK} and \ref{fig_hubbles_GP_Individuals}.
}
\end{table*}
\renewcommand{\tabcolsep}{6pt}
\renewcommand{\arraystretch}{1}

\renewcommand{\arraystretch}{0.001}
\renewcommand{\tabcolsep}{3pt}
\begin{table*}
\begin{center}
\caption{\snIa{} $\yjhk$ Distance Moduli from Gaussian-process Method at $B$ max}
\scriptsize
\begin{tabular}{l c cccc c}
\hline \hline
 & & $\hat{\mu}^{\rm Y}$ & $\hat{\mu}^{\rm J}$  & $\hat{\mu}^{\rm H}$ & $\hat{\mu}^{\rm K}$ & $\hat{\mu}$  \\
SN name & Source & (mag)   & (mag)  & (mag)  & (mag) & (mag) \\
\hline
SN1998bu     & CfA & ... & $30.11 \pm 0.02$ & $30.04 \pm 0.02$ & $29.90 \pm 0.02$ & $29.98 \pm 0.014$ \\
SN1999ee     & CSP & ... & $33.31 \pm 0.02$ & $33.28 \pm 0.02$ & ... & $33.29 \pm 0.013$ \\
SN1999ek     & Others & ... & $34.26 \pm 0.02$ & $34.31 \pm 0.02$ & ... & $34.30 \pm 0.013$ \\
SN2000ca     & CSP & ... & $34.86 \pm 0.02$ & ... & ... & $34.86 \pm 0.019$ \\
SN2000E      & Others & ... & $31.77 \pm 0.04$ & $31.96 \pm 0.05$ & $31.68 \pm 0.06$ & $31.85 \pm 0.041$ \\
SN2001ba     & CSP & ... & $35.43 \pm 0.09$ & $35.54 \pm 0.07$ & $35.47 \pm 0.09$ & $35.51 \pm 0.055$ \\
SN2001bt     & Others & ... & $33.86 \pm 0.06$ & $34.02 \pm 0.05$ & $33.81 \pm 0.06$ & $33.94 \pm 0.039$ \\
SN2001cz     & Others & ... & $33.81 \pm 0.05$ & $34.01 \pm 0.07$ & $33.90 \pm 0.08$ & $33.97 \pm 0.058$ \\
SN2001el     & Others & ... & $31.36 \pm 0.02$ & $31.30 \pm 0.03$ & $31.20 \pm 0.04$ & $31.25 \pm 0.023$ \\
SN2002dj     & Others & ... & $32.97 \pm 0.02$ & $33.00 \pm 0.02$ & $32.87 \pm 0.03$ & $32.95 \pm 0.018$ \\
SN2004eo     & CSP & $33.96 \pm 0.01$ & $33.94 \pm 0.01$ & $33.99 \pm 0.01$ & ... & $33.98 \pm 0.009$ \\
SN2004ey     & CSP & $34.06 \pm 0.02$ & $33.92 \pm 0.02$ & $34.07 \pm 0.04$ & ... & $34.08 \pm 0.027$ \\
SN2005cf     & CfA & ... & $32.23 \pm 0.01$ & $32.22 \pm 0.01$ & $32.28 \pm 0.01$ & $32.24 \pm 0.009$ \\
SN2005el     & CSP & $33.96 \pm 0.01$ & $33.97 \pm 0.02$ & $34.01 \pm 0.02$ & ... & $33.99 \pm 0.017$ \\
SN2005iq     & CSP & $35.70 \pm 0.02$ & $35.76 \pm 0.03$ & $35.80 \pm 0.06$ & ... & $35.76 \pm 0.042$ \\
SN2005kc     & CSP & $33.90 \pm 0.01$ & $33.88 \pm 0.02$ & $33.89 \pm 0.02$ & ... & $33.90 \pm 0.013$ \\
SN2005ki     & CSP & $34.59 \pm 0.02$ & $34.61 \pm 0.03$ & $34.58 \pm 0.03$ & ... & $34.58 \pm 0.022$ \\
SN2006ax     & CSP & $34.20 \pm 0.01$ & $34.13 \pm 0.01$ & $34.21 \pm 0.01$ & ... & $34.21 \pm 0.010$ \\
SN2006bh     & CSP & $33.28 \pm 0.03$ & $33.34 \pm 0.09$ & $33.31 \pm 0.05$ & ... & $33.30 \pm 0.037$ \\
SN2006bt     & CSP & ... & $35.43 \pm 0.04$ & ... & ... & $35.43 \pm 0.041$ \\
SN2006D      & CfA & ... & $32.89 \pm 0.03$ & $32.91 \pm 0.05$ & $32.94 \pm 0.05$ & $32.92 \pm 0.035$ \\
SN2006kf     & CSP & $34.68 \pm 0.01$ & $34.73 \pm 0.01$ & $34.76 \pm 0.03$ & ... & $34.73 \pm 0.022$ \\
SN2006lf     & CfA & ... & $33.45 \pm 0.03$ & $33.57 \pm 0.05$ & ... & $33.54 \pm 0.038$ \\
SN2007A      & CSP & $34.34 \pm 0.01$ & $34.19 \pm 0.04$ & ... & ... & $34.30 \pm 0.015$ \\
SN2007af     & CSP & $31.95 \pm 0.01$ & $31.96 \pm 0.01$ & $31.96 \pm 0.01$ & ... & $31.95 \pm 0.007$ \\
SN2007ai     & CSP & $35.57 \pm 0.03$ & ... & ... & ... & $35.57 \pm 0.027$ \\
SN2007as     & CSP & ... & ... & $34.30 \pm 0.04$ & ... & $34.30 \pm 0.036$ \\
SN2007bc     & CSP & ... & $34.75 \pm 0.04$ & ... & ... & $34.75 \pm 0.037$ \\
SN2007bd     & CSP & $35.56 \pm 0.02$ & $35.60 \pm 0.04$ & ... & ... & $35.57 \pm 0.020$ \\
SN2007ca     & CSP & ... & ... & $34.02 \pm 0.02$ & ... & $34.02 \pm 0.021$ \\
SN2007jg     & CSP & $36.08 \pm 0.02$ & $36.24 \pm 0.03$ & ... & ... & $36.12 \pm 0.018$ \\
SN2007le     & CSP & $32.41 \pm 0.01$ & $32.24 \pm 0.01$ & $32.26 \pm 0.01$ & ... & $32.32 \pm 0.005$ \\
SN2008ar     & CSP & $35.37 \pm 0.03$ & $35.41 \pm 0.05$ & $35.34 \pm 0.08$ & ... & $35.35 \pm 0.053$ \\
SN2008bc     & CSP & $34.14 \pm 0.02$ & $34.08 \pm 0.03$ & $33.93 \pm 0.03$ & ... & $34.00 \pm 0.023$ \\
SN2008bf     & CSP & $35.01 \pm 0.02$ & $34.98 \pm 0.02$ & ... & ... & $35.00 \pm 0.012$ \\
SN2008gb     & CfA & ... & ... & ... & $36.28 \pm 0.20$ & $36.28 \pm 0.200$ \\
SN2008gp     & CSP & $35.59 \pm 0.01$ & $35.50 \pm 0.03$ & $35.67 \pm 0.06$ & ... & $35.65 \pm 0.044$ \\
SN2008hj     & CSP & $35.99 \pm 0.05$ & $36.05 \pm 0.10$ & ... & ... & $36.01 \pm 0.044$ \\
SN2008hs     & CfA & ... & $34.96 \pm 0.09$ & ... & $34.60 \pm 0.13$ & $34.64 \pm 0.115$ \\
SN2008hv     & CSP & $33.88 \pm 0.01$ & $33.83 \pm 0.02$ & $33.72 \pm 0.03$ & ... & $33.78 \pm 0.022$ \\
SN2009aa     & CSP & $35.22 \pm 0.02$ & $35.18 \pm 0.02$ & $35.22 \pm 0.03$ & ... & $35.23 \pm 0.023$ \\
SN2009ad     & CSP & $35.28 \pm 0.01$ & $35.24 \pm 0.02$ & $35.35 \pm 0.04$ & ... & $35.33 \pm 0.027$ \\
SN2009ag     & CSP & $33.11 \pm 0.00$ & $33.12 \pm 0.00$ & $33.09 \pm 0.00$ & ... & $33.10 \pm 0.003$ \\
SN2009al     & CfA & ... & $35.02 \pm 0.04$ & $34.86 \pm 0.04$ & ... & $34.90 \pm 0.033$ \\
SN2009an     & CfA & ... & $33.48 \pm 0.02$ & $33.37 \pm 0.03$ & $33.49 \pm 0.05$ & $33.42 \pm 0.026$ \\
SN2009bv     & CfA & ... & $35.94 \pm 0.08$ & $36.03 \pm 0.07$ & ... & $36.01 \pm 0.058$ \\
SN2009cz     & CSP & $34.72 \pm 0.08$ & $34.77 \pm 0.12$ & $34.72 \pm 0.06$ & ... & $34.71 \pm 0.054$ \\
SN2009D      & CSP & $34.96 \pm 0.02$ & $34.90 \pm 0.01$ & $34.90 \pm 0.02$ & ... & $34.92 \pm 0.015$ \\
SN2009Y      & CSP & $33.01 \pm 0.01$ & $32.91 \pm 0.02$ & $32.98 \pm 0.01$ & ... & $32.99 \pm 0.011$ \\
SN2010ai     & CfA & ... & $35.02 \pm 0.03$ & $34.91 \pm 0.09$ & $34.89 \pm 0.08$ & $34.89 \pm 0.065$ \\
SN2010kg     & CfA & ... & $34.42 \pm 0.08$ & $34.13 \pm 0.08$ & $34.26 \pm 0.25$ & $34.17 \pm 0.113$ \\
SN2011ao     & CfA & ... & $33.35 \pm 0.03$ & $33.30 \pm 0.03$ & ... & $33.31 \pm 0.022$ \\
SN2011B      & CfA & ... & $31.78 \pm 0.43$ & ... & ... & $31.78 \pm 0.434$ \\
SN2011by     & CfA & ... & $31.67 \pm 0.06$ & $31.64 \pm 0.04$ & ... & $31.65 \pm 0.034$ \\
SN2011df     & CfA & ... & $33.99 \pm 0.01$ & $33.93 \pm 0.03$ & ... & $33.94 \pm 0.022$ \\
SNf20080514-002 & CfA & ... & $35.03 \pm 0.04$ & $34.99 \pm 0.07$ & ... & $35.00 \pm 0.054$ \\

\hline
\end{tabular}
\label{tab_distanceMu_GaussianProc_Bmax}
\end{center}
\tablecomments{
Same as Table~\ref{tab_distanceMu_GaussianProc_NIRmax} but using the Gaussian-process method at $B$ max. The Hubble diagrams from these data are shown in Figs.~\ref{fig_hubbles_GP_AnyYJHK_Bmax} and \ref{fig_hubbles_GP_Individuals_Bmax}.
}
\end{table*}
\renewcommand{\tabcolsep}{6pt}
\renewcommand{\arraystretch}{1}

\renewcommand{\arraystretch}{0.001}
\renewcommand{\tabcolsep}{3pt}
\begin{table*}
\begin{center}
\caption{\snIa{} distance moduli from the optical $BVR$ bands}
\scriptsize
\begin{tabular}{l c c c }
\hline \hline
 & & SALT2 & \snoopy{}  \\
SN name & Source & (mag)  & (mag)   \\
\hline
SN1998bu           & CfA     &  $30.214 \pm 0.052$  &  $30.562 \pm 0.009$ \\ 
SN2005cf           & CfA     &  $32.327 \pm 0.056$  &  $32.355 \pm 0.017$ \\ 
SN2006D            & CfA     &  $32.923 \pm 0.060$  &  $32.877 \pm 0.015$ \\ 
SN2006lf           & CfA     &  $33.153 \pm 0.196$  &  $33.254 \pm 0.013$ \\ 
SN2008gb           & CfA     &  $36.062 \pm 0.079$  &  $35.920 \pm 0.042$ \\ 
SN2008hs           & CfA     &  $34.721 \pm 0.072$  &  $34.649 \pm 0.036$ \\ 
SN2009al           & CfA     &  $34.916 \pm 0.064$  &  $35.065 \pm 0.045$ \\ 
SN2009an           & CfA     &  $33.236 \pm 0.059$  &  $33.225 \pm 0.013$ \\ 
SN2009bv           & CfA     &  $36.292 \pm 0.064$  &  $36.204 \pm 0.036$ \\ 
SN2010ai           & CfA     &  $35.051 \pm 0.062$  &  $34.895 \pm 0.029$ \\ 
SN2010kg           & CfA     &  $33.814 \pm 0.062$  &  $33.999 \pm 0.018$ \\ 
SN2011ao           & CfA     &  $33.130 \pm 0.057$  &  $33.359 \pm 0.056$ \\ 
SN2011B            & CfA     &  $31.585 \pm 0.055$  &  $31.678 \pm 0.014$ \\ 
SN2011by           & CfA     &  $31.898 \pm 0.052$  &  $31.953 \pm 0.012$ \\ 
SN2011df           & CfA     &  $33.909 \pm 0.060$  &  $33.999 \pm 0.022$ \\ 
SN1999ee           & CSP     &  $33.138 \pm 0.048$  &  $33.454 \pm 0.010$ \\ 
SN2000ca           & CSP     &  $34.933 \pm 0.058$  &  $34.910 \pm 0.014$ \\ 
SN2001ba           & CSP     &  $35.607 \pm 0.100$  &  $35.609 \pm 0.020$ \\ 
SN2004eo           & CSP     &  $33.884 \pm 0.054$  &  $33.823 \pm 0.017$ \\ 
SN2004ey           & CSP     &  $34.096 \pm 0.057$  &  $33.991 \pm 0.006$ \\ 
SN2005el           & CSP     &  $34.070 \pm 0.057$  &  $33.853 \pm 0.015$ \\ 
SN2005iq           & CSP     &  $35.903 \pm 0.057$  &  $35.771 \pm 0.011$ \\ 
SN2005kc           & CSP     &  $33.936 \pm 0.057$  &  $34.038 \pm 0.013$ \\ 
SN2005ki           & CSP     &  $34.589 \pm 0.057$  &  $34.479 \pm 0.007$ \\ 
SN2006ax           & CSP     &  $34.329 \pm 0.053$  &  $34.284 \pm 0.010$ \\ 
SN2006bh           & CSP     &  $33.292 \pm 0.055$  &  $33.178 \pm 0.011$ \\ 
SN2006bt           & CSP     &  $35.672 \pm 0.052$  &  $35.429 \pm 0.060$ \\ 
SN2006kf           & CSP     &  $34.686 \pm 0.077$  &  $34.602 \pm 0.017$ \\ 
SN2007A            & CSP     &  $34.447 \pm 0.057$  &  $34.436 \pm 0.025$ \\ 
SN2007af           & CSP     &  $31.993 \pm 0.053$  &  $32.041 \pm 0.015$ \\ 
SN2007ai           & CSP     &  $35.695 \pm 0.081$  &  $35.742 \pm 0.023$ \\ 
SN2007as           & CSP     &  $34.848 \pm 0.049$  &  $34.273 \pm 0.015$ \\ 
SN2007bc           & CSP     &  $34.588 \pm 0.057$  &  $34.645 \pm 0.020$ \\ 
SN2007bd           & CSP     &  $35.554 \pm 0.059$  &  $35.467 \pm 0.027$ \\ 
SN2007ca           & CSP     &  $34.343 \pm 0.053$  &  $34.462 \pm 0.012$ \\ 
SN2007jg           & CSP     &  $36.189 \pm 0.062$  &  $36.014 \pm 0.039$ \\ 
SN2007le           & CSP     &  $32.117 \pm 0.053$  &  $32.330 \pm 0.014$ \\ 
SN2008ar           & CSP     &  $35.274 \pm 0.057$  &  $35.279 \pm 0.010$ \\ 
SN2008bc           & CSP     &  $34.074 \pm 0.047$  &  $33.990 \pm 0.010$ \\ 
SN2008bf           & CSP     &  $35.083 \pm 0.054$  &  $34.979 \pm 0.010$ \\ 
SN2008gp           & CSP     &  $35.659 \pm 0.055$  &  $35.584 \pm 0.012$ \\ 
SN2008hj           & CSP     &  $35.943 \pm 0.056$  &  $35.905 \pm 0.013$ \\ 
SN2008hv           & CSP     &  $33.715 \pm 0.054$  &  $33.691 \pm 0.011$ \\ 
SN2009aa           & CSP     &  $35.318 \pm 0.056$  &  $35.284 \pm 0.007$ \\ 
SN2009ad           & CSP     &  $35.328 \pm 0.056$  &  $35.289 \pm 0.011$ \\ 
SN2009ag           & CSP     &  $33.219 \pm 0.064$  &  $33.261 \pm 0.025$ \\ 
SN2009cz           & CSP     &  $34.797 \pm 0.057$  &  $34.803 \pm 0.017$ \\ 
SN2009D            & CSP     &  $34.965 \pm 0.054$  &  $34.930 \pm 0.015$ \\ 
SN2009Y            & CSP     &  $32.700 \pm 0.053$  &  $32.650 \pm 0.034$ \\ 
SN1999ek           & Others  &  $34.380 \pm 0.092$  &  $34.517 \pm 0.017$ \\ 
SN2000E            & Others  &  $31.655 \pm 0.060$  &  $31.756 \pm 0.028$ \\ 
SN2001bt           & Others  &  $33.691 \pm 0.057$  &  $33.836 \pm 0.016$ \\ 
SN2001cz           & Others  &  $33.964 \pm 0.056$  &  $34.026 \pm 0.011$ \\ 
SN2001el           & Others  &  $31.457 \pm 0.055$  &  $31.479 \pm 0.017$ \\ 
SN2002dj           & Others  &  $32.810 \pm 0.056$  &  $32.899 \pm 0.014$ \\ 
SNf20080514-002    & Others  &  $35.102 \pm 0.061$  &  $34.933 \pm 0.017$ \\ 
\hline
\end{tabular}
\label{tab_distanceMu_Optical}
\end{center}
\tablecomments{
Distance moduli estimated by fitting the optical $BVR$ bands for the same \nAnyYJHKgp{} supernovae listed in Table~\ref{tab_distanceMu_GaussianProc_NIRmax} using the SALT2 and \snoopy{} fitters. The Hubble diagrams for these 2 cases are shown in Fig.~\ref{fig_hubbles_SALT2_Snoopy}.
}
\end{table*}
\renewcommand{\tabcolsep}{6pt}
\renewcommand{\arraystretch}{1}

\section{Discussion}
\label{sec_disc}

Tables~\ref{Tab_scatter_GPSubsample} and \ref{Tab_IntDispersion_PecVel} summarize the scatter in the Hubble residuals measured with the either the intrinsic scatter $\sigmaint$, the wRMS, or the RMS. We compute these both for our fiducial peculiar velocity uncertainty of $\sigmaVpec = 150$ km/s as well as the value $\sigmaVpec =250$ km/s used in \citet{scolnic18}.

While the formula for RMS in Eq.~\ref{eq_RMS} does not depend on the assumed value of $\sigmaVpec$ (see Appendix \ref{sec_WRMS_IntDisp}), the value of $\sigmaint$ is quite sensitive to the assumed value of $\sigmaVpec$. In particular, larger assumed values of $\sigmaVpec$ yield smaller inferred values of $\sigmaint$ (see columns 4 and 5 of Tables~\ref{Tab_scatter_GPSubsample} and \ref{Tab_IntDispersion_PecVel}).
The assumption of $\sigmaVpec =  150$ km/s in this work therefore yields a more conservative estimate of $\sigmaint$ compared with larger values of $\sigmaVpec$ because, in the latter case, most of the scatter in the Hubble residuals can be explained as arising solely from peculiar velocities. For instance, the Hubble residuals using only $H$-band \lcs{} from the GP (NIR max) method produce an intrinsic scatter of zero when assuming $\sigmaVpec =  250$ km/s.

We found that wRMS is less sensitive than $\sigmaint$ to the assumed value of $\sigmaVpec$, producing differences of $\sim 0.001$ mag between $\sigmaVpec =  150$ and $\sigmaVpec =  250$ km/s.

Of the three NIR methods used to derive distance moduli, the GP method at NIR max yields smaller RMS, wRMS, and intrinsic scatter in the Hubble residuals than the template and GP methods at $B$ max methods applied to the same \nAnyYJHKgp{} \snIa{} with data from any of the $\yjhk$ bands. When we combine the GP distance moduli for these same \snIa{} referenced to the NIR maxima, we find an RMS $=\rmsAnyYJHKgp$, wRMS $=\wrmsAnyYJHKgp $, and intrinsic scatter of $\sigmaint=\sigmaIntAnyYJHKgp $ mag. Using the GP method instead referenced to $B$-max for the same \snIa{} yields RMS = $\rmsAnyYJHKgpBmax$, wRMS = $\wrmsAnyYJHKgpAtTBmaxGPsubsample $, and $\sigmaint=\sigmaIntAnyYJHKgpAtTBmaxGPsubsample $ mag. The NIR maxima thus yield comparable dispersion in the Hubble residuals than $B$-max for each individual NIR band subset with the GP method (see Table~\ref{Tab_NIRmax_vs_Bmax}).

By comparison, when using the NIR template method referenced to $B$-max for these same \snIa, we find a larger value of RMS $=\rmsAnyYJHKTempGPsample$, wRMS $=\wrmsAnyYJHKTempGPsubsample$, and $\sigmaint = \sigmaIntAnyYJHKTempGPsubsample$ mag.

When we create the Hubble diagram using optical-only \lcs{} of the same \nAnyYJHKgp{} supernovae, we find RMS $=\rmsSALT$, wRMS $=\wrmsSALT$, and $\sigmaint = \sigmaIntSALT$ mag when using SALT2, and RMS $=\rmsSnoopy$, wRMS $=\wrmsSnoopy $, and $\sigmaint = \sigmaIntSnoopy$ mag with \snoopy{}.

Overall, as shown in Table~\ref{Tab_IntDispersion_PecVel1}, depending on the NIR $\yjhk$ subset, the NIR-only GP method yields a RMS in the Hubble residuals that is as much as $\sim \optminusnirRMSnsSmallestCombined $-$\optminusnirRMSnsLargestCombined \sigma$ {\it smaller} than the SALT2 and \snoopy{} fits using optical-only $BVR$ data. Furthermore, our ``any $\yjhk$'' set of \nAnyYJHKgp{} \snIa{} yields a RMS for our GP method at NIR max that is $\snoopyminusnirgpAnyYJHKrms $ mag smaller than \snoopy{} and $\saltminusnirgpAnyYJHKrms $ mag smaller that SALT2 applied to the corresponding $BVR$ data, again at the $\sim \saltminusnirgpAnyYJHKrmsNS{} \sigma$ level. We interpret the smaller  intrinsic scatter as additional evidence, at the \optminusnirsigma{} level, that NIR \snIa{} \lcs{} at NIR maximum, without \lc{} shape or dust corrections, are already {\it better} standard candles than optical-only \snIa{} \lcs{} referenced to $B$-max that apply such corrections. In addition, it is possible that NIR data or a combination of NIR and optical could yield even smaller intrinsic scatter if employing a method that applies \lc{} shape and dust corrections, for example, using a hierarchical Bayesian approach like \bayesn{} (\citealt{mandel09,mandel11}).

In Table~\ref{Tab_IntDispersion_PecVel1}, we note that the uncertainty on the difference in the dispersion estimates between any two methods has been computed conservatively.  The uncertainty of the dispersion of each individual method has been computed independently, and then the uncertainty in the difference is found by adding in quadrature, assuming the independence of the samples and therefore the individual uncertainties. However, this ignores the fact that the supernovae in our optical sample are exactly the same ones as those in our NIR sample.  Therefore, the actual peculiar velocity-distance errors must be the same in each sample (and not just the variance of these errors).  Because of this common component of scatter, the dispersion estimate for the optical Hubble Diagram is (positively) correlated with that for the NIR Hubble Diagram in each comparison.  The effect of this positive correlation is to reduce the variance in the differences in dispersion.  Using our estimates of $\sigma_\text{int}$, $\sigma_{\text{fit},s}$ and $\sigma_{\mu_\text{pec},s}$ for the sample and each method, we have run simulations to account for this correlation and quantify this effect.  For example, we find that the uncertainty in $\Delta$RMS for ''SNooPy - any $YJHK_s$'' is $\sim 30\%$ smaller than naive uncertainty assuming independent samples, resulting in a significance greater than $3 \sigma$.

For the Hubble diagrams created using just one of the $\yjhk$ bands, when using the GP method at NIR max, the \rmsGPNIRmaxOneBandSmallestBand{} band has the smallest scatter with a RMS of $\rmsGPNIRmaxOneBandSmallest $ mag. When using the template method, the \rmsTempOneBandSmallestBand{} band has also the smallest scatter with RMS $=\rmsTempOneBandSmallest$.

For every individual band and subset of NIR bands shown in Table~\ref{Tab_NIRmax_vs_Bmax}, the GP method yields \textit{smaller} intrinsic scatter when referencing to NIR max instead of $B$ max, by mean amounts of up to $\sim 0.03$ mag for the same \snIa{} at up to the $\sim$$1.0$$\sigma$ level. While not as statistically significant as the NIR vs. optical comparison in Table~\ref{Tab_IntDispersion_PecVel1}, we note that the NIR maxima yield smaller intrinsic scatter $\sigmaint$ and wRMS than $B$ max {\it for all subsets} of the NIR data except for $K_s$.\footnote{The only exception we tested is the $K_s$-band, which has only $\nKgp$ \snIa{} \lcs{}, where we find an essentially equivalent wRMS $\sim \wrmsNIRmaxK$ mag when referenced to either NIR max or $B$-max.} While NIR data at NIR max are {\it better} standard candles in comparison to optical data, they are also {\it at least as good or better} than when referenced to $B$-max. Therefore, future analyses should consider using $\tnirmaxx$ as the reference time instead of the traditional $\tbmax$.

\begin{figure*}
\centering
\includegraphics[width=15cm]{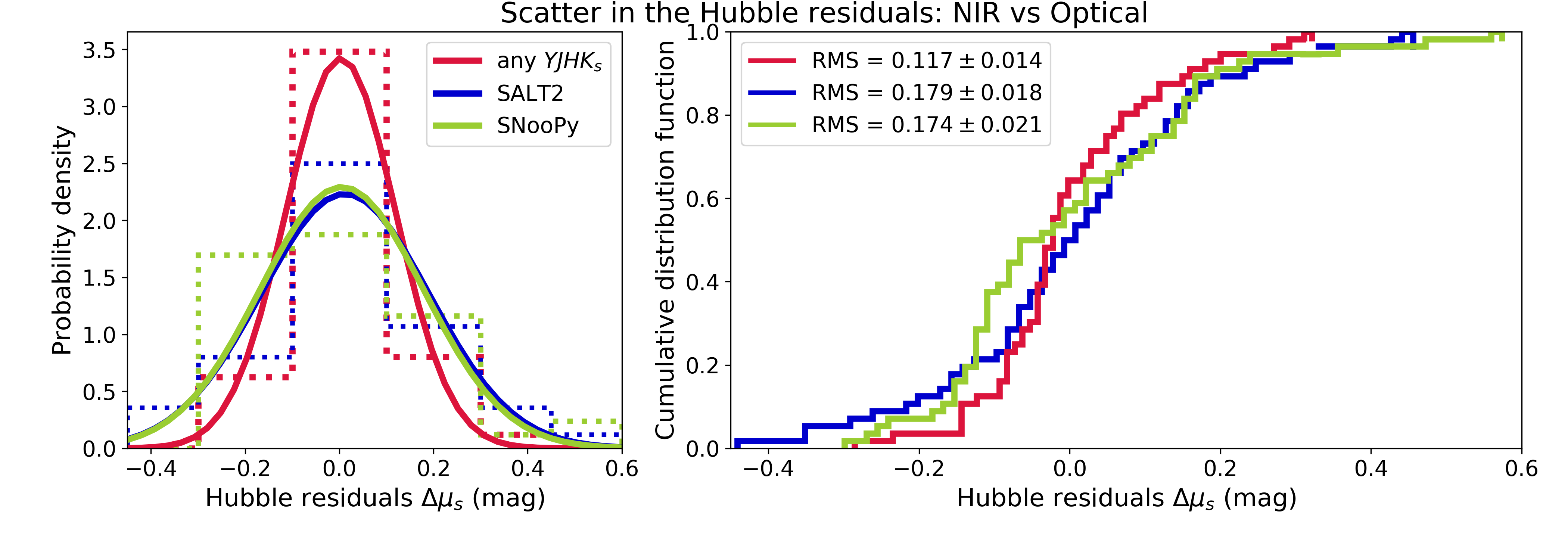}
\caption{Comparing the scatter in the Hubble residuals, $\{\Delta \mu_s \}$, as defined in Eq. (\ref{eq_HubbleResidual1}), using NIR and optical methods for the same \nAnyYJHKgp{} \snIa. The red, green, and blue colors correspond to the Hubble residuals from the ``any $YJHK_s$'' GP (NIR max) method (lower left panel of Fig.~\ref{fig_hubbles_GP_AnyYJHK}), SALT2 (lower left panel of Fig.~\ref{fig_hubbles_SALT2_Snoopy}), and \snoopy{} (lower right panel of Fig.~\ref{fig_hubbles_SALT2_Snoopy}), respectively. The left panel shows
histograms (dashed lines) and Gaussian approximation to the histograms (solid lines) of the Hubble residuals, where we observe that the  distribution of the NIR Hubble residuals (red) is \textit{narrower} than either optical distribution (blue or green). The right panel shows the corresponding cumulative probability distribution functions, where we also note that the slope of the NIR curve is \textit{steeper} than the optical curves, asymptotic to 1 at a smaller value of $\Delta \mu$, again indicating that the Hubble residual scatter is smaller in the NIR compared to the optical.
}
\label{Fig_histo_HubbleResiduals}
\end{figure*}

As an additional comparison between NIR and optical Hubble residuals, in Fig.~\ref{Fig_histo_HubbleResiduals}, we plot the histograms (dashed lines) with their Gaussian approximation (left panel), and the cumulative distribution function (right panel) for Hubble residuals using the same \nAnyYJHKgp{} \snIa{} used for the ``any \yjhk{}'' GP method at NIR max (lower left panel on Fig.~\ref{fig_hubbles_GP_AnyYJHK}), SALT2 (lower left panel of Fig.~\ref{fig_hubbles_SALT2_Snoopy}), and \snoopy{} (lower right panel of Fig.~\ref{fig_hubbles_SALT2_Snoopy}). The Gaussian approximations of the histograms in the left panel of Fig.~\ref{Fig_histo_HubbleResiduals} show that the Hubble residuals are more narrowly distributed for the NIR data (solid red curve) compared to both optical methods (solid green and blue), while in the right panel of Fig.~\ref{Fig_histo_HubbleResiduals} the cumulative distribution function curve for the NIR Hubble residuals is steeper than for either optical curve. Both approaches suggest that the Hubble residual scatter is smaller in the NIR compared to the optical. A larger sample of \snIa{} in the NIR would strengthen the evidence for this conclusion.

\section{Conclusions}
\label{sec_concl}

This work bolsters and confirms a growing body of evidence that \snIa{} in NIR are excellent standard candles in the $\yjhk$ bands in comparison to the optical $BVR$ bands.
Depending on the NIR data subset, our GP method performs $\snoopyminusnirgpAnyYJHKrmsNS $-$\saltminusnirgpAnyYJHKrmsNS \sigma$ better in RMS than either the SALT2 or \snoopy{} \lc{} fitters for the same \nAnyYJHKgp{} \snIa{} using $BVR$ data and applying \lc{} shape and color corrections. Using a suitable~subset of the existing low-redshift sample including \nsnIa{} spectroscopically normal \snIa{} with NIR data, $\yjhk$ photometry alone already provides a simple means to estimate accurate and precise host galaxy distances in each band, without the \lc{} shape or host galaxy dust reddening corrections required for optical data.

In this work, we employed a hierarchical Bayesian model, combined with a Gaussian process \lc{} fitter, to construct new mean NIR \lc{} templates. We then used these templates, along with Milky Way dust corrections, NIR $K$-corrections, and the measured spectroscopic redshifts (corrected for local velocity flows), and redshift independent distance information (e.g. Cepheids) for special cases, to estimate host galaxy distances and uncertainties and construct Hubble diagrams in each of the individual $\yjhk$ bands. When considering NIR-only methods, our GP method referenced to the time of NIR maximum yields slight smaller Hubble diagram intrinsic scatter and error weighted RMS than when referenced to $B$ max and significantly smaller intrinsic scatter compared to the template method.

Our approach is intermediate in complexity between earlier analyses by our group by \citealt{woodvasey08} and the \bayesn{} approach detailed in \citealt{mandel09,mandel11}. The \bayesn{} methodology presents a coherent, principled, hierarchical Bayesian model that takes into account the full correlation structure between all the input optical and NIR bandpasses, both in color and phase, in order to determine the posterior distributions for distance moduli $\mu$, host galaxy dust estimates $A_V$, and separate $R_V$ values for each supernova. Nevertheless, \bayesn{} is considerably more complex to implement than the simpler analysis methods in this work, which perform quite well for our sample of NIR data.

Compared to optical \lcs, NIR \snIa{} \lcs{} have a narrow luminosity distribution, and are less sensitive to host galaxy dust extinction. This could help to limit systematic galaxy distance errors that arise from the degeneracy between the intrinsic supernova colors and reddening of light by dust, that affects optical-only \snIa{} cosmology (\citealt{krisciunas04a,woodvasey08,folatelli10,BurnsEtal2011Snoopy,burns14,kattner12,mandel09,mandel11,mandel17,scolnic14,scolnic17}). Studies combining NIR and optical \snIa{} photometry have already shown that the addition of NIR data is an extremely promising way to break the degeneracy between intrinsic color and dust reddening, allowing distance estimates to become increasingly insensitive to the assumptions behind individual \lc{} fitting models (\citealt{mandel11,mandel14a}).

We have recently begun to augment the existing low-$z$ \snIa{} in NIR sample from the CfA, CSP, and other groups using the Hubble Space Telescope RAISIN program in Cycles 20 and 23 (\citealt{kirshner12,foley13a,kirshner14,kirshner14a}). In RAISIN1, we observed 23 \snIa{} at $z\sim0.35$ in the rest-frame NIR with WFC3/IR, followed by observations of 24 additional \snIa{} at $z\sim0.5$ for RAISIN2. Each of these HST NIR observations was accompanied by well-sampled ground based optical photometry from Pan-STARRS  (PS1; \citealt{rest14,DavidJones2018,scolnic18}) and the Dark Energy Survey (DES; \citealt{des16,des18,DES_Brout2018b}). Analysis of the RAISIN data will be presented in future work.

The evidence from this work further emphasizes the promise of NIR wavelength observations not only for the ongoing HST RAISIN project, but also for future space studies of cosmic acceleration and dark energy (\citealt{gehrels10,beaulieu10,astier11,hounsell17,riess18a}). Upcoming missions that could exploit nearby NIR data as a low-$z$ anchor include the Large Synoptic Survey Telescope (LSST; \citealt{ivezic08}), the NASA Wide-Field Infrared Survey Telescope (WFIRST-AFTA; \citealt{gehrels10,spergel15}), the European Space Agency's EUCLID mission \citep{beaulieu10,wallner17}, as well as the NASA James Webb Space Telescope (JWST; \citealt{clampin11,greenhouse16}).

NIR photometry can also augment our knowledge of the spectral energy distribution of \snIa{}, for example the Type Ia parametrized SALT2 model, which is currently poorly constrained at infrared wavelengths (\citealt{pierel18a,pierel18b}). This will dovetail nicely with the NIR capabilities of JWST and WFIRST and be useful for future \snIa{} surveys.

Methods such as \bayesn{} (\citealt{mandel09,mandel11}), \snoopy{}, and SALT2ext (\citealt{pierel18a,pierel18b}) that use empirical \lc{} fitters and provide host galaxy distance estimates using both optical and NIR data can be extended to obtain cosmological inferences and dark energy constraints using both low-$z$ and high-$z$ samples. Combining the growing low-redshift \snIa{} in NIR samples from the CfA, CSP, and other samples in the literature with higher redshift optical and NIR data sets will continue to lay the foundation for ongoing and future, ground and space-based, supernova cosmology experiments, which seek to further test whether dark energy is best described by Einstein's cosmological constant $\La$ or some other physical mechanism that varies on cosmic timescales.

\acknowledgments

{\centerline {\normalsize \bf Acknowledgments}}
The authors would like to thank Michael Wood-Vasey and Anja Weyant for help compiling redshifts for the nearby sample. We would also like to thank Dan Scolnic and Michael Foley for help determining the local flow corrected redshifts for our sample. We further thank Saurabh Jha, Suhail Dhawan, and Gautham Narayan for useful conversations. A.A. acknowledges support from the Harvard-Mexico fellowship sponsored by Fundaci\'on M\'exico en Harvard and CONACyT. A.S.F. acknowledges support from NSF Awards SES 1056580 and PHYS 1541160. R.P.K. acknowledges NSF Grants AST-1516854, AST 12-11196, AST 09-097303, and AST 06-06772. R.P.K, A.S.F, K.M. and A.A acknowledge Hubble Space Telescope Awards HST GO-14216 and HST GO-13046 supporting the HST RAISIN program. D.O.J. is supported by a Gordon and Betty Moore Foundation postdoctoral fellowship at the University of California, Santa Cruz. We gratefully made use of the NASA/IPAC Extragalactic Database (NED). The NASA/IPAC Extragalactic Database (NED) Is operated by the Jet Propulsion Laboratory, California Institute of Technology, under contract with NASA. This publication makes use of data products from the 2MASS Survey, funded by NASA and the US National Science Foundation (NSF). IAUC/CBET were very useful.

\appendix
\section{Gaussian process regression}
\label{Sec_GaussianProc}

Given the dataset of observations in an absolute magnitude NIR \lc{}, $(\mathbf{M}, \mathbf{t})$ for a given supernova, we want to use this information to estimate the latent absolute magnitudes $\latMVec^*$ at a grid of phases $\mathbf{t}^*$ described in Section \ref{sec_gaussProcess}. To do so, we define a Gaussian process with these data and variables.

To model the covariance  $\cov[\latMVec^*,{\latMVec^*}^\top]$ we choose the squared exponential GP kernel that is defined as
\begin{equation}\label{EqGPKernel}
k(t_i,t_j) = \sigmakGPkernel^2 \exp\left[ - \frac{(t_i-t_j)^2}{2l^2} \right],
\end{equation}
where $\sigmakGPkernel$ and $l$ are the GP kernel hyperparameters that we explain how to compute at the end of this section. We choose the GP kernel of Eq. (\ref{EqGPKernel}) because it is simple, produces smooth curves, and has the general properties we need to model the observed shapes of the NIR \lcs{}: for two phases very close to each other, $t_i \sim t_j$, their covariance is close to 1, and for distant phases, $t_i \ll t_j$, then $k(t_i, t_j) \sim 0$, such that they are almost uncorrelated.

We also take into account the uncertainty associated with each datum $M(t_i)$ in the variance $\sigma^2_M = \sigma^2_m + \sigma^2_A + \sigma^2_{\rm Kcorr} + \smupecNoS^2 $ [see Eq. (\ref{eq_err_AbsMag}) for details], by defining the total covariance between two data points as
\begin{equation}\label{eq_covGPscalar}
    \cov[M_i, M_j] = k(t_i,t_j) + \delta_{ij}(\sigma^2_m + \sigma^2_{\rm Kcorr}) + \smupecNoS^2  + \sigma^2_A
\end{equation}
where $ \delta_{ij}$ is the Kronecker delta function, and we assume that the measurement and $K$-correction errors are independent between two different $M_i$ and $M_j$, but that both the peculiar velocity-distance error and the Milky Way extinction error are not independent at different times because they are the same over the whole \lc{} in a single filter for a given supernova.
In matrix notation we can write Eq. (\ref{eq_covGPscalar}) for all the data $\mathbf{M}$ in a \lc{} as
\begin{equation}\label{eq_covGPmatrix}
    \cov[\mathbf{M},\mathbf{M}^\top] = \mathbf{K}(\mathbf{t}, \mathbf{t}) + \mathbf{W} + (\smupecNoS^2 + \sigma^2_A ) \Ivector \cdot \Ivector^\top,
\end{equation}
where  $\mathbf{K}(\mathbf{t}, \mathbf{t})$ is a square matrix with elements given by Eq. (\ref{EqGPKernel}),  $\mathbf{W}$ is a diagonal matrix of dimension $\nLCs \times \nLCs$ with elements given by
\begin{equation}
W_{ij} =  \delta_{ij} \left( \sigma^2_m + \sigma^2_{\rm Kcorr} \right),
\end{equation}
and $\Ivector$ is a vector of ones, so that the term $(\smupecNoS^2 + \sigma^2_A ) \Ivector \cdot \Ivector^\top$, is a square matrix of dimension $\nLCs \times \nLCs$ with elements all equal to $(\smupecNoS^2 + \sigma^2_A )$.

Following the standard GP formalism (e.g., Chapter 2 of \citet{Rasmussen_Williams_GP}),  we first write the joint distribution of the observed absolute magnitudes, $\mathbf{M}$, and latent absolute magnitudes, $\latMVec^*$, with a constant prior mean as
\begin{equation}\label{eq_jointdistGP}
\begin{bmatrix}
\mathbf{M} \\
\latMVec^*
\end{bmatrix}
\sim
\GaussDistr
\begin{pmatrix}
\begin{bmatrix}
a \Ivector \\
a \Ivector^*
\end{bmatrix}
,
\begin{bmatrix}
\mathbf{K}(\mathbf{t}, \mathbf{t}) + \mathbf{W} + (\smupecNoS^2 + \sigma^2_A) \Ivector \cdot \Ivector^\top & \qquad \mathbf{K}(\mathbf{t}, \mathbf{t}^*) \\
\mathbf{K}(\mathbf{t}^*, \mathbf{t}) & \qquad \mathbf{K}(\mathbf{t}^*, \mathbf{t}^*)
\end{bmatrix}
\end{pmatrix}
\end{equation}
where $\Ivector$ and $\Ivector^*$ are vectors of ones and of dimensions $\nLCs$ and $\nGPgrid$ respectively, and $a$ is a scalar that we assign the value of $-17.5, -17, -18$ and $-18$ mag for the $Y$, $J$, $H$ and $K_s$ bands, respectively. We assume these values of $a$ just for computational convenience in the GP fitting, and verified that the final templates are insensitive to these choices over a wide range of values for $a$.
The matrices $\mathbf{K}(\mathbf{t}, \mathbf{t})$, $\mathbf{K}(\mathbf{t}^*, \mathbf{t})$, $\mathbf{K}(\mathbf{t}, \mathbf{t}^*)$, and $\mathbf{K}(\mathbf{t}^*, \mathbf{t}^*)$, are of dimensions $\nLCs \times \nLCs $, $\nGPgrid \times \nLCs $, $\nLCs \times \nGPgrid $ and $\nGPgrid \times \nGPgrid $ respectively, with elements defined by Eq. (\ref{EqGPKernel}).

The conditional distribution of $\latMVec^*$ given $\mathbf{t}, \mathbf{t}^*$ and $\mathbf{M}$, can be written as
\begin{equation}\label{eq_condDist}
    \latMVec^* | \mathbf{t}, \mathbf{t}^*, \mathbf{M} \; \sim \; \GaussDistr \left( \postGPmeanVec, \postGPcovMatrix \right)
\end{equation}
where the posterior mean $\postGPmeanVec$ and posterior covariance $\postGPcovMatrix$ are given as
\begin{equation}\label{EqMeanGP}
\postGPmeanVec \equiv \mean[\latMVec^* | \mathbf{t}, \mathbf{t}^*, \mathbf{M}] = a\Ivector +  \mathbf{K}(\mathbf{t}^*, \mathbf{t})  \left[ \mathbf{K}(\mathbf{t}, \mathbf{t}) + \mathbf{W} + (\smupecNoS^2 + \sigma^2_A) \Ivector \cdot \Ivector^\top \right]^{-1} \; \left(\mathbf{M}-a\Ivector \right)
\end{equation}
\begin{equation}\label{EqCovarianceGP}
\postGPcovMatrix \equiv \cov \left[ \latMVec^*, {\latMVec^*}^{\top} | \: \mathbf{t}, \mathbf{t}^*, \mathbf{M} \right] = \mathbf{K}(\mathbf{t}^*, \mathbf{t}^*) -  \mathbf{K}(\mathbf{t}^*, \mathbf{t})  \left[ \mathbf{K}(\mathbf{t}, \mathbf{t}) + \mathbf{W} + (\smupecNoS^2 + \sigma^2_A) \Ivector \cdot \Ivector^\top \right]^{-1} \mathbf{K}(\mathbf{t}, \mathbf{t}^*)
\end{equation}
The final values we obtain from the GP regression are the vector $\postGPmeanVec$ and the matrix $\postGPcovMatrix$, that we estimate using Eqs. (\ref{EqMeanGP}) and (\ref{EqCovarianceGP}) respectively.

The coefficients $\sigmakGPkernel$ and $l$ in Eq. (\ref{EqGPKernel}) are called the \textit{hyperparameters} of the GP kernel that we determine by assuming that the \lcs{} for all the SN in a given NIR band are {independent} of each other, and that the GP hyperparameters describe the \textit{population} of the SN \lcs{} in a given band rather than each individual LC. With these assumptions, we can write the global marginal likelihood distribution
\begin{equation}\label{EqGlobalLike}
   p\Bigl( \{ \mathbf{M} \}_s | \{ \mathbf{t} \}_s, \sigmakGPkernel, l \Bigr) \; =\;
    \prod^{\nSNT}_{s=1} \GaussDistr \left(\mathbf{M}_s \; | \; a \, \Ivector_s, \;
    \mathbf{K}_s(\mathbf{t}_s, \mathbf{t}_s) + \mathbf{W}_s(\mathbf{t}_s,\mathbf{t}_s) + (\smupec^2 + \sigma^2_{A,s}) \Ivector_s \cdot \Ivector_s^\top
    \right),
\end{equation}
where the subindex $s$ refers to quantities for supernova $s$, $\nSNT$ is the number of \snIa{} used to construct the normalized LC template in a given NIR band, and ``$\{ \; \}_s$'' means the collection of values from all the $\nSNT$ \snIa{}.
To compute the MLE values for $(\sigmakGPkernel, l)$, we minimize the negative of the logarithm of Eq. (\ref{EqGlobalLike}), obtaining the values shown in Table \ref{tab_GP_hyperpars}.

\begin{table}
\begin{center}
\caption{Values of the GP hyperparameters}
\begin{tabular}{  c c  c}
\hline \hline \\[-0.3cm]
band & $l$ & $\sigmakGPkernel$  \\
\hline \\[-0.3cm]
$Y$ & 7.90 & 0.70 \\
$J$ & 7.02 & 0.95 \\
$H$ & 9.81 & 0.75 \\
$K_s$ & 8.19 & 0.55 \\[0.2cm]
\hline
\end{tabular}
\label{tab_GP_hyperpars}
\end{center}
\end{table}

\subsection{Normalization of the GP light curves}
\label{Sec_NormalizationGP_Appendix}

In Section \ref{Sec_NormalizationGP}, we explained that we are primarily interested in the \textit{shape} of the light curves. For this reason, after determining the posterior light curve described by ($\postGPmeanVec, \postGPcovMatrix$), we \textit{normalize} the \lc{} using $\tbmaxx$ as the reference time where the light curve will have a value of zero.

First, for computational convenience, we rewrite the linear transformation of Eq. (\ref{eq_defNormaLC}) as the matrix operation
\begin{equation}
     \normaLCVec = \mathbf{A} \latMVec^*
\end{equation}
where $\mathbf{A}$ is a $\nGPgrid \times \nGPgrid$ square matrix defined as $\mathbf{A} \equiv  \mathbf{I} - \mathbf{V}_k$, where $\mathbf{I}$ is the identity matrix, and $\mathbf{V}_k$ is a matrix containing only 1s in the $k$th column and zeros everywhere else, assuming that the $k$th element of $\mathbf{t}^*$ correspond to phase $t^*_k = \tbmaxx$.

We compute the mean of the normalized \lc{} as, $\meanNormaLCVec = \mean[\normaLCVec | \data] = \mathbf{A} \; \mean[\latMVec^* | \data] =  \mathbf{A} \postGPmeanVec $,
where $\data \equiv (\mathbf{t}, \mathbf{t}^*, \mathbf{M})$ is the conditional data in Eq (\ref{eq_condDist}). And the covariance is given by
\begin{equation}\label{eq_covNormaLCMatrix_1}
    \covNormaLCMatrix = \mathbf{A} \postGPcovMatrix \mathbf{A}^\top.
\end{equation}
From these expressions at $t^*_k = \tbmaxx$, the posterior mean and variance of the normalized \lcs{} are both identically zero:
\begin{equation}
    \mean [L_k | \data] = 0, \qquad \var [L_k, L_k | \data] = 0,
\end{equation}
which is required for self-consistency with the definition of the normalized \lc{}.

\section{Hierarchical Bayesian Model}
\label{Section_HierarBayes}

Using Bayes' theorem, applying the product rule for probability, and assuming conditional independence of the means of the normalized LCs, $\meanNormaLC_s$'s, with respect to the population mean and variance $(\hypermeanHBM, \hyperStdDevHBM^2)$, we can write the joint posterior distribution in our hierarchical model as
\begin{eqnarray}
\label{EqJointPosterior1}
p \left( \{ \AbsMagTilde \}, \hypermeanHBM, \hyperStdDevHBM | \{\meanNormaLC_s , \AbsMagTildeSigma \} \right) \propto
p \left( \hypermeanHBM, \hyperStdDevHBM  \right)  \times p \left( \{ \AbsMagTilde \} |\hypermeanHBM, \hyperStdDevHBM \right)  \times
p \left( \{ \meanNormaLC_s \} | \{ \AbsMagTilde \}, \{\AbsMagTildeSigma \} \right).
\end{eqnarray}
Inserting Eqs. (\ref{EqGaussianParams1}) and (\ref{EqGaussianHyperpars1}) into  Eq. (\ref{EqJointPosterior1}), we obtain,
\begin{eqnarray}
\label{EqJointPosterior2}
p \left( \{ \AbsMagTilde \}, \hypermeanHBM, \hyperStdDevHBM | \{ \meanNormaLC_s, \AbsMagTildeSigma \}  \right)  \propto
 p\left( {\hypermeanHBM}, \hyperStdDevHBM \right) \times \prod^{\nSNTast}_{s=1} \GaussDistr\left( \AbsMagTilde | {\hypermeanHBM}, \sigma^2_{\hypermeanHBM} \right) \times
  \prod^{\nSNTast}_{s=1} \GaussDistr \left( \meanNormaLC_s | \AbsMagTilde, \AbsMagTildeSigmaSq \right).
\end{eqnarray}
\noindent where $\nSNTast$ is the number of supernovae for which we have determined the best fitting function at phase $t^*$.
Note that since each \lc{} has a \textit{different} number of photometric data points over \textit{different} phase ranges, this implies that $\nSNTast$ is different for each phase  $t^*$.

For computation convenience, following \citet{GelmanEtal_BayesianDataAnalysis}, we decompose the joint posterior distribution using the product rule  as
\begin{eqnarray}
\label{EqJointPosteriorFinal}
p \left( \{ \AbsMagTilde \},  \hypermeanHBM,  \hyperStdDevHBM  |  \{ \meanNormaLC_s, \AbsMagTildeSigma \}  \right)  \propto
p \left( \{ \AbsMagTilde \} | \hypermeanHBM,  \hyperStdDevHBM, \{ \meanNormaLC_s, \AbsMagTildeSigma \} \right) \times
p \left( \hypermeanHBM |  \hyperStdDevHBM ,  \{ \meanNormaLC_s, \AbsMagTildeSigma \} \right) \times
p \left( \hyperStdDevHBM | \{ \meanNormaLC_s, \AbsMagTildeSigma \}  \right)\, ,
\end{eqnarray}
where the first factor to the right of the proportionality sign of Eq. (\ref{EqJointPosteriorFinal}) can be written for the supernova $s$ as
\begin{equation}\label{Eq_PosteriorDecomp1}
p \left( \AbsMagTilde | \hypermeanHBM,  \hyperStdDevHBM,  \meanNormaLC_s, \AbsMagTildeSigma \right) = \GaussDistr \left( \AbsMagTilde | \rho_s, R_s \right)\, ,
\end{equation}
where
\begin{align}
\rho_s & \equiv \frac{ \meanNormaLC_s / \AbsMagTildeSigmaSq  + \hypermeanHBM / \sigma^2_{\hypermeanHBM} }{ 1 / \AbsMagTildeSigmaSq  + 1 / \sigma^2_{\hypermeanHBM} }\, ,
\end{align}
and
\begin{align}
R_s & \equiv \frac{1}{ 1 / \AbsMagTildeSigmaSq  + 1 / \sigma^2_{\hypermeanHBM} }.
\end{align}
The middle factor to the right of the proportionality sign of Eq. (\ref{EqJointPosteriorFinal}) can be written as
\begin{equation}\label{Eq_PosteriorDecomp2}
p \left( \hypermeanHBM \, | \, \hyperStdDevHBM ,  \{ \meanNormaLC_s, \AbsMagTildeSigma \} \right) = \GaussDistr \left( \hypermeanHBM  \, | \, \hat{\hypermeanHBM}, R \right)\, ,
\end{equation}
where
\begin{align}\label{Eq_PosteriorDecomp3}
\hat{\hypermeanHBM} & \equiv \frac{ \sum^{\nSNTast}_{s=1}  \meanNormaLC_s \left( \AbsMagTildeSigmaSq + \sigma^2_{\hypermeanHBM} \right)^{-1} }{ \sum^{N_{\rm SN}}_{s=1} \left( \AbsMagTildeSigmaSq + \sigma^2_{\hypermeanHBM} \right)^{-1}  } \, ,
\end{align}
and
\begin{align}\label{Eq_PosteriorDecomp3a}
R^{-1} & \equiv \sum^{\nSNTast}_{s=1} \frac{1}{\AbsMagTildeSigmaSq + \sigma^2_{\hypermeanHBM} }.
\end{align}
Finally, the last term to the right of the proportionality sign can be written as
\begin{eqnarray}\label{Eq_PosteriorDecomp4}
p \left( \hyperStdDevHBM | \{ \meanNormaLC_s, \AbsMagTildeSigma \}  \right) \propto
R^{1/2} \prod^{\nSNTast}_{s=1}  \left( \AbsMagTildeSigmaSq  + \sigma^2_{\hypermeanHBM}  \right)^{-1/2}
\exp \left(\frac{-(\meanNormaLC_s - \hat{\hypermeanHBM})^2 }{2(\AbsMagTildeSigmaSq + \sigma^2_{\hypermeanHBM} )} \right) \, ,
\end{eqnarray}
where we are assuming a uniform prior distribution $p(\hyperStdDevHBM ) \propto 1$.

We use Eq. (\ref{EqJointPosteriorFinal}) combined with Eqs. (\ref{Eq_PosteriorDecomp1})-(\ref{Eq_PosteriorDecomp4}) to simultaneously determine the posterior best estimates of ($\{ \AbsMagTilde \}, {\hypermeanHBM}, \hyperStdDevHBM$) at phase $t^*$, given the data $\{ \meanNormaLC_s, \AbsMagTildeSigma \}$, following the computational procedure described in Appendix C.3, subsection {``Marginal and conditional simulation for the normal model''}, of \citet{GelmanEtal_BayesianDataAnalysis}. We use the R code presented there to build our R code to make the computations described in this work.

\section{RMS, weighted RMS, and the intrinsic scatter}
\label{sec_WRMS_IntDisp}

We use the RMS to quantify the scatter in the Hubble residuals because it is simple and straightforward to compute and compare with the Hubble residuals reported by other authors. The definition we use is
\begin{equation}\label{eq_RMS}
    \text{RMS} = \sqrt{N^{-1}_{\rm SN}  \left( \sum_{s=1}^{N_{\rm SN}} \Delta \mu^2_s \right)} \, ,
\end{equation}
where $N_{\rm SN}$ is the total number of \snIa{} in the Hubble diagram. We compute the uncertainty on RMS using bootstrap resampling.

To weight the root mean square (RMS) by the uncertainties in each SN distance modulus estimate in each NIR band, we compute the inverse-variance weighted root mean square (wRMS) of the residuals as
\begin{equation}\label{eq_wRMS}
\text{wRMS} = \sqrt{ \left( \sum_{s=1}^{N_{\rm SN}}  w_s \right)^{-1} \; \; \sum_{s=1}^{N_{\rm SN}}  w_s \, \Delta \mu^2_s } \, ,
\end{equation}
where $w_s \equiv 1/(\sigmafithat^2 + \sigmainthat^2 + \smupec^2)$ and $\Delta \mu_s$ is defined in Eq. (\ref{eq_HubbleResidual1}). We also compute the uncertainty on wRMS using bootstrap resampling.

We determine the \textit{intrinsic scatter}, $\sigmaint$, in the Hubble residual following the procedure described in Eqs.~(B.6)-(B.7) in Appendix B of \citet{BlondinEtal2011}. This dispersion tries to quantify the scatter due to intrinsic differences in the NIR \snIa{} absolute magnitudes only and \textit{not} due to the peculiar-velocity uncertainty of each SN.
The intrinsic scatter corresponds to the remaining dispersion observed in the Hubble-diagram residuals \textit{after} accounting for the uncertainty in distance modulus due to the peculiar-velocity uncertainty, $\smupec^2$, and the photometric errors $\{ \sigmafithat \}$. When comparing our notation to Eqs.~(B.6)-(B.7) of \citet{BlondinEtal2011}, note that where we use $\sigmafit$, $\sigmaint$ and $\smupec$, \citet{BlondinEtal2011} instead uses the notation $\sigma_{m,s}$, $\sigma_{\rm pred}$, and $\sigma_{{\rm pec},s}$, respectively.

\section{Covariance matrix $C_\mu$ of Hubble residuals}
\label{sec_covmatrix}

In this section we provide the numerical values for different cases of the covariance matrix $C_\mu$.

For the template method, we find the following values of the sample covariance matrix $C_\mu$ for the $\yjh$ bands:
\begin{equation}\label{MatrixCov_yjh_tm}
C_{\mu} =
\begin{pmatrix}
0.0227 & 0.0192 & 0.0167 \\
0.0192 & 0.0246 & 0.0201 \\
0.0167 & 0.0201 & 0.0211
\end{pmatrix}\, ,
\end{equation}
and for the $\jhk$ bands:
\begin{equation}\label{MatrixCov_jhk_tm}
C_{\mu} =
\begin{pmatrix}
0.0356 & 0.0276 & 0.0202 \\
0.0276 & 0.0317 & 0.0237 \\
0.0202 & 0.0237 & 0.0426
\end{pmatrix}.
\end{equation}

For the GP method, we find the following values for the sample covariance matrix for the $\yjh$ bands:
\begin{equation}\label{MatrixCov_yjh_GP}
C_{\mu} =
\begin{pmatrix}
0.0109 & 0.0110 & 0.0080 \\
0.0110 & 0.0133 & 0.0084 \\
0.0080 & 0.0084 & 0.0080
\end{pmatrix}\, ,
\end{equation}
and for the $\jhk$ bands:
\begin{equation}\label{MatrixCov_jhk_GP}
C_{\mu} =
\begin{pmatrix}
0.0279 & 0.0217 & 0.0213 \\
0.0217 & 0.0238 & 0.0192 \\
0.0213 & 0.0192 & 0.0283
\end{pmatrix}.
\end{equation}

\begin{figure*}
\centering
\begin{center}
\includegraphics[width=18cm]{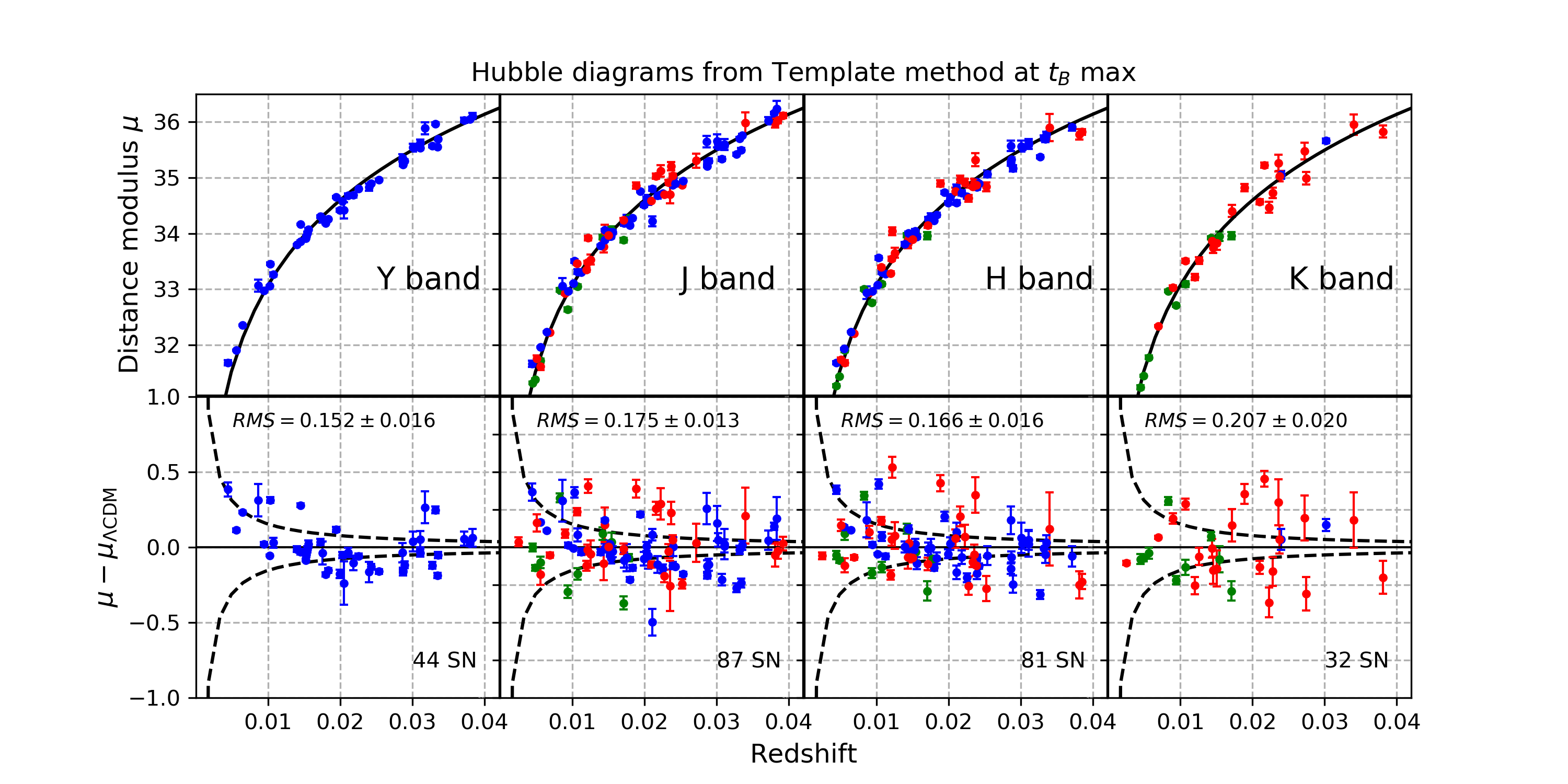}
\caption{
Individual $\yjhk$ Hubble diagrams (top row) and residuals (bottom row) using the template method. Points are color coded by NIR photometric data source, including the CfA (red; \citealt{woodvasey08,friedman15}), the CSP (blue; \citealt{krisciunas17}), and other data from the literature (green; see Table~\ref{Table_LC_params}). Note that only the CSP used a $Y$-band filter. In Table~\ref{tab_distanceMu_template}, we report the numerical values of the distance moduli shown in this figure. Table~\ref{Tab_IntDispersion_PecVel} shows the intrinsic scatter in the Hubble diagram.
}
\label{fig_hubbles_Template_Individuals}
\end{center}
\end{figure*}

\begin{figure*}
\begin{center}
\includegraphics[width=18cm]{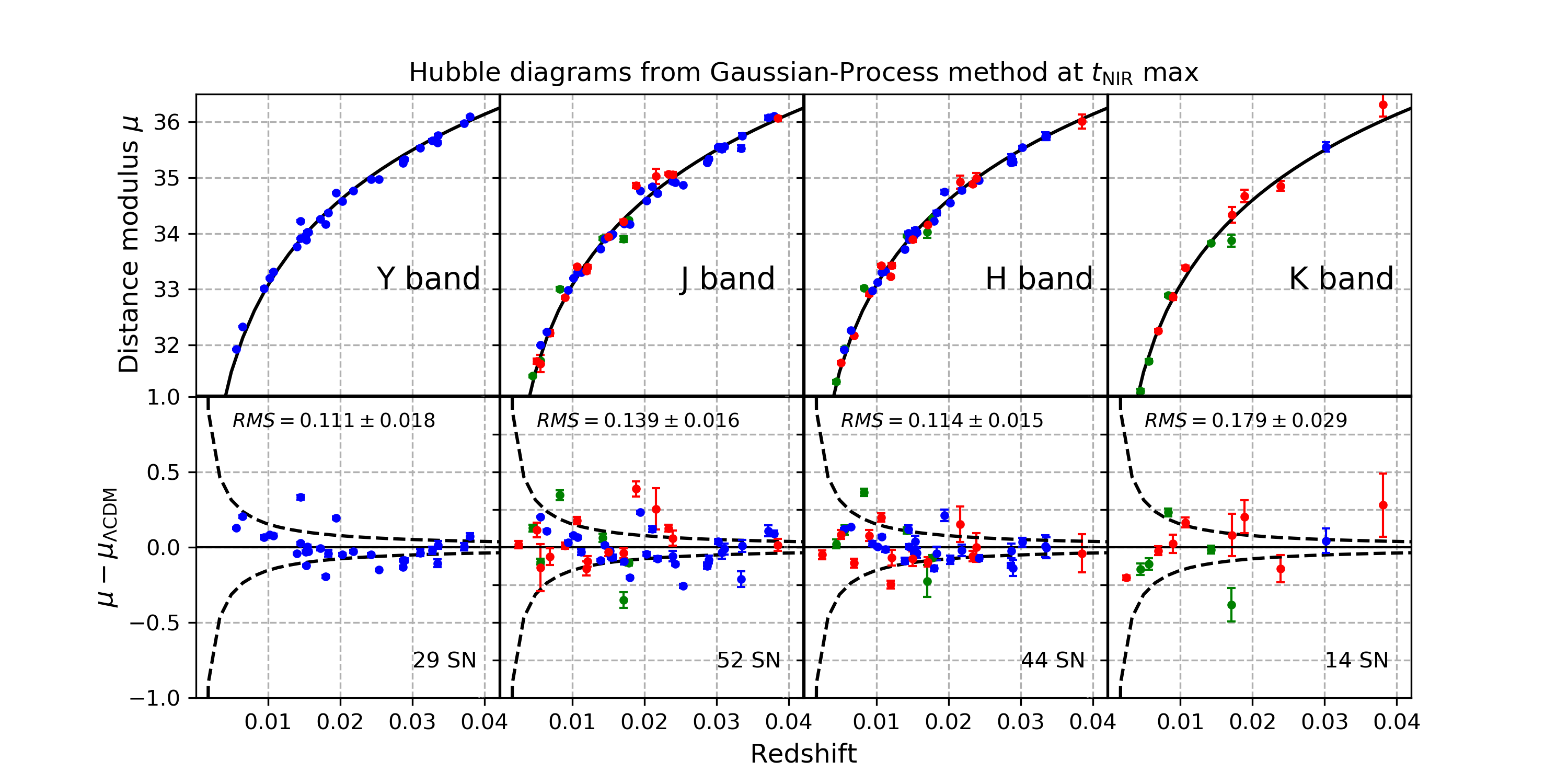}
\caption{Individual $\yjhk$ Hubble diagrams (top row) and residuals (bottom row) using the Gaussian-process method at NIR max. See the caption of Fig.~\ref{fig_hubbles_Template_Individuals}. In Table~\ref{tab_distanceMu_GaussianProc_NIRmax}, we report the numerical values of the distance moduli shown in this figure.
}
\label{fig_hubbles_GP_Individuals}
\end{center}
\end{figure*}

\begin{figure*}
\begin{center}
\includegraphics[width=18cm]{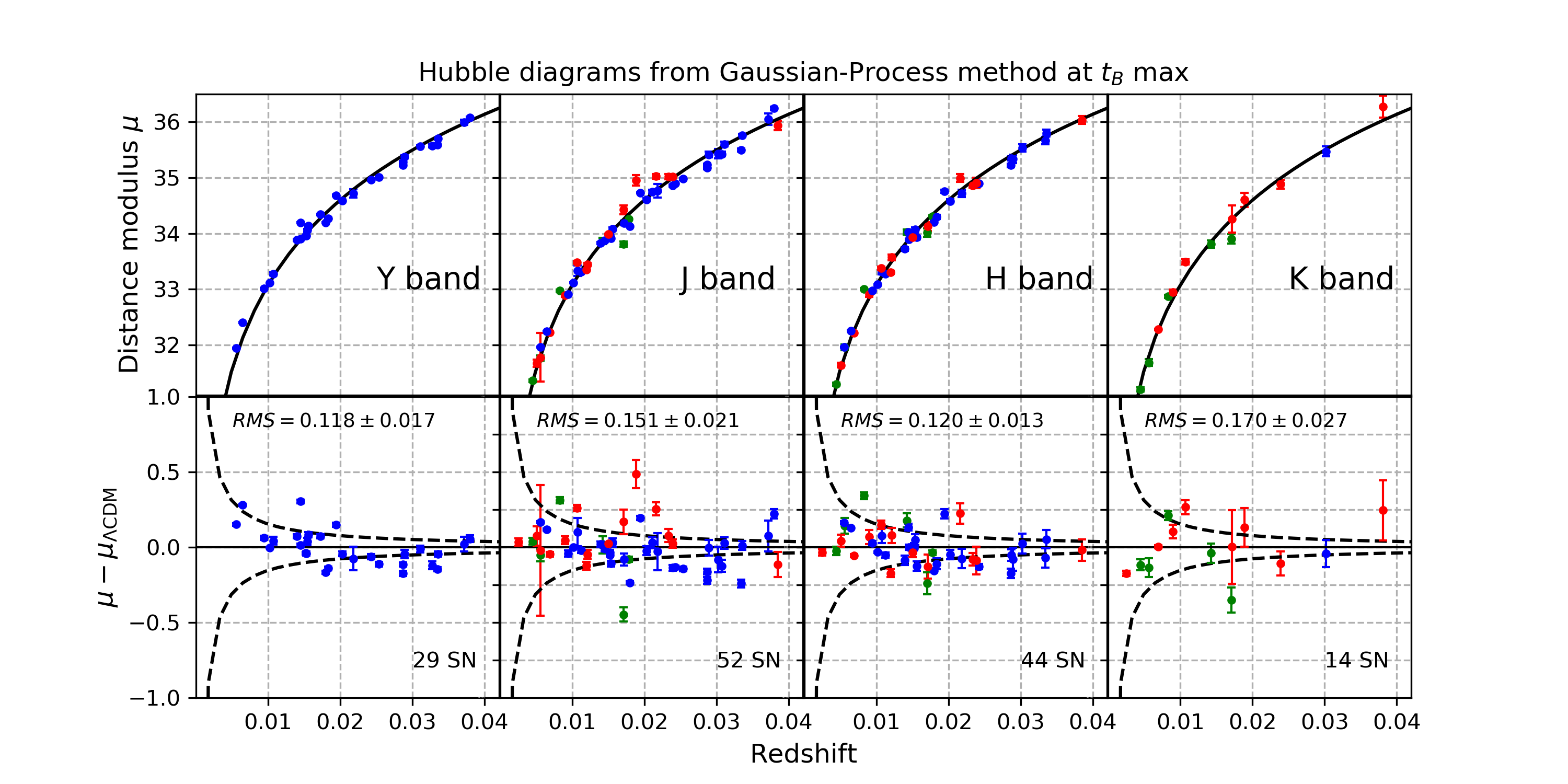}
\caption{
Individual $\yjhk$ Hubble diagrams (top row) and residuals (bottom row) using the Gaussian-process method at $B$ max. See the caption of Fig.~\ref{fig_hubbles_Template_Individuals}. In Table~\ref{tab_distanceMu_GaussianProc_Bmax}, we report the numerical values of the distance moduli shown in this figure.
}
\label{fig_hubbles_GP_Individuals_Bmax}
\end{center}
\end{figure*}

\newpage


\end{document}